\documentclass[aps,superscriptaddress,twocolumn,showpacs,preprintnumbers,amsmath,amssymb,showkeys]{revtex4}
\usepackage[cp1251]{inputenc}
\usepackage[english]{babel}
\usepackage{amssymb,amsmath}
\usepackage{amstext,textcomp} 
\usepackage[dvips]{hyperref}
\usepackage[mathscr]{eucal}
\usepackage{longtable}
\usepackage{graphicx}
\begin{document}
\title{Classical and Quantum Electrodynamics  Concept Based on Maxwell Equations' Symmetry}
\author{Dmitri Yerchuck (a),  Alla  Dovlatova (b), Andrey Alexandrov (b)\\
\textit{(a) Heat-Mass Transfer Institute of National Academy of Sciences of RB, Brovka Str.15, Minsk, 220072, dpy@tut.by\\ 
(b) M.V.Lomonosov Moscow State University, Moscow, 119899}}
\date{\today}%
\begin{abstract}The symmetry studies of Maxwell equations gave new insight on  the nature of electromagnetic (EM) field. Tey are reviewed in the work presented. It is drawing the attention on the following aspects.

EM-field has in general case quaternion structure,
consisting of four independent field constituents, which differ from each other by the parities under space inversion
and time reversal.  

Any complex relativistic field   has the gauge invariant
conserving quantity - two-component scalar or pseudoscalar value - complex charge.

There exists physical conserving
quantity, which is simultaneously invariant under  both Rainich dual and additional hyperbolic dual symmetry transformation of Maxwell
equations. It is spin in general case or  spirality in the corresponding geometry.

EM-field is  described  instead  of unobservable vector and scalar potentials by observable electric field 4-vector-function with the components $E_\alpha(\vec{r},t) = \{E_x(\vec{r},t), E_y(\vec{r},t), E_z(\vec{r},t), i \frac{c \rho_e(\vec{r},t)}{\lambda}\}$ and (or in the case of free EM-field) by means of magnetic field 4-vector-function $H_\mu(\vec{r},t) = \{H_x(\vec{r},t), H_y(\vec{r},t), H_z(\vec{r},t), i \frac{c \rho_m(\vec{r},t)}{\lambda}\}$, where  $i c \rho_e(\vec{r},t)$, $ic\rho_m(\vec{r},t)$ are the $j_4(\vec{r},t)$-component of 4-current density, corresponding to contribution of electric and magnetic component  of charge densities correspondingly, $\lambda$ is  conductivity, which for the case of EM-field propagation in vacuum is  $\lambda_v$ = $\frac{1}{120\pi}$ $(Ohm)^{-1}$.

Generalized
Maxwell equations for quaternion four-component EM-field are obtained.

Invariants for EM-field,
consisting of dually symmetric parts are found. 

The main postulate of quantum mechanics: "To any mechanical quantity can be set
up in the correspondence the Hermitian
matrix by quantization" was proved.

Canonical Dirac quantization method was developed in
two aspects. The first aspect is its application the only to observable quantities. The second aspect
is the realization along with well known time-local quantization of space-local quantization and
space-time-local quantization. It is also shown, that Coulomb field can be quantized in 1D and 2D
systems. 

New model of photons is proposed. The photons in quantized EM-field are main excitations  in oscillator structure of EM-field, which is equivalent to spin S = 1 "boson-atomic"  1D lattice structure, consisting of  the "atoms"  with zeroth rest mass.   The photons of the first kind and of the second kind represent themselves respectively neutral chargeless spin 1/2 EM-solitons and charged spinless EM-solitons of  Su-Schrieffer-Heeger family. 
\end{abstract}  

\maketitle                         
\section{Introduction}
Electromagnetic (EM) field is well studied, however some new theoretical and experimental results, discussed further, indicate on existence of earlier unknown new general properties, which extend the notion,  concerning EM-field nature itself. Basis equations in electrodynamics (ED) are well known Maxwell equations, which are unique in the sence, that they have the most rich symmetry among all fundamental equations of theoretical physics. The symmetry studies of Maxwell equations have started in 1892, when Heaviside \cite{Heaviside} has paid attention to the symmetry between electrical and magnetic quantities in given equations. So, the symmetry studies of EM-field have a long history, which was starting already in 19-th century and what is interesting, it is continuing hitherto. Let us give very briefly some moments in given studies.

 Mathematical formulation of given symmetry, consisting in  invariance of Maxwell equations for free EM-field under the duality transformations
\begin{equation} 
\label{eq1a}
\vec {E} \rightarrow \pm\vec {H}, \vec {H} \rightarrow \mp\vec {E},
\end{equation}
gave Larmor \cite{Larmor}. Duality transformations (\ref{eq1a}) are particular case of the more general dual transformations, established by Rainich \cite{Rainich}.  Dual transformations produce oneparametric abelian  group $U_1$, which is subgroup  of the group of chiral transformations.  Dual transformations correspond to irreducible representation of  the group of chiral transformations in particular case of quantum number $j = 1$ \cite{Tomilchick} and they are 
\begin{equation} 
\label{eq1b}
\begin{split}
\raisetag{40pt}                                                    
\vec {E'} \rightarrow \vec {E} \cos\theta + \vec {H} \sin\theta\\
\vec {H'} \rightarrow \vec {H} \cos\theta - \vec {E} \sin\theta,
\end{split}
\end{equation}
where parameter $\theta$ is arbitrary continuous variable,
$\theta \in [0,2\pi]$. 
Then it has been found, that maximal local symmetry group of Maxwell equations with sources is fifteen-parametric conform group $C(1,3)$ \cite{Bateman}.  $C(1,3)$ group includes linear 10-parametric Poincare group and linear scale transformations, which produce together the maximal local group of linear  transformations of coordinates and time, under which Maxwell equations with sources are invariant. It includes also local nonlinear conform transformations according to mapping
\begin{equation}
\label{eq1z}
 x_{\mu}: x_{\mu} \to x'_{\mu} = \frac{x_{\mu} - b_{\mu} x_{\nu} x^{\nu}}{1 - 2b_{\nu}x^{\nu} +  x_{\rho}x^{\rho}b_{\tau}b^{\tau}}, 
\end{equation}
where $b_{\mu}$ are parameters, which belongs to the set of real numbers, $b_{\mu} \in R$. 
At the same time the maximal local symmetry group of Maxwell equations without sources is 16-parametric group, which is direct product of $C(1,3)$ group and one-parametric group $U_1$ of Rainich dual transformations, that is $C(1,3) \otimes U_1$. It was proved relatively recently (1967) \cite{Ibragimov}. Some time later (1979) it was established the existance of nonlocal symmetry of Maxwell equations under transformations of 23-dimensional Lee algebra, which is direct product of $C(1,3)$ group and $A_8$-algebra, that is $C(1,3) \otimes A_8$ \cite{Fushchich}. Lee algebra $A_8$ is isomorphous to Lee algebra $U(2) \otimes U(2)$.
Symmetry  studies of Maxwellian electrodynamics were continued in the works \cite{Yearchuck_Alexandrov_Dovlatova} and \cite{Dovlatova_Yerchuck}, which will be reviewed
in present paper. So, additional gauge invariance of Maxwell equations has been established in \cite{Yearchuck_Alexandrov_Dovlatova}, indicating on the  presence of new independent  observable  charateristics of EM-field - the scalar charge function, that, in its turn, speaks,  minimum, on complex character of EM-field. 

The role of symmetry of Maxwell equations under group $U_1$ of Rainich dual transformations was studied \cite{Dovlatova_Yerchuck}, where additional symmetry of Maxwell equations under hyperbolic dual transformations were found and studied.
 It was also shown, that dual symmetry of Maxwell equations leads to conclusion on compound character of EM-field, consisting in that,  that vector quantities, which characterize EM-field are in general case the four vector-component objects - quaternions - with different time t- and space P-parities. We have to say,  that the idea, that vector-functions $\vec {E}(\vec{r},t)$, $\vec {H}(\vec{r},t)$, $\vec {D}(\vec{r},t)$, $\vec {B}(\vec{r},t)$, which characterize EM-field, are compound quantities and that they include both gradient and solenoidal parts, that is uneven and even parts relatively space inversion  transformation, was put forward earlier in \cite{Tomilchick}. There is also theoretical assumption in \cite{Berezin}, where along with usual choice, that is, that electric field  $\vec {E}$ is polar vector, magnetic field  $\vec {H}$ is axial vector, the alternative choice is provided. 
 At the same time given ideas were unknown (and they are almost unknown at present) to general circle  of the researchers in the field. It was also true for us. To conclusion on compound character of EM-field, in particular, to  conclusion, that along with part with polar electrical vector symmetry there is part with the axial symmetry of electrical vector characteristics, which are responsible for optical infrared (IR) resonance absorption, Raman scattering,  we came from analysis of experimental results independently, see, for instance the works   \cite{Yearchuck_Doklady}, \cite{Yearchuck_Yerchak}, \cite{Yearchuck_PL} and  \cite{D_Yearchuck_A_Dovlatova}, where the experimental confirmation of the phenomenon of dual symmetry of EM-field vector-functions under space inversion is presented (some details are given also in Section II). Moreover, the conclusion was arisen, that the exhibition of polar or axial properties of EM-field vector-functions depends on the experimental situation.  Given conclusion was to some extent unexpected, since it is considered always in voluminous literature, that electric field vector has to be the only polar vector and magnetic field vector has to be the only axial vector. Given possibility corresponds to known field theory consideration, in which is concluded, that 4-vector of EM-field potential has to be polar and t-even 4-vector \cite{Akhiezer} and, consequently, three-dimensional electric vector quantities have to be polar vectors. It has to remark, that uneven space inversion parity of electric vector quantities has very good experimental confirmation too. For instance, all electric charge transport phenomena can be well described in the suggestion, that electric field vector is polar vector. 
 At the same time mathematically the  idea of  compound character of EM-field in the equations of dual electrodynamics in \cite{Tomilchick}  is presented the only in implicit form.

The conclusions, which follow from experimental data and above cited theoretical ideas and assumptions  were mathematically proved more carefully in\cite{Yearchuck_Alexandrov_Dovlatova} and \cite{Dovlatova_Yerchuck}, that will be realized in given paper.  So, the first task of presented paper is to review  the results of the study of the dual symmetry  of classical EM-field, leading to existence of electrical and magnetic vector quantities with both even and uneven parities under improper rotations.

The second task for classical electrodynamics is concerned of more detailed theoretical substantiation for the existence of  two type of scalar and pseudoscalar quantities - electric and magnetic charge functions of EM-field on the one hand and description of the EM-field in condensed matter containing along with the particles with electric charge the particles with magnetic charge, including dually charged particles,  on the other hand. In fact the proof of existence of independent observable scalar characteristics, that is charge scalar (pseudoscalar) function will allow to describe the EM-field the only within the framework of really observable quantities and the descriptions by means of unobservable quantities - vector and scalar potentials to  refer to  the history.
So, the aim of presented paper is the review  of field theory proof for given general properties of  EM-field.

It represents also the interest the  quantum description of EM-field. The well known now
simple formula $E = h \nu$, $h = const > 0$, proposed right at the beginning of the 20th century by Planck \cite{Planck} and  by Einstein \cite{Einstein} became epoch-making and  a real symbol of the substantial progress in the science. The interpretation, given by    Einstein indicates
straight on real existence of light quanta of frequency $\nu$  with the total energy $E$, which in its turn has led to 
 a new understanding of the nature of the electromagnetic field.  In fact, it was the indication on oscillatory, that is, discrete in space structure of EM-field, which in the same time is the set  of 
physical objects  strongly connected to some periodic with
time period $T = 1/\nu$ process, being to be intrinsic for given  objects, at that  Lorentz invariant
product $E T$ is equal to $h$.  Similar interpretation of 
De Broglie’s relationship \cite{Broglie}  leads to conclusion on quantum nature of the charges.
Although there are the great  achievements in quantum theory, the great challenge to give  an adequate description of  light quanta, which were called by Lewis photons \cite{Lewis}, 
still has not brought satisfactory results. Even  Einstein himself  has recognized \cite{Speziali},
 that ”the whole fifty years of conscious brooding have not brought me nearer to the answer
to the question what are light quanta”, and now, more than half a century later, theoretical physics still
needs progress to present a satisfactory explanation of the   photon nature. We consider
the corresponding theoretically directed efforts to be  necessary and even urgent in view of requirements of the modern science  and engineering, in particular, in
connection with rash progress in nanotechnogy. The structure of photons will be proposed.

The paper is organized in the following way. In Sec.2, Rainich dual symmetry of Maxwell equations is analysed and its experimental confirmation by comparison with literature data is given. In Sec.3, the algebraic properties of 
EM-field functions are summarized. In Sec.4, theoretical foundation of the existence of dually charged particles or quasiparticles is given. Comparison with experiment is also presented.  In Sec.5, the quaternion structure of EM-field is argued in details. In Sec.6 the symmetry of mechanics and electrodynamics differential  equations in application to  quantum theory is analysed. In Sec.7, the cavity classical and quantum electrodynamics is considered by taking into account the dual symmetry of  EM-field,  the connection between gauge invariance of  EM-field and analicity of its  vector-functions is considered. In Sec.8, the effect of spin-charge separation in quantized EM-field structure is described and structure of photons is proposed. In Sec.9, the conclusions are formulated.

\section{Rainich dual symmetry oF Maxwell equations and its experimental confirmation}
Rainich dual symmetry oF Maxwell equations was analysed in \cite{Dovlatova_Yerchuck}, and it will be reviewed in details. It is accentuated in 
\cite{Dovlatova_Yerchuck}, that dual transformations (\ref{eq1b}) have fundamental significance for the symmetry properties of vector-functions $\vec {E}, \vec {H}$. Dual symmetry  of Maxwell equations indicates, that both electric and magnetic vector force characteristics $\vec {E}$ and $\vec {H}$ of EM-field have to be peer in the sense of their symmetry properties including the symmetry properties under improper rotations. Really,  the expression (\ref{eq1b}) will be mathematically correct, if vector-functions $\vec {E}\cos\theta, \vec {H}\sin\theta$ (and correspondingly $\vec {H} \cos\theta,  \vec {E} \sin\theta$) in upper (bottom) line will   have the same symmetry, that is both polar or axial ones. It is not evident in given representation, if parameter $\theta$ is arbitrary variable. Let us clarify given situation. 
In matrix form the transformations (\ref{eq1b}) are
\begin{equation}
\label{eq1bc}
 \left[\begin{array} {*{20}c}  \vec {E'} \\ \vec {H'} \end{array}\right] = \left[\begin{array} {*{20}c} \cos\theta&\sin\theta  \\-\sin\theta&\cos\theta \end{array}\right]\left[\begin{array} {*{20}c}  \vec {E} \\ \vec {H} \end{array}\right].
\end{equation}
 At the same time to any complex number $a + ib$ can be set up in conformity the $[2 \times 2]$-matrix according to biective mapping
\begin{equation}
\label{eq1abcd}
 f : a + ib \to \left[\begin{array} {*{20}c} a&-b  \\ b&a \end{array}\right].
\end{equation}
Bijectivity of mapping (\ref{eq1abcd}) indicates on the existence of inverse mapping, that is to any matrix, which has the structure, given by right side in relation (\ref{eq1abcd}), corresponds the complex number, determined by left side. 
Consequently, we have
\begin{equation}
\label{eq1bcd}
 \left[\begin{array} {*{20}c}  \vec {E'} \\ \vec {H'} \end{array}\right] = e^ {-i\theta}\left[\begin{array} {*{20}c}  \vec {E} \\ \vec {H} \end{array}\right],
\end{equation}
that is 
 \begin{equation} 
\label{eq1c}
\begin{split}
\raisetag{40pt}                                                    
&\vec {E'} = \vec {E} \cos\theta - i\vec {E} \sin\theta\\
&\vec {H'} = \vec {H} \cos\theta - i\vec {H} \sin\theta.
\end{split}
\end{equation}
  It means, that to real planes, which are determined  by the vectors  $\vec {E}$ and $\vec {H}$  can be set in conformity the complex planes for the vectors $\vec {E'}$ and $\vec {H'}$.
                    
It is clear, that    the vectors $\vec {E'}$ and $\vec {H'}$ will consist both,  of even  and uneven components under improper rotations. So, it is evident from the relation (\ref{eq1c}), that if one component of, for instance, $\vec {E'}$ will be even under reflection in the plane, situated transversely to abscissa-axis, then the second component will be uneven. 

Therefore, dual transformation symmetry of Maxwell equations, established by Rainich \cite{Rainich}, indicates simultaneously on both complex nature of EM-field in general case, and that both electric and magnetic fields are consisting in general case of the components with various parity under improper rotations. In other words, the rotation of  EM-field vectors, determined by (\ref{eq1bcd}) is accompanied by appearance of axial component for initially polar electric field and polar component for initially axial magnetic field. 

It has to be taken into account,  that in the case $\theta = 0$ we have well known electrodynamics with odd parity of electric field and even parity of magnetic field. In given case both the vectors are real vectors. Then, by the applications  of the method, in which the real EM-field vector and scalar characteristics  are replaced by corresponding  complex  quantities [given method  was considered earlier to be a formal, but  convenient mathematical technique], is always the addition of corresponding complex conjugate  quantities was used. It is correct for  the case $\theta = 0$ and will be incorrect in the case  $\theta \neq 0$, since, although  adding of corresponding complex conjugate  quantities allows to obtain Re$\vec {E}$ and Re$\vec {H}$, however significant information, which is given by Im$\vec {E}$ and Im$\vec {H}$ remains  to be unclaimed.

It is easily to show, that in the case $\theta \neq 0$ the following relationship is taking place
\begin{equation}
\label{eq1bcde}
 \left[ \vec {E}^2 - \vec {H}^2 + 2i(\vec {E}\vec {H})\right]  e^ {-2i\theta} =inv,
\end{equation}
that is, we have at fixed parameter $\theta \neq 0$  two real EM-field invariants
\begin{equation} 
\label{eq1cde}
\begin{split}
\raisetag{40pt}                                                    
&(\vec {E}^2 - \vec {H}^2) \cos 2\theta + 2(\vec {E}\vec {H}) \sin 2\theta = I'_1 = inv \\
&2(\vec {E}\vec {H}) \cos 2\theta  - (\vec {E}^2 - \vec {H}^2) \sin 2\theta = I'_2 = inv.
\end{split}
\end{equation}
  It follows  from  relation (\ref{eq1cde}) that, in particular,  at $\theta = 0$ we have well known EM-field invariants
\begin{equation} 
\label{eq1cdef}
\begin{split}
\raisetag{40pt}                                                    
&(\vec {E}^2 - \vec {H}^2)  = I_1 = inv \\
&(\vec {E}\vec {H}) = I_2 = inv.
\end{split}
\end{equation}
  It is interesting, that at $\theta =  \frac {\pi}{4}$  and at $\theta = \frac {\pi}{2}$ EM-field invariants are determined by the same relation (\ref{eq1cdef}) and by arbitrary  $\theta$ we have two linearly independent combinations of given known invariants. At the same time is it evident, that the following relationship takes place
\begin{equation} 
\label{eq1cdeff}
\begin{split}
\raisetag{40pt}                                                    
 {I'_1}^2 + {I'_2}^2 = {I_1}^2 + 4{I_2}^2 = K = inv. 
\end{split}
\end{equation}
It means, that quantity $K$ is not depending on $\theta$, that is $K$ is invariant of dual transformations. The invariance of quantity $K = {I_1}^2 + 4{I_2}^2$ under dual transformations was found earlier \cite{Tomilchick}, where the physical meaning of given invariance was also explained - the  quantity, equaled to $\frac{1}{4} K$, is square of 4-impulse density.

The examples, where the dual symmetry of  EM-field becomes apparent experimentally, are well known.  For instance, the equality of magnetic and electric energy values  for free electromagnetic wave means the invariance of total energy of free electromagnetic wave under Larmor  transformation on the one hand and  the invariance of magnetic and electric energy components, being to be taken separately, under the same  transformations on the second hand. Quite similar situation takes place for  EM-field  in the cavity or in LC-tank. 

Subsequent extension of dual symmetry for the EM-field with sources leads also to requirement of the existence of two type of other physical quantities - two type of charges and currents and two type of  intrinsic moments of the particles or the absorbing (dispersive) centers in condensed matter.  Relatively recently some new theoretical and experimental results were obtained, which concern the dual symmetry of EM-field in the matter with two type of charges and intrinsic moments of quasiparticles. Let us review given results in some details. 
So, the operator equation, describing the optical transition dynamics, has been obtained by using of transition operator method \cite{Yearchuck_Doklady}, \cite{D_Yearchuck_A_Dovlatova}. It has been shown, that given equation is operator equivalent to Landau-Lifshitz (L-L) equation \cite{Landau} in its 
difference-differential form, which takes usual differential form in continuum limit. In view of isomorphism of algebras of 
transition operators $\hat {\vec {\sigma }}_k, k \in N $ and components of the spin $S = 1/2$, the symmetry of Bloch vector $\vec {P}$ under improper rotations  and  physical meaning of all its components in optical Bloch equations  have been established. Let us remember, that optical Bloch equations are the essence of gyroscopic model for spectroscopic transitions proposed for the first time formally \cite{Abragam} by
F.Bloch \cite{Bloch}. It was concluded in \cite{Yearchuck_Doklady}, that Bloch vector is electrical dipole moment, however it is axial vector. In particular, Bloch vector is electrical intrinsic moment of (quasi) particles in condensed matter, which is proportional to spin moment  (electrical "spin" moment), which was predicted by Dirac already in 1928 \cite{Dirac1928} and which  was discovered experimentally in \cite{Yearchuck_PL}.  
The nature of the second vector, that is, $\vec {E}$, entering into
optical Bloch equations, was also clarified. It is usual electric field vector, and it is still and all axial vector.  Given conclusions and the model used upon the whole were experimentally confirmed in \cite{Yearchuck_Yerchak} and \cite{Yearchuck_PL}, in which there was reported on the discoveries of ferroelectric  and antiferroelectric  spin wave resonances. They were predicted earlier \cite{Yearchuck_Doklady} on the base of  semiclassical models  for the systems EM-field plus the chain of electrical dipole moments, interacting between themselves, correspondingly, by a mechanism like to Heisenberg exchange or antiferroelectrically. It was found, that the mechanism of interaction of electrical dipole moments like to Heisenberg exchange is  realized in carbon 1D chains in  carbynoids by means of formation of  the lattice of spin-Peierls $\pi$-solitons, being to be electrical dipole moment carriers, in $\pi$-subsystem of valence bonds. Simultaneously in $\sigma$-subsystem of valence bonds Su-Schrieffer-Heeger $\sigma$-polarons produce also the lattice, consisting of  two Su-Schrieffer-Heeger $\sigma$-soliton sublattices,  resulting in realization of antiferroelectrical interaction between intrinsic electrical moments (electrical "spin" moments) of $\sigma$-polarons. Especially interesting, that in \cite{Yearchuck_PL} was experimentally proved, that really purely imaginary electrical "spin" moments, in full correspondence with Dirac discovery \cite{Dirac1928}, are responsible for the phenomenon of antiferroelectric  spin wave resonance in the samples studied. It seems to be understandable, that in general case electric dipole moment can be represented by the complex vector-function, including along with imaginary part, which corresponds to electrical intrinsic moment, the real part, corresponding  to orbital  motion.  
Therefore, it was shown, that the carbon in 1D allotropic form possesses by very interesting quantum physical properties. Let us remark, that strict experimental proof of the existence of carbon in 1D allotropic form was obtained by electron spin resonance method, that is by one of the most powerful [and, consequently, the most reliable] experimental methods for the studies of quantum physics phenomena \cite{Ertchak_J_Physics_Condensed_Matter}. Physical explanation of the possipility of existence of carbon in 1D allotropic form is rather simple. The Moscow chemicists group [Kudryavtsev Yu.P and coauthors, see for details, for instance, the book \cite{Heimann}] are succeeded in remoteness of individual carbon chains in graphene plane from  each other by means of placing in interchain space of simple molecules, which do not produce any chemical bonds with main carbon chains. Given conclusion was strictly proved in \cite{Ertchak_J_Physics_Condensed_Matter},  and it is  the consequence of symmetry requirements for the chain, where  the topological solitons are produced and propagated. In fact, the  structure of carbynoids has the resemblance with the structure of the other well known carbon allotropic form - graphite, in which individual graphene planes are remoted from each other and do not produce any chemical bonds between themselves and with  simple molecules like to $O_2$ in interplane space.

\section{Algebraic properties of EM-field}

Some useful results from algebra of the complex numbers were summarized in the paper \cite{Dovlatova_Yerchuck}. They will be
reproduced here and will be completed by some results from algebra of the
hypercomplex numbers.

 The numbers $1$ and $i$ are usually used to be basis of the linear space of complex numbers over the field of real numbers. At the same time to any complex number $a + ib$ can be set up in conformity the $[2 \times 2]$-matrix aforegoig (see relation (\ref{eq1abcd})).
 The matrices
\begin{equation}
\label{eq61}
\left[\begin{array} {*{20}c} 1&0  \\ 0&1 \end{array}\right], 
\left[\begin{array} {*{20}c} 0&-i  \\ i&0 \end{array}\right]
\end{equation}
produce basis for complex numbers $\{a + ib\}$, $a,b \in R$  in the linear space  of $[2 \times 2]$-matrices, defined over the field of real numbers. It is convenient often to define the space of complex numbers over the group of real positive numbers, then the dimensionality of the matrices and basis has to be duplicated, since to two unities - positive $1$ and negative $-1$ 
can be set up in conformity  the $[2 \times 2]$-matrices according to biective mapping
\begin{equation}
\label{eq62}
 \xi : 1 \to \left[\begin{array} {*{20}c} 1&0  \\ 0&1  \end{array}\right], -1 \to \left[\begin{array} {*{20}c} 0&1  \\ 1&0 \end{array}\right],
\end{equation}
 which allow to recreate the  operations with negative numbers without recourse of negative numbers themselves. Consequently, the following $[4 \times 4]$-matrices, so called [0,1]-matrices, can be basis of complex numbers
\begin{equation}
\label{eq63}
\begin{split}
\raisetag{40pt}
&\zeta : 1 \to [e_1]=\left[\begin{array} {*{20}c}1&0&0&0 \\  0&1&0&0  \\ 0&0&1&0 \\  0&0&0&1 \end{array}\right], \\
&i \to  [e_2]=\left[\begin{array} {*{20}c}  0&1&0&0  \\ 0&0&1&0 \\0&0&0&1\\ 1&0&0&0 \end{array}\right],\\
&-1 \to [e_3]=\left[\begin{array} {*{20}c} 0&0&1&0 \\0&0&0&1\\ 1&0&0&0\\0&1&0&0  \end{array}\right],\\
&-i \to  [e_4]=\left[\begin{array} {*{20}c}  0&0&0&1\\ 1&0&0&0\\0&1&0&0  \\ 0&0&1&0 \end{array}\right].
\end{split}
\end{equation}
The choise of basis is ambiguous. Any four  $[4 \times 4]$  [0,1]-matrices, which satisfy the rules of cyclic recurrence
\begin{equation}
\label{eq64}
i^1 = i, i^2 = -1, i^3= -i, i^4 = 1
\end{equation}
 can be basis of complex numbers.
 In particular, the following 
$[4 \times 4]$  [0,1]-matrices
\begin{equation}
\label{eq65}
\begin{split}
\raisetag{40pt}
&[e'_1]=\left[\begin{array} {*{20}c}1&0&0&0 \\  0&1&0&0  \\ 0&0&1&0 \\  0&0&0&1\end{array}\right], [e'_2]=\left[\begin{array} {*{20}c}  0&0&1&0  \\ 0&0&0&1 \\0&1&0&0\\ 1&0&0&0 \end{array}\right],\\
&[e'_3]=\left[\begin{array} {*{20}c} 0&1&0&0 \\ 1&0&0&0 \\0&0&0&1 \\ 0&0&1&0 \end{array}\right],
[e'_4]=\left[\begin{array} {*{20}c} 0&0&0&1 \\ 0&0&1&0 \\ 1&0&0&0 \\ 0&1&0&0
\end{array}\right]
\end{split}
\end{equation}
can also be basis of complex numbers. Naturally, the set of [0,1]-matrices, given by (\ref{eq65}) is isomorphous to the set, which is given by (\ref{eq63}).
It is evident, that the system of complex numbers can be constructed by infinite number of the ways, at that cyclic basis can consist of $m$ units, $ m \in N$, starting from three. It is remarkable, that the conformity between complex numbers and matrices is realized by biective mappings. It means, that there are also to be existing the inverse mappings, by means of which to any   squarte matrices, belonging to the linear space with a basis given by (\ref{eq63}), or (\ref{eq65}), or any other, satisfying the rules of cyclic recurrence like to (\ref{eq64}) can be set up in conformity the complex numbers. In particular, to any Hermitian  matrix $H$ can be set up in conformity the following complex number
\begin{equation}
\label{eq66}
  \zeta: H \to  S + iA = \left[\begin{array} {*{20}c} S&-A  \\ A&S  \end{array}\right], 
\end{equation}
where $S$  and $A$  are symmetric and antisymmetric parts of Hermitian  matrix. 

In the works \cite{Dovlatova_Yearchuck}, \cite{Yerchuck_Dovlatova} was used the apparatus of hypercomplex $n$-numbers  by
calculation of the optical properties of 1D carbon zigzag shaped nanotubes, taking
into account quantum nature of EM-field. Hypercomplex $n$-numbers were defined to
be the elements of commutative ring $Z_n$, being to be the direct sum of $n$ fields $C$ of complex numbers, $n\in N$, that is

\begin{equation}
\label{Eq5}
Z_n = C \oplus {C} \oplus{...}\oplus {C}.
\end{equation}
 It means, that any hypercomplex $n$-number $z \in Z_n$ is $n$-dimensional quantity with the components $k_\alpha \in C, \alpha = \overline{0, n-1}$, It can be represented in row matrix form $z \in Z_n$   
\begin{equation}
\label{Eq6}
z = [k_\alpha] = [k_0, k_1, ..., k_{n-1}].
\end{equation}
On the other hand, hypercomplex $n$-number $z \in Z_n$  can also be represented in the form
\begin{equation}
\label{Eq7}
z = \sum_{\alpha = 0}^{n-1}k_\alpha\pi_\alpha,
\end{equation}
where $\pi_\alpha$,$\alpha = \overline{0, n-1}$, are basis elements of $Z_n$. They are the following
\begin{equation}
\label{Eq8}
\begin{split}
&\pi_0 = [1,0, ...,0,0],  \pi_1 =[0,1, ...,0,0],\\
&..., \pi_{n-1} = [0,0, ...,0,1].
\end{split}
\end{equation}
Basis elements $\pi_\alpha$ possess by projection properties
\begin{equation}
\label{Eq9}
\pi_\alpha\pi_\alpha = \pi_\alpha\delta_{\alpha\beta}, \sum_{\alpha = 0}^{n-1}\pi_\alpha = 1, z \pi_\alpha = k_\alpha\pi_\alpha.
\end{equation}
Consequently, the set of $k_\alpha \in C, \alpha = \overline{0, n-1}$ is the set of eigenvalues of hypercomplex $n$-number $z \in Z_n$, the set  of $\{\pi_\alpha\}$, $\alpha = \overline{0, n-1}$ is eigenbasis of $Z_n$-algebra.

The simplest hypercomplex number is quaternion.
Any quaternion number $x$ can be determined according to relation
\begin{equation} 
\label{eq1Acacdegh}
\begin{split}
\raisetag{40pt}
&x = (a_1 + ia_2)e + (a_3 + ia_4)j,
\end{split}
\end{equation}
where $\{a_m\}\in R, m = \overline{1,4}$ and
 $e, i, j, k$ produce basis, elements of which are satisfying the conditions  
\begin{equation} 
\label{eq15Acacdegh}
\begin{split}
\raisetag{40pt}\\
&(ij) = k, (ji) = -k, (ki) = j, (ik)= -j,\\
&(ei)= (ie) = i, (ej)= (je) = j, (ek) = (ke) = k.
\end{split}
\end{equation}
Quaternions in given representation were used in
\cite{Dovlatova_Yerchuck} and in \cite{Yerchuck_Dovlatova_Alexandrov} to formulate the Maxwell equation in general form and to establish the properties of corresponding EM-field operators by its quantisation (see further). 

Quaternion set $\{Q\}$ can be considered to be the generalization of the set $Z_n$ for n = 2 in accordance with the relation
\begin{equation}
\label{Eq5a}
Z^Q_2 = C e \oplus {C} j,
\end{equation}
that is, it represents itself noncommutative ring.     Here $e$ and $j$ are real and imaginary unities correspondingly. Geometrically  the  set $\{Q\}$   can be interpreted to be representing  the numbers, which are determined by four points in two mutually perpendicular  planes in the space $R^{Q}_4$ with  the frame of reference, consisting of  orthogonal one real axis and three  imaginary axes, at that $i, j, k$ are imaginary unities  along coordinate axes in three dimensional imaginary $Z^i_3$ space, being to be the subspace of $R^{Q}_4$. The result of the multiplication of two imaginary unities is equivalent to $\pi/2$ rotation in the corresponding  pure imaginary plane. It seems to be evident, that the space $R^{Q}_4$ is isomorphous to Minkowski space $R_4$.
Taking into account given interpretation, to any quaternion number $x = (a_1 + ia_2)e + (a_3 + ia_4)j$ can be set up in conformity the $[2\times 2]$-matrix according to biective
mapping
\begin{equation}
\label{eq1ABabcd}
\begin{split}
\raisetag{40pt}
&q : (a_1 + ia_2)e + (a_3 + ia_4)j \to\\
&\left[\begin{array} {*{20}c} a_1&-a_2 \\ a_2&a_1 \end{array}\right]E+\left[\begin{array} {*{20}c} a_3&-a_4 \\ a_4&a_3 \end{array}\right]J,
\end{split}
\end{equation}
where $E$ is $[2 \times 2]$ unity matrix, $J$ is the following  $[2 \times 2]$ matrix
\begin{equation}
\label{eqA61}
J = \left[\begin{array} {*{20}c} 0&1  \\ -1&0 \end{array}\right]. 
\end{equation}
 The matrices
\begin{equation}
\label{eqAB61}\begin{split}
\raisetag{40pt}
&E = \left[\begin{array} {*{20}c} 1&0  \\ 0&1 \end{array}\right], 
I = \left[\begin{array} {*{20}c} 0&-i  \\ i&0 \end{array}\right],\\
&J = \left[\begin{array} {*{20}c} 0&1  \\ -1&0 \end{array}\right],
K = \left[\begin{array} {*{20}c} i&0  \\ 0&i \end{array}\right]
\end{split}
\end{equation}
produce basis for  the set $\{x\}$  of quaternion numbers $\{x\} = \{(a_1 + ia_2)e + (a_3 + ia_4)j\}$   in the linear space  of $[2 \times 2]$-matrices, defined over the field of real numbers. It is evident, that the matrices (\ref{eqAB61}) produce anticommutative multiplicative group. It is also clear,  that the system of quaternion numbers like to complex numbers can be built by infinite number of the ways.

Given short consideration has  allowed to formulate the following statements.

1.\textit{Two real vector-functions can be set up in the correspondence to Hermitian operator vector-function of quantized  EM-field in general case}.

Proof is evident and it is based on  (\ref{eq66}), if to take into account, that quantized free EM-field  can be  determined by Hermitian operator vector functions $\hat{\vec{E}}(\vec{r},t)$ and $\hat{\vec{H}}(\vec{r},t)$ in Hilbert space, representing themselves the full set of quantized free EM-field operator vector-functions, that is, they can serve for basis in corresponding operator vector-functional space (see Sec.IV). We consider here for definiteness one set of possible four sets (see further) of Hermitian operator vector functions $\hat{\vec{E}}(\vec{r},t)$ and $\hat{\vec{H}}(\vec{r},t)$ with fixed parities under improper rotations. Given statement can be generalized.

2.\textit{Any quantumphysical Hermitian operator in Hilbert space, which is set up in conformity to corresponding classical physical quantity, defines two real connected between themselves sets of observables (amplitude and phase  sets or average value and its dispersion), that is  complex quantity in general case, which allows to describe correctly the real noninstantaneous processes}.

 Proof is evident and it is based on the same relationship, since any Hermitian operator can be represented by Hermitian matrix in corresponding space, which, in its turn, is complex number in corresponding basis or produces biective mapping with complex numbers in usual form $z = a+ib$, or $z = |z|e^{i\phi}$.  
Therefore, two sets of observables, which are determined by real functions, correspond to any quantumphysical operator quantity. Given conclusion seems to be substantial. Physically it is understandable, if to take into account that, for instance, the processes of interaction by their quantum description are characterized by interaction time. It means in its turn, that along with amplitude, any physical quantity, which belongs to the set of quantities, describing interaction, has to be characterized by phase. It corresponds mathematically to well known trigonometrical form of complex numbers. Equivalent representation, corresponding to algebraic form of complex numbers, characterizes the physical quantity itself and its dispersion. It is remarkable, that the representation of physical quantities in complex form is widely used also in classical physics, however it is considered to be the only formal, although  convenient mathematical technique, see for example \cite{Angot}. Taking into account the sibling connection between quantum and classical physics given technique is taken on clear physical meaning and it  becomes strict mathematical substantiation. 
Practical consequence of the statement number two for quantumphysical tasks is the following. In particular, by the solution of main quantummechanical equation - Schr\"{o}dinger  equation $\hat{H}\psi_n = E_n\psi_n$, that is, by finding of the eigenvalues $\{E_n\}$, $n \in N$, of Hamilton operator, $E_n$ has to be represented in complex form. In  the application to atom physics $ReE_n$ is the energy of $n$th atomic level, $ImE_n$ is its dispersion, that is width of given level.  Usually used solution by $E_n \in R$ leads to the lost of the half of the possible information. The conclusion is similar for any other quantumphysical task.

It seems to be also remarkable, that the representation  of eigenvalues of Hamilton operator  in complex form like to aforesaid classical physics technique \cite{Angot}, is also used in a practice of theoretical calculations by a number of authors to be self-evident without mathematical ground. For instance, in \cite{Baryshevsky} the  eigenvalues of Hamilton operator are represented in the form $E_n = E^{(0)}_n - \frac{1}{2} i \Gamma_n$, where $E^{(0)}_n$ is the atom level energy, $\Gamma_n$ is the atom level width, in full correspondence with  the statement number two. The authors of the work \cite{Schmidt} have represented the generalization of linear Luttinger liquid theory. Along with energetic characteristics of quasiparticles (spinons and holons) spectra, given by  dynamic response functions, in particular, by the
spectral function $A(k, \omega)$, they related for arbitrary momenta and Galilean invariant
systems  the phase shifts to another
set of measurable properties. They took into consideration, that unitary transformation, which removes the interaction term between quasiparticles, is characterized by phase shifts, which are different for different quasiparticle modes. In other words, the eigenvalues of Hamilton operator were also represented in complex form in correspondence with simple, however, fundamental conclusion number two.

Let us summarise  now the general algebraic properties of EM-field to confirm  the conclusion concerning the symmetry properties under improper rotations first of all of the quantities $\vec {P}$, $\vec {E}$, entering optical Bloch equations, that is, to confirm  the conlusions of the works \cite{Yearchuck_Doklady}, \cite{D_Yearchuck_A_Dovlatova} from algebraic viewpoint. It was considered in details in the work \cite{Yearchuck_Alexandrov_Dovlatova}. 

It is well known, that real
EM-field can be characterized by both contravariant tensor 
$F^{\mu \nu }$ (or covariant $F_{\mu \nu })$ and contravariant pseudotensor 
$\tilde {F}^{\mu \nu }$, which is dual to $F_{\mu \nu }$ (or 
covariant $\tilde {F}_{\mu \nu }$, which is dual to $F^{\mu \nu })$. For example, $\tilde {F}^{\mu \nu }$ is determined by the relationship $\tilde {F}^{\mu \nu }=\frac{1}{2}e^{\mu \nu \alpha \beta }F_{\alpha \beta }$, where $e^{\mu \nu \alpha \beta}$ is Levi-Civita 4-tensor. It seems to be understandable, that field tensors and pseudotensors  are peer and independent characteristics of EM-field. It follows immediately from general consideration 
of the geometry of Minkowski space $^{1}R_4$. Really, the geometric structure of  pseudo-Euclidean abstract space of index 1, to which Minkowski space is 
isomorphic, determines unambiguously 3 kinds of peer, independent sets of linear centereuclidian
geometrical objects - tensors, pseudotensors and spinors (spin-tensors) \cite{Rashevskii}. The simplest example of practical usage of independency of EM-field tensors and pseudotensors from each other is  the method of
 obtaining  of 
EM-field invariants, see for instance \cite{Landau_Lifshitz_Field_Theory}, \cite {Stephani}, where the independence of EM-field tensors and  pseudotensors is considered to be  going without saying. Algebraic properties of union of two sets of EM-field tensor  and pseudotensor functions of 4-radius-vector $x$
 $\left\{{F}^{\mu \nu }(x)\right\}$ and   $\left\{\tilde {F}^{\mu \nu }(x)\right\}$ respectively  
are summarized in the following statements.

3.\textit{Union of contravariant (or covariant) EM-field tensors and pseudotensors} [\textit{tensor functions of 4-radius-vector} $x$, \textit{determined on some set} $S$ $\subseteq {^{1}R_4}$ \textit{in general case}] \textit{produces linear space} $\left\langle F,+,\cdot \right\rangle$ \textit{over a field of scalars} $P$, \textit{consisting of two invariant subspaces (correspondingly, tensor  and pseudotensor ones).}

4.\textit{Union of contravariant (or covariant) EM-field tensors and pseudotensors} [\textit{tensor functions of 4-radius-vector} $x$, \textit{determined on some set} $S$ $\subseteq {^{1}R_4}$ \textit{in general case}] \textit{produces linear algebra} $\mathfrak F$ = $\left\langle \mathfrak F,+,\cdot, \ast \right\rangle$ \textit{with algebraic operations of proceeding to dual elements by means of convolution with Levi-Civita 4-tensor, composition (addition) of "vectors" and multiplication of "vectors"  by scalar}.

Term "vector"   means  an element of  tensor-pseudotensor union. The proofs of statements 3 and 4 are simple, and they do not reproduce here.
  Let us  define on the space $\left\langle F,+,\cdot \right\rangle$ the functional $\Phi(x)$ by the following relationships
\begin{equation}
\label{eq4aa}
 \Phi[{F}^{\mu \nu}(x)] \equiv \left\langle {F}^{\mu \nu}(x) | \Phi\right\rangle = F^{\mu \nu}(x)  F_{\mu \nu}(x^*), \end{equation} and 
\begin{equation}
\label{eq4bb}
 \Phi [\tilde {F}^{\mu \nu}(x)] \equiv \left\langle \tilde {F}^{\mu \nu}(x) | \Phi \right\rangle = \tilde {F}^{\mu \nu}(x) \tilde{F}_{\mu \nu }(x^*). 
\end{equation}
We can also define on the space $\left\langle F,+,\cdot \right\rangle$ the functional $\tilde{\Phi}(x)$ by the following relationships
 \begin{equation}
\label{eq4cc}
 \tilde{\Phi}[{F}^{\mu \nu}(x)] \equiv \left\langle F^{\mu \nu}(x) | \Phi\right\rangle = F^{\mu \nu}(x) \tilde {F}_{\mu \nu}(x^*) \end{equation} and 
\begin{equation}
\label{eq4dd}
 \tilde{\Phi}[ \tilde {F}^{\mu \nu}(x)] \equiv \left\langle \tilde {F}^{\mu \nu}(x) | \Phi \right\rangle = \tilde {F}^{\mu \nu}(x) F_{\mu \nu}(x^*). 
\end{equation} 
 In Eq.(\ref{eq4aa} to \ref{eq4dd}) $x^*$ is fixed value of 4-radius-vector $x$.
It is clear, that $\Phi$ is antilinear functional on the space $F$, if  field of scalars is  field of complex numbers $C$. In  particular, if  field of scalars is  field of real numbers $R$, $\Phi$ is linear functional. Then the statement 5 takes place.

5.\textit{The set of  functionals} $\left\{\Phi [\tilde {F}^{\mu \nu}(x)]\right\}$, $\left\{\tilde{\Phi}[\tilde {F}^{\mu \nu}(x)]\right\}$, \textit{preassigned on the space} $\left\langle F,+,\cdot \right\rangle$, \textit{produces  linear space} $\left\langle \Phi',+,\cdot \right\rangle$ \textit{over a field of scalars} $P$, \textit{which is dual to the space} $\left\langle F,+,\cdot \right\rangle$, \textit{however it is nonselfdual}.

Here $\Phi'$ is union of $\Phi$ and $\tilde{\Phi}$. The proof of statement is simple, and it does not reproduce.
Quite analogous statement can be formulated, if instead of a field of scalars $P$ some set of pseudoscalars $\tilde P$ will be taken into consideration.  
Statement 5 (for both the cases) can be expressed shortly by the relationship 
\begin{equation}
\label{eq5a}
 \left\langle \Phi',+,\cdot \right\rangle = \left\langle {F}^{\times},+,\cdot \right\rangle.
 \end{equation} 
 The nonselfduality of  ${F}^{\times}$ seems to be substantional, and it determines the practical significance of given statement. Really, from the statement 5 follows the
necessity to take into consideration always both the spaces, that is   $\left\langle F,+,\cdot \right\rangle$ and  $\left\langle \Phi' + \cdot \right\rangle$, by the study of any physical process with participation of EM-field. More strictly, known Gelfand triple, which includes together with spaces $\left\langle F + \cdot \right\rangle$ and  $\left\langle \Phi' + \cdot \right\rangle$ the Hilbert space with topology, determined in the proper way, has to be taken into consideration, see, for example \cite{Bohm}. In other words, for full physical description of dynamical systems, interacting with EM-field, and for description of any physical phenomena, where the interaction with EM-field presents, on the whole, it 
is necessary to study the response to two Gelfand triples, determined  correspondingly over the scalar $P$ field and over pseudoscalar $\tilde P$ set. 

It is advisable to indicate, that pseudoscalars' set $\tilde P$ does not produce a field, although  given set  
produces an additive group. It is evident, that the union of sets of scalars $P$ and pseudoscalars $\tilde P$ produces the ring ${P^{'}}$ without unit.  It leads to union of linear space $\left\langle F,+,\cdot \right\rangle$ over a field of scalars  $P$ and linear space $\left\langle F,+,\cdot \right\rangle$ over a group of pseudoscalars $\tilde P$, if we define both the tensor functions ${F}^{\mu\nu}$ and $\tilde {F}^{\mu \nu}$ over a ring of scalar and pseudoscalar union $\tilde {P^{'}}$. Let us designate  linear space obtained $\left\langle \mathcal F,+,\cdot \right\rangle$.  The union of sets of scalars $P$ and pseudoscalars $\tilde P$  leads also to the union of linear algebras $\mathfrak F$ = $\left\langle \mathfrak F,+,\cdot, \ast \right\rangle$ prescribed over scalar $P$ field and pseudoscalar $\tilde P$ set   by means of  their definition over a ring of scalar and pseudoscalars $\tilde {P^{'}}$. It is clear, that convolution of algebra elements with Levi-Civita 
4-tensor, that is   proceeding to dual elements, realizes automorphism. 

It is easily to show, that the space $\left\langle \mathcal F,+,\cdot \right\rangle$ is selfdual. Then, foregoing practical remark can be reformulated -

 \textit{The solution of one or another task with EM-field participation has to be performed in the  space $\left\langle \mathcal F,+,\cdot \right\rangle$ over a ring of scalars and pseudoscalars $\tilde {P^{'}}$}.

  It is also evident, that partition of given space $\left\langle \mathcal F,+,\cdot \right\rangle$ into  four invariant subspaces $\left\langle\mathcal F^{(i)},+,\cdot \right\rangle$, $i = \overline {1,4}$,  can be realized. Elements of the first subspace $\left\langle \mathcal F^{(1)},+,\cdot \right\rangle$ are genuine EM-field tensors (tensor function in general case), determined over a scalar field $P$. Elements of the second subspace $\left\langle \mathcal F^{(2)},+,\cdot \right\rangle$ are also genuine EM-field tensors (tensor function in general case), determined however over a pseudoscalar $\tilde P$ additive group. Elements of the third subspace $\left\langle \mathcal F^{(3)},+,\cdot \right\rangle$ are EM-field pseudotensors (pseudotensor function in general case), determined over a scalar field $P$. Elements of the fourth subspace $\left\langle \mathcal F^{(4)},+,\cdot \right\rangle$ are EM-field pseudotensors (pseudotensor function in general case), determined now over a pseudoscalar $\tilde P$ additive group. Let us characterize the symmetry kind under improper rotations of the components of tensor elements for each subspace, that is, let us thereby establish the symmetry kind of vectors of electric field $\vec {E}$ and magnetic field $\vec {H}$. In subspaces  $\left\langle \mathcal F^{(1)},+,\cdot \right\rangle$,  $\left\langle \mathcal F^{(2)},+,\cdot \right\rangle$  vector $\vec {E}$ is polar vector and vector $\vec {H}$ is axial. Given conclusion is evident for the first subspace. Vector $\vec {E}$ in the second subspace is dual vector to antisymmetric 3-\textit{pseudo}tensor, that determines its polar character and vector $\vec {H}$, respectively, is axial. At the same time, in contrast to the case 1, the components of vector $\vec {E}$ correspond now to pure space components of EM-field  4-\textit{pseudo}tensor ${\tilde {F}^{\mu \nu}}$ in given case. The components of vector $\vec {H}$ correspond for the case 2 to time-space mixed components of given  4-\textit{pseudo}tensor, that determines  the axial symmetry of vector $\vec {H}$. Arbitrary element of subspace  $\left\langle \mathcal F^{(3)},+,\cdot \right\rangle$ can be represented in the form of
 $ \alpha {F}^{\mu \nu }(x_1) + \beta {F}^{\mu \nu }(x_2)$, where
 $\alpha, \beta \in \tilde {P}$ and $x_1, x_2 \in S \subseteq {^{1}R_4} $. It is 4-\textit{pseudo}tensor, since $\alpha, \beta \in \tilde {P}$. Its space components are in fact the components of antisymmetric 3-\textit{pseudo}tensor, which determine dual to given tensor polar magnetic field vector $\vec {H}$, while time-space mixed components are the components of axial 3-vector $\vec {E}$ of electric field.   
Therefore, the symmetry properties of the components of vectors $\vec {E}$ and $\vec {H}$ under  improper rotations in the case 3 will be opposite to the case 1. It is evident, that in the 4-th case the symmetry properties of the components of $\vec {E}$ and $\vec {H}$ under  improper rotations will be opposite to the case 2. 
 
 Given consideration has clear mathematical and physical meaning. If electric field strength vectors correspond to above considered case 3 or 4, that is they are pseudovectors
(and, consequently, electric dipole moments are also pseudovectors), the equation of dynamics of optical transitions will have the structure, which is
mathematically equivalent to the structure of the equation for 
dynamics of magnetic resonance transitions (by which magnetic field vector-function components correspond to the case 1 or 2). In other words, mathematical abstractions in optical Bloch equation
 become, in agreement with results \cite{Yearchuck_Doklady}, \cite{Yearchuck_Yerchak}, real physical meaning, that is really $\vec{E}$ is the part of intracrystalline and 
external electric field, which has axial vector symmetry, $\vec{P}$ is electrical moment, which seems to be built like 
to magnetic moment with the same axial symmetry.

The result obtained allows to suggest, that free EM-field is 4-fold degenerated under improper rotations. The interaction of EM-field with device (or, generally, with some substance) can relieve degeneracy and can lead, especially by interaction with extended centers, to unusual, above discussed, symmetry of field vector-functions. On the other hand, there is CPT-invariance requirement, and optical resonance system has to satisfy to it too. It will take place, since in compaund 4-component EM-field structure always will be presenting the component with the symmetry properties, which is necessary to ensure CPT-invariance. At that we have to take into account, that spectroscopic transitions are not instantaneous, that is, it seems to be taking place the formation of resonance state of united {field + matter} system or, in particular, of united {field + device} system, for which CPT-invariance has to be taking place. Let us remark that the resonance state of united {field + matter} system can have very long life time in comparison with field mode period, determined, for instance, by the time of Rabi-oscillation amplitude damping. Given conclusion becomes especially actual, if to take into account relatively recent theoretical discovery of new quantum optics phenomenon -  quantum Rabi  corpuscular-wave packets   formation and propagation \cite{Slepyan_Yerchak} for the systems with strong electron-photon interaction. By Rabi corpuscular-wave packets    formation the interaction process of the photons with matter is so long, that it can be visible by the detection with usual stationary spectroscopy methods by means of appearance of additional spectral lines, that has to be taken into consideration by interpretation of spectroscopic results, \cite{Yerchuck_Dovlatova}. In given case the symmetry of responce function under improper rotations can be changed with the change of propagation distance and time growth (for instance by contribution of space- and/or time dispersion in initial absorption responce signal, taking place in conducting media). Then, it is reasonable to suggest, that autotuning of the symmetry of EM-field component is also realized.

Therefore, foregoing simple algebraic consideration of the properties of EM-field vector-functions leads in fact to conclusion, that free EM-field is complicated dynamical system and consist of two independent fields with various P-parity, which in turn  consist of two independent fields with various  t-parity. Given conclusion has to be taken into account by interpretation of experimental data.

It seems to be reasonable along with considered symmetry of EM-field under improper rotations to consider in more details the gauge symmetry of EM-field.

\section{Dually charged quasiparticles - theory and its experimental confirmation}
It is remarkable, that the 
 concept to take into consideration the complex charge in EM-field theory was proposed  already in 1981 in \cite{Tolkachev} by the description of electrodynamics of dually charged particles (then, naturally, hypothetical ones), and it was done  for the first time (to our knowledge).  At the same time, the  viewpoint,  that EM-field is vector real field only and that  it cannot be characterized by any charges at all, is dominating upto now, and we understand well, that  the algebraic properties of EM-field, described in previous Section, have to be theoretically argued in more details.

\subsection{Additional gauge invariance of complex relativistic fields}

 We have above argued, that along with quantized EM-field, the classical EM-field  in the matter has to be considered also in general case  to be complex field. Let us remember, that quantized EM-field has to be considered always complex, if operators for $\vec{E}$, $\vec{H}$ have   antisymmetric parts. At the same time, the  complex character of classical EM-field takes place by $\theta \neq 0$  in Rainich dual transformations. So, for correct description of the system  \{EM-field + matter\} two sets of real EM-field vector-functions have to be taken into consideration. The simplest example is description of optical absorption, transmission, reflection, scattering or luminescence experiments by its homodyne or heterodyne detection, when along with amplitude the phase of EM-field has to be taken into account.  In  the work \cite{Yearchuck_Alexandrov_Dovlatova} the idea, that for any complex field the conserved quantity, corresponding to its gauge symmetry, that is charge field function, can be in general case also complex, was proved. Since given result seems to be substantial, we reproduce given fragment 
in given Section.

 Let  $\left\{ {\,u_{i}(x) \,} \right\}$, $i = \overline{1,n}$, the set of the functions of some complex relativistic field, that is, scalar, vector or spinor field, given in some space of Lorentz group representations. It is well known, that Lagrange equations for any complex relativistic field can be represented in the form of one matrix relativistic differential equation of the first order in partial derivatives, that is in the form of so called generalized relativistic equation, and analogous equation for the field with Hermitian conjugated (complex conjugated in the case of scalar fields) functions $\left\{ {\,u_{i}^{+}(x) \,} \right\}$. The equation for the set $\left\{ {\,u_{i}(x) \,} \right\}$ of field functions, represented in one column matrix form $\|{u(x)}\|$ is
\begin{equation}
\label{eq4a}
(\|\alpha_{\mu}\| \partial_{\mu} + \kappa \|\alpha_{0}\|) \|{u(x)}\| = 0.
\end{equation}
Similar equation for the field with Hermitian conjugated (complex conjugated in the case of scalar fields) functions, that is, for the functions
   $\left\{ {\,u_{i}^{+}(x) \,} \right\}$,   $i = \overline{1,n}$, is
\begin{equation}
\label{eq4b}
\partial_{\mu} \|u^{+}(x)\| \|\alpha_{\mu}\| + \kappa \|{u^{+}(x)}\| \|\alpha_{0}\| = 0.
\end{equation}

In equations (\ref{eq4a}, \ref{eq4b}) $\|\alpha_{\mu}\|, \|\alpha_{0}\|$ are matrices with constant numerical elements. They have an order, which coincides with dimension of corresponding space of Lorentz group representation, realized by $\left\{ {\,u_{i}(x) \,} \right\}$, $i = \overline{1,n}$. In particular, they are $[n\times{n}]$- matrices, if $\left\{ {\,u_{i}(x) \,} \right\}$, $i = \overline{1,n}$ are scalar functions. It is evident, that the transformation
\begin{equation}
\label{eq10}
\|{u'(x)}\| = \beta exp(i \alpha)\|{u(x)}\|,
\end{equation}
where $\alpha,\beta\in R$,  and analogous transformation for Hermitian conjugate functions (or complex conjugate functions in the case of scalar fields) 
\begin{equation}
\label{eq11}
\|{u'^{+}(x)}\| = \beta exp(-i \alpha) \|{u^{+}(x)}\|
\end{equation}
keep Lagrange equations $(\ref{eq4a}, \ref{eq4b})$ to be invariant. It is understandable, that transformation of field functions by relationships  $(\ref{eq10}), (\ref{eq11})$ is equivalent to multiplication of field functions  by arbitrary complex number. It is well known, that given linear transformation  is the simplest example of isomorphism of corresponding linear space, which is given over the field of complex numbers, onto itself, that is, in the case considered  the relationships  $(\ref{eq10}, \ref{eq11})$ give  automorphism of the space of  field functions.  Automorphism of any linear  space leads to some useful properties of the objects, which belong to given space. For instance, if to set up in a correspondence to the space of field functions the affine space, then conservation laws of collinearity of the points and of simple relation of the triple of collinear points will be fulfilled by automorphism in given affine space. Consequently,  we have to expect the physical consequences of given algebraic property in the case of physical spaces. Conformably to the case considered
 we have in fact gauge transformation of field functions, which is more general in comparison with usually used. The set $(\beta exp(-i \alpha)$ for all possible $\alpha$, $\beta \in R$ produces the group $\Gamma$, which is direct product of known symmetry group $U_1$, and multiplicative group $\mathfrak R$ of all real numbers (without zero). Therefore, in the case considered the symmetry group of given complex field acquires additional parameter. So, we  have
\begin{equation}
\label{eq12}
\Gamma(\alpha, \beta) = U_1(\alpha) \otimes \mathfrak R (\beta)
\end{equation}
  In the work \cite{Yearchuck_Alexandrov_Dovlatova} the irreducible representations of the group $\mathfrak R(\beta)$ were found. They are
\begin{equation}
\label{eq18a}
T(\beta) = \beta^{2k+1}= exp[{(2k+1) ln\beta}],
\end{equation}
where $k \in N$. Then irreducible representations of the group $\Gamma(\alpha, \beta)$ represent
 direct product of irreducible representations of the groups $U_1(\alpha)$ and $\mathfrak R(\beta)$ 
\begin{equation}
\label{eq18b}
T(U_1(\alpha)) \otimes T(\mathfrak R(\beta)) = exp(-i m \alpha) exp[{(2k+1) ln\beta}],
\end{equation}
where
$m, k = 0, \pm1, \pm2, ...$ .

It is clear, that some conserved quantity has to correspond to gauge symmetry of the field, which is determined by the group $\mathfrak R(\beta)$. Thus, the  formulation of the following statement was done in \cite{Yearchuck_Alexandrov_Dovlatova}.

6.\textit{Conserving quantity - complex charge field function, which is invariant under total gauge transformations $\|{u'(x)}\| = \beta exp(i \alpha)\|{u(x)}\|$, $\|{u'^{+}(x)}\| = \beta exp(-i \alpha) \|{u^{+}(x)}\|$ corresponds to any complex relativistic field (scalar, vector, spinor).}
 
\textbf{Proof}.
 Really, since generalized relativistic equations are invariant under transfomations ($\ref{eq10}, \ref{eq11}$) and, consequently, the variation of action integral with starting Lagrangian is equal to zero, then the variation of action integral with transformed Lagrangian in accordance with ($\ref{eq10}, \ref{eq11}$) will also be zero. Therefore, all the conditions of the applicability of N\"{o}ther theorem, by proof of which the only invariance under Lagrange equations is sufficient, \cite{Noether}, are held true. According to N\"{o}ther theorem, the conserved quantity, corresponding to $\nu -th$ parameter ($\nu = \overline{1,k}$) by invariance of field under some $\textit{k}$-parametric symmetry group, is
\begin{equation}
\label{eq19a}
Q_\nu(\sigma) = \int\limits_{(\sigma)}\theta_{\mu \nu}d\sigma_\mu = const,
\end{equation}
where
\begin{equation}
\label{eq19b}
\theta_{\mu \nu} = \frac{\partial{L}}{\partial(\partial_{\mu}u_i)} [\partial_{\rho}u_i X_{\rho \nu} - Y_{i \nu}] - L X_{\mu\nu},
\end{equation}
$L$ is field Lagrangian, $\sigma$ is any spacelike hypersurface, $\sigma$ $\subset {^{1}R_4}$.  We have to draw attention to typical mistake,  which is abundant in the literature, consisting in that, that for  the applicability of N\"{o}ther theorem the Lagrangian invariance under corresponding symmetry transformations is required. At the same time, the only invariance of Lagrange equations under corresponding symmetry transformations, which certainly takes place in given case, is necessary (see proof of N\"{o}ther theorem). The matrices $X_{\rho \nu}, \text{Y}_{i \nu}$ are determined by matrix representations $\left\|(I_{\nu})_{\mu\ \nu}\right\|$ and $\left\|(J_{\nu})_{i k}\right\|$ of infinitesimal operators of symmetry group in coordinate space and in the space of field functions respectively in accordance with the following relationships
\begin{equation}
\label{eq19c}
X_{\rho \nu} = (I_{\nu})_{\mu \alpha} x_\alpha,
Y_{i \nu} = (J_{\nu})_{i k} u_{k}.
\end{equation}
So, using N\"{o}ther theorem, we obtain for 4-vector $\theta_\mu$ the following expression
\begin{equation}
\label{eq19}
\theta_{\mu} = -\frac{\partial{L}}{\partial(\partial_{\mu}u_i)} u_{i} -\frac{\partial{L}}{\partial(\partial_{\mu}u_i^{*})} u_{i}^*,
\end{equation}
Components of 4-vector $\theta_\mu$ satisfy to continuity equation
\begin{equation}
\label{eq20}
\partial_{\mu}{\theta_{\mu}} = 0.
\end{equation}
Conserving quantity, corresponding to (\ref {eq19}), is the charge field function, which is equal to
\begin{equation}
\label{eq21}
Q^{'}_{2} = iQ_{2} = -i \int\theta_{4}d^{3}x.
\end{equation}
So $iQ_{2}$ is determined by relationship 
\begin{equation}
\label{eq22}
iQ_{2} = i\int[\frac{\partial{L}}{\partial(\partial_{4}u_i)} u_{i} + \frac{\partial{L}}{\partial(\partial_{4}u_i^{*})} u_{i}^*]d^{3}x.
\end{equation}

It is seen from relationship (\ref{eq22}), that obtained additional charge function to well known charge function of complex fields, which is real quantity, is purely imaginary quantity. Let us remember, that known conserved gauge invariat quantity for any complex field, for instance, for Dirac field, that  is  real guantity - charge $Q_{1}$, is the consequence of gauge symmetry, consisting in the invariance of Lagrange equations under the transformations
\begin{equation}
\label{eq22a}
\|{u'(x)}\| = exp(i \alpha) \|{u(x)}\| 
\end{equation}
and
\begin{equation}
\label{eq22b}
\|{u'^{+}(x)}\| = exp(-i \alpha)\|{u^{+}(x)}\|. 
\end{equation}
In general case $Q_{1}$, see, for instance, \cite{Bogush}, is 
\begin{equation}
\label{eq23}
Q_{1} = -\int[\frac{\partial{L}}{\partial(\partial_{4}u_i)} u_{i} - \frac{\partial{L}}{\partial(\partial_{4}u_i)^{*}} u_{i}^*]d^{3}x. 
\end{equation}
 
 Therefore,
 any relativistic complex field can be characterized by complex conserving quantity $Q$, that is by complex charge function, which can be represented in the form
\begin{equation} 
\label{eq23a} 
Q = Q_{1} + iQ_{2}. 
\end{equation} 
The  statement is proved.

It is remarkable, that, like to mechanics, a number of conservation laws, which can have  EM-field, are optional in their simultaneous fulfilment. In particular, it is evident, that by automorphic transformation of the space of EM-field functions by relationship (\ref{eq10}) the conservation law for charge will  always take place. At the same time the energy conservation law and the conservation of Poynting vector will be fullfilled, if given transformation is applied to EM-field potentials.  The force characteristics, that is $\vec{E}$-, $\vec{H}$-vector functions can be used to be basis for free EM-field description, since they will represent the full set in free  EM-field case. However, the energy conservation law and the conservation of Poynting vector, that is mathematical constructions, to which enter $\vec{E}$-, $\vec{H}$-vector functions, will not  be fullfilled by transformation (\ref{eq10}) at arbitrary $\beta$.  We see, that the charge conservation law for EM-field is fullfilled  even in the case, when the energy conservation law does not take place.  Therefore, the charge conservation law can be considered in given meaning to be more fundamental.

Taking into account the connection between complex electromagnetic  charge function and its real and imagine components, which are according to \cite{Tolkachev} electric and magnetic charges correspondingly, the existence in the matter of the dually charged particles, which have simultaneously electric and magnetic charges becomes to be understanable. Let us remark, that the possibility of their existence was discussed a long time ago, see, for instance, \cite{Tomilchick}.
 
There seems to be essential, that the existence of dually charged particles in condensed matter was  experimentally confirmed. Their discovery was described in \cite{D_Yearchuck_A_Dovlatova} from the comparison of experimental results, reported in \cite{Ertchak_J_Physics_Condensed_Matter}, \cite{Ertchak_Carbyne_and_Carbynoid_Structures} and in \cite{Yearchuck_Yerchak}. Let us give some details, concerning given conclusion. Carbynoid samples, studied in above  cited works, were active both in electron spin resonance and in optical spectroscopy [infrarot (IR) absorption, reflection and Raman scattering (RS) spectroscopy]. Carbynoid samples, representing themselves the systems of quasi-1D carbon chains, can be considered to be the most simple modelling systems for verification of theoretically predicted effects for quasionedimensional structures with very many practical applications. The samples (designated "samples A" in \cite{Yearchuck_Yerchak}), in which the ferromagnetic and ferroelectric spin wave resonances (FMSWR and FESWR) were observed on the same chain structures, have been used. Both the resonances can be described by in fact the same equations, obtained in \cite{Yearchuck_Doklady} and modified by taking into consideration 
the relaxation processes in  \cite{Yearchuck_Yerchak}. So, the equation for the description of optical transition dynamics in a chain  
is 
\begin{equation}
\label{eq1}
\begin{split}
\raisetag{40pt}
&\frac{\partial \vec {S}(z)}{\partial t} = \left[ {\vec {S}(z)\times \gamma_{E} 
\vec {E}} \right] -\\
&\frac{4a^2J_{E}}{\hbar ^2}\left[ {\vec {S}(z)\times \nabla 
^2\vec {S}(z)} \right] + \frac{\vec {S}(z)-\vec {S}_0 (z)}{\tau },
\end{split}
\end{equation}
where $\vec {S}(z)$ is electric analogue of intrinsic magnetic moment, $\gamma_{E}$, $J_{E}$ are optical analogues of gyromagnetic ratio and exchange 
interaction constant respectively, $\hbar$ is Planck's constant, $a$ is lattice 
spacing, $\vec {E}$ is electric field, 
vector-function $\vec {S}_0 (z)$ is equilibrium value of axially symmetric electrical dipole moment (in particular, electrical "spin" moment), $\tau$ is relaxation time. 
Vector-functions $\vec {S}(z)$, $\vec {S}_0 (z)$ acquire in the case of FMSWR the meaning of magnetic  moment vector-functions, $E_{1} $ is replaced in FMSWR-case by $H_{1}$, that is, by amplitude of magnetic component of external oscillating EM-field, $J_{E}$ is replaced by the exchange interaction constant 
$J_{H}$, $\gamma_{E}$ by $\gamma_{H}$. For the agreement with experiments, in which the values of oscillating external electric field 
components $E^x, E^y$ in $(\ref{eq1})$ are greatly less in 
comparison with the value of intracrystalline electric field component 
$E^z$, and consequently under analogous relationships between corresponding  components of total 
electrical dipole moment (since x-, y-dipole components are induced by weak external electric field $E^x, E^y$)
 the linearized equation in \cite{Yearchuck_Yerchak} was obtained. It was solved under  additional assumption, that equilibrium distribution of $\vec {S}_0 (z)$ along the 
chain is homogeneous. All the assumptions were entirely correct for IR measurements. The solution of linearized equation gives 
 the relationships for a shape and 
amplitudes of resonance modes and for dispersion law. They are
\begin{equation}
\label{eq2}
a_{n} = \left\{ {{\begin{array}{*{40}c}
 {-\frac{i \gamma_{E} S \tau^2 E_{1}}{\pi n} \frac{\left[{(\omega_n - \omega
) - \frac{i}{\tau}} \right]}{\left[ {1 + (\omega_{n} - \omega)^2 \tau^2} 
\right]},\,\,n = 1,\,3,\,5,... \hfill} \\
{\,0,\,\,\,\,\,\,\,\,\,\,\,\,\,\,\,\,\,\,\,\,\,\,\,\,\,\,\,\,\,\,\,\,\,\,\,\,\,n = 2,\,4,\,6,...} 
\hfill, \\
\end{array} }} \right. 
\end{equation}
\begin{equation}
\label{eq3}
\nu _n =\nu _0 - \mathfrak A n^2,
\end{equation}
where $n\in N$ including zero, $\nu_{n}$ is a frequency of \textit{n-th} mode, $\mathfrak A$ is 
a material parameter ($\mathfrak A = \frac{2 \pi {a}^{2} S \left|J_E\right|}{\hbar^2 L^{2}}  > 0, L$ is chain length).
Like to magnetic resonance measurements, $Re\,a_n$ is proportional to absorption signal, $Im\,a_n$ is proportional to dispersion signal. 
It was found the following.

1.Dispersion law (\ref{eq3})  holds true both by IR- and RS-detection of FESWR. 

2.The excitation of the only uneven modes by IR-detection (by which the experimental conditions were corresponding to applicability of linearized equation) takes place in accordance with (\ref{eq2}). 

3.Inversely proportional dependence  of the amplitudes of modes (at resonance)  on mode number $n$ holds also true in accordance with (\ref{eq2}), however, also the only by IR-detection of FESWR. 

4.Splitting of Raman active vibration modes is characterized by the value of parameter $\mathfrak{A}$, being greater
approximately by factor 2, than parameter $\mathfrak{A}$, which characterizes IR FESWR spectra 
(by the frequencies of zero modes reduced by means of linear approximation 
procedure to the same value),  that  confirms well the theoretical prediction in \cite{Yearchuck_Doklady}.

 Moreover, the value of spin $S$, which is equal for optically active local centers studied to 1/2, has been obtained   for the first time  in optical spectroscopy from pure optical experiment, demonstrating the advantages of transition operator method in comparison with density matrix formalism in description of dynamics of spectroscopic transitions, which was discussed in details in  \cite{D_Yearchuck_A_Dovlatova}. 
The agreement between theory and experiment obtained denotes unambiguously  the reliability of identification of  FESWR phenomenon  on the one hand and  the validity of the conclusion on the existence of even parity of electric vector-functions under space inversion transformations in spectroscopic transition phenomena on the other hand.  Analogous conclusions on coincidence between  theory and experiment  take place for AFESWR phenomenon. We wish to draw attention, that both the phenomena are rather general and there are many experimental works, where FESWR and especially AFESWR were registered, however they were unidentified. For instance, we can point out  the work \cite{Winter_Kuzmany} with unidentified FESWR registration. In \cite{Winter_Kuzmany} the characteristic splitting of two lowest (Hg)-modes, corresponding  to  the phenomenon of FESWR, has been observed  by Raman scattering study in a single crystal of metallic potassium fulleride at 80 K. Our analysis of given results gives the value of splitting parameter $\sim 1.6 cm^{-1} $ for (Hg(2))-mode, that is substantially smaller in comparison with splitting parameter in $ 62.2 cm^{-1}$, which was found in  carbynoid samples (in suggestion of comparable thickness of the samples used in both the systems).  It allows therefore to compare quantitatively the effectivity of exchange interaction between electric dipole moments in carbynoids and potassium fulleride. We see, that it differs by more than order of value in above suggestion. 

The most substantional  is that, that  the ratio of two exchange constants $J_{E }/J_{H}$  
was obtained from the ratio of  the values of splitting parameters  $\mathfrak{A}^E$ in IR FESWR spectra and $\mathfrak{A}^H$ in the spectra of FMSWR in A-sample reported earlier in \cite{Ertchak_J_Physics_Condensed_Matter}, \cite{Ertchak_Carbyne_and_Carbynoid_Structures}. The range of given ratio is $(1.2 - 1.6)10^{4}$. The 
appearance of two exchange constants, which differ from  each other more 
than 4 order for the same chain structures is in fact the direct indication, that EM-field in the matter has along with complex vector characteristics the complex scalar characteristic - charge,  consisting of elecric and magnetic components,  corresponding to real and imaginary charge function parts correspondingly. Moreover, the values of  the ratio $J_{E }/J_{H}$ of exchange constants, observed in the same sample on the same carbon chains, allows  to evaluate the ratio of electric and magnetic $e_{H} \equiv g$ to $e_{E}\equiv e$  components of complex charge of simultaneously optically and magnetically active centers, that is dually charged centers - spin-Peierls solitons. It is sufficient to take into account the relation  $\frac{g}{e} \sim \sqrt{\frac{J_{E}}{J_{H}}} $, which is valid in the suggestion, that it can be neglected  by the change of the space distribution of the wave function  of given local quasiparticles  (characteristic size of spin-Peierls solitons is evaluated in 20 interatomic units) in magnetic field and by its absence. Hence it has been obtained, that $\frac{g}{e}\approx (1.1 - 1.3)\times 10^{2}$.    It is obviously, that given numerical evaluation is agreeing well with Dirac charge quantization theory \cite{Dirac P.A M}, that is, with relationship $\frac{g}{e} = 68.5 e n$, where $n = ±1, ±2,...,$ at n = 2. Some deviation is quite acceptably, since the guess is rather crude. Given result seems to be strong  experimental proof of existence of both types of charges (electric and magnetic ones) and dually charged carriers in condensed matter.  

Therefore, given comparison allows to conclude on experimental discovery of dually charged particles, the existence of which and their properties were intensively theoretically discussed in a number of publications of Belarusian theoretical physics branch of V.Fock schools, see for instance, monographs \cite{Tomilchick}, \cite{Berezin} and many references therein.

Although spin-Peierls solitons are quasiparticles, which are relatively strong localized   in comparison with chain length, their characteritic localization length (and, consequently, upper boundary for characteritic localization length of magnetic charge) is exceeding in more than eight order the very strong pointlike localization degree of electric charge, realized  by electrons. Electron size does not exceed according to existing at present viewpoint the value $~10^{-16}$ cm \cite {Ph.Enc.}. At the same time  experimental data above discussed, indicating on the existence of magnetic charge, do not indicate lower space localization boundary for given two magnetic charge quanta. It means, that  the existence of Dirac monopoles, that is, the particles with strong pointlike localization degree of magnetic charge, being to be comparable with the localization degree of electric charge in electrons remains to be open. Let us remark, that quite recently appears a new viewpoint on degree of electron localization \cite{Burinskii}.
"The observable gravitational and electromagnetic parameters of an
electron: mass m, spin $J = \frac{\hbar}{2}$, charge $e$ and magnetic moment $e a = e\frac{\hbar}{2m}$
indicate that the consistent with Einstein-Maxwell gravity electron should lead to
a Kerr-Newman background geometry. Contrary to the widespread opinion that
gravity plays essential role only on the Planck scales, the Kerr-Newman gravity
displays a new dimensional parameter $a = \frac{\hbar}{2m}$ which, for parameters of an
electron, corresponds to its Compton wavelength and is very far from the Planck
scale." It is further argued in \cite{Burinskii} "that, although gravitational field of an electron is extremely weak,
extremely large spin of the electron produces the Kerr geometry, which has very
essential topological changes at the Compton distance. The Kerr geometry is
formed by a congruence of twistor null lines, which are focused on the Kerr
singular ring, forming a branch line of Kerr background on two sheets. The
gravitational and electromagnetic fields are also focused on the Kerr ring, forming
a sort of a closed string, structure of which is close to described by Sen field
model of a heterotic string. The stringlike Kerr-Newman electron contradicts
to the claims on a structureless electron, but it confirms peculiar role of the
Compton area of a “dressed” electron in QED and matches with the known limit
of localization of the Dirac electron." 

Therefore, from analysis of known experimental and theoretical works we have two reasonable  conclusions of  the real existence of  new general peculiatities of  EM-field. They are 1) EM-field has compound character, that is, it can be represented to be the superposition of components with various time inversion (t) and space inversion (P) parities.  On the other hand, above considered analysis shows, that dual transformations, determined by (\ref{eq1bcd}) convert real classic EM-field into complex (and vice versa). It means, that  2) classic EM-field itself has to possess in general case by the charge, which seems  to be also complex in general case.

\section{Quaternion Nature  of EM-Field}

The existing interpretation of Maxwell equations with sources have some dissatisfaction of a following nature. In fact, every equation in mathematical form A = B means that A and B are different notations for the
same quantity. In  existing interpretation of Maxwell equations are absent the indications concerning the physical quantity, which has to characterise simultaneously the local properties of the field and  local properties  of the charged particles,
that is, which has to be  represented simultaneously through the field derivatives, and through the characteristics of
the charged particles.  If the EM-field, considered to be a physical
object, cannot carry  charge, then from
physical point of view both  two quantities have
different physical nature and cannot be equivalent. 
The presence of complex charge to be characteristic of EM-field allows to remove given contradiction. It means that 4-vector of current $j_\mu$ for any complex field is complex vector. In its turn, it means, that independently on starting origin of the charges and current carriers in the matter [they can be result of presence of Dirac field or another complex field] all the characteristics of EM-field in the matter have also to be complex-valued. Given conclusion follows immediately from Maxwell equations, since complex current $j_\mu$ enters explicitly in Maxwell equations. It is also substantial, that Maxwell equations themselves along with Lagrange equations for EM-field are invariant under the transformation of EM-field functions by relationship (\ref{eq10}). 

Now we will review in details the results of \cite{Dovlatova_Yerchuck}, indicating that generalization of Maxwell equations by means of their representation in quaternion form, which includes four independent constituents of EM-field, is direct consequence of dual invariance symmetry. 

\subsection{Generalized  Maxwell Equations}

 We have to remark, that setting into EM-field theory of two type of charges and two type of  intrinsic moments is actual, since recent above cited discovery of dually charged quasiparticles \cite{Yearchuck_Yerchak} and the particles with pure imaginary own electric  moments \cite{Yearchuck_PL} in condensed matter, - respectively, spin-Peierls $\pi$-solitons and Su-Schrieffer-Heeger $\sigma$-solitons in carbynoids and polyvinilidenhalogenides, is clear experimental proof, that  EM-field theory with complex charges and complex  electric dipole moments has physical content. EM-field theory in the matter with complex charges and complex electric dipole  moments ceases, consequently, to be the only formal  model, which  although is very suitable for  many technical calculations, but was considered upto now to be  the mathematical abstraction, in which magnetic charges and magnetic currents are fictitious quantities. Similar conclusion concerns
 the conception ofthe other electric and magnetic characteristics  of  the matter.  Given conception agrees well with all practice of electric circuits' calculations. Very fruitfull above indicated mathematical method for electric circuits' calculations, which uses all complex electric characteristics, see for example \cite {Angot}, was also considered earlier the only to be  formal, but  convenient mathematical technique (haw it was accentuated abov). Informality of given technique gets now natural explanation.

Let us designate the terms in (\ref{eq1c})
 \begin{equation} 
\label{eq10c}
\begin{split}
\raisetag{40pt}                                                    
& \vec {E} \cos\theta = \vec{E}^{[1]},  \vec {E} \sin\theta = \vec{E}^{[2]} \\
&\vec {H} \cos\theta = \vec{H}^{[1]}, \vec {H} \sin\theta =  \vec{H}^{[2]}.
\end{split}
\end{equation}
The Maxwell equations for the EM-field ($\vec {E'}$, $\vec {H'}$) in the matter in general case of both type of charged particles (that is electrically and magnetically charged), including dually charged particles are
\begin{equation}
\label{eq4abc}
\left[\nabla\times\vec{E'}(\vec{r},t)\right] = -\mu_0 \frac{\partial \vec{H'}(\vec{r},t)}{\partial t} - \vec{j'_g}(\vec{r},t), 
\end{equation}
\begin{equation}
\label{eq2abc}
\left[\nabla\times\vec{H'}(\vec{r},t)\right] = \epsilon_0 \frac{\partial\vec{E'}(\vec{r},t)}{\partial t } + \vec{j'_e}(\vec{r},t), 
\end{equation}
\begin{equation}
\label{eq4abcd}
(\nabla \cdot \vec{E'}(\vec{r},t)) = \rho'_e(\vec{r},t),
\end{equation}
\begin{equation}
\label{eq4abcde}
(\nabla \cdot \vec{H'}(\vec{r},t)) = \rho'_g(\vec{r},t),
\end{equation}
where $\vec{j'_e}(\vec{r},t)$,  $\vec{j'_g}(\vec{r},t)$ are respectively electric  and magnetic current densities,  $\rho^{'}_e(\vec{r},t)$, $\rho^{'}_g(\vec{r},t)$ are respectively electric  and magnetic charge densities.
Taking into account the relation (\ref{eq1c}) and (\ref{eq10c}) the system (\ref{eq4abc}), (\ref{eq2abc}), (\ref{eq4abcd}), (\ref{eq4abcde}) can be rewritten
\begin{equation}
\label{eq5abc}
\begin{split}
\raisetag{40pt}
&\left[\nabla\times(\vec{E}^{[1]}(\vec{r},t) - i\vec{E}^{[2]}(\vec{r},t)) \right] = \\
&- \mu_0 \left[\frac{\partial \vec{H}^{[1]}(\vec{r},t)}{\partial t} - i \frac{\partial \vec{H}^{[2]}(\vec{r},t)}{\partial t}\right]\\ 
&- \vec{j_g}^{[1]}(\vec{r},t) + i \vec{j_g}^{[2]}(\vec{r},t), 
\end{split}
\end{equation}
\begin{equation}
\label{eq6bcd}
\begin{split}
\raisetag{40pt}
&\left[\nabla\times(\vec{H}^{[1]}(\vec{r},t) - i\vec{H}^{[2]}(\vec{r},t)) \right] = \\
&\epsilon_0 \left[\frac{\partial \vec{E}^{[1]}(\vec{r},t)}{\partial t} - i \frac{\partial \vec{E}^{[2]}(\vec{r},t)}{\partial t}\right]\\ 
&+ \vec{j_e}^{[1]}(\vec{r},t) - i \vec{j_e}^{[2]}(\vec{r},t), 
\end{split}
\end{equation}
\begin{equation}
\label{eq7abcd}
(\nabla \cdot (\vec{E}^{[1]}(\vec{r},t) - i\vec{E}^{[2]}(\vec{r},t))) = \rho^{[1]}_e(\vec{r},t) - i\rho^{[2]}_e(\vec{r},t),
\end{equation}
\begin{equation}
\label{eq8abcd}
(\nabla \cdot (\vec{H}^{[1]}(\vec{r},t) - i\vec{H}^{[2]}(\vec{r},t))) = \rho^{[1]}_g(\vec{r},t) - i\rho^{[2]}_g(\vec{r},t),
\end{equation}
where $\vec{j_e}^{[1]}(\vec{r},t)$, $\vec{j_e}^{[2]}(\vec{r},t)$, $\vec{j_g}^{[1]}(\vec{r},t)$,  $\vec{j_g}^{[2]}(\vec{r},t)$ are correspondingly electric  and magnetic current densities, which by dual transformations are obeing to  the relation like to (\ref{eq1c}) \cite{Tomilchick}, they  are designated like to (\ref{eq10c}),
$\rho^{[1]}_e(\vec{r},t)$, $\rho^{[2]}_e(\vec{r},t)$, $\rho^{[1]}_g(\vec{r},t)$,  $\rho^{[2]}_g(\vec{r},t)$ are correspondingly electric  and magnetic charge densities, which transformed and designated like to field strengths and currents. In fact, the system of equations (\ref{eq5abc}), (\ref{eq6bcd}), (\ref{eq7abcd}), (\ref{eq8abcd}) represent the integrated Maxwell equations for two kinds of 
EM-fields (photon fields in quantum case), which differ by parities of vector and scalar quantities, entering in equations, under space inversion transformations. 
 So, the components $\vec{E}^{[1]}(\vec{r},t)$, $\vec{H}^{[2]}(\vec{r},t)$, $\vec{j_e}^{[1]}(\vec{r},t)$ have uneven parity, $\vec{E}^{[2]}(\vec{r},t)$, $\vec{H}^{[1]}(\vec{r},t)$,$\vec{j_e}^{[2]}(\vec{r},t)$ have even parity, $\rho^{[1]}_e(\vec{r},t)$, $\rho^{[2]}_g(\vec{r},t)$ are scalars, $\rho^{[2]}_e(\vec{r},t)$, $\rho^{[1]}_g(\vec{r},t)$ are pseudoscalars. In the case, when $\vec{j'_g}(\vec{r},t) = 0$, $\rho'_g(\vec{r},t) = 0$ we obtain the equations of usual singly charge electrodynamics for compound EM-field in mathematically correct form, which allows to separate the components of EM-field with various parities $P$ under space inversion. Let us accentuate once again, that the idea, that vector quantities, which characterize EM-field, are compound quantities and inlude both gradient and solenoidal parts, that is uneven and even parts under space inversion was put forward earlier in implicit form in \cite{Tomilchick}.  The representation in explicit form by equations (\ref{eq5abc}), (\ref{eq6bcd}), (\ref{eq7abcd}), (\ref{eq8abcd}) seems to be actual, since field vector and scalar functions with various $t$- and $P$-parities are mathematically heterogeneous and, for instance, their simple linear combination, for instance, for $P$-uneven and $P$-even  electric field vector-functions $\vec{E}^{[1]}(\vec{r},t)$, $\vec{E}^{[2]}(\vec{r},t)$
\begin{equation}
\label{eq5abcdef}
\alpha_1 \vec{E}^{[1]}(\vec{r},t)  + \alpha_2\vec{E}^{[2]}(\vec{r},t)
\end{equation}
with coefficients $\alpha_1$, $\alpha_2$ from the field of real numbers, which is taking place in a number of  theoretical and experimental works, is collage. Similar situation was discussed in \cite{D_Yearchuck_A_Dovlatova} by analysis of Bloch vector symmetry under improper rotations. 
 Mathematically the objects,
which are like to (\ref{eq5abcdef}) can exist. Actually, to the set
of $ \{\vec{E}^{[1]}(\vec{r},t)\}$
and to the set of $\{\vec{E}^{[2]}(\vec{r},t)\}$
can  be put in
correspondence the affine space. However, given affine space corresponds
to direct sum of two usual vector spaces, consisting of
different physically objects, that is, it represents in fact also collage. Really,
given direct sum can be represented by direct sum of linear capsule
\begin{equation}
\label{eq3a}
\left\{\alpha^k_1\vec{E}^{[1]}(\vec{r},t)| \alpha^k_1 \in R, k \in N \right\},
\end{equation}
representing itself three-dimensional
vector space of the set $\{\vec{E}^{[1]}(\vec{r},t)\}$
and linear capsule 
\begin{equation}
\label{eq3b}
\left\{\alpha^l_2\vec{E}^{[2]}(\vec{r},t)| \alpha^l_2 \in R, l \in N \right\},
\end{equation}
representing itself three-dimensional
vector space of the set $\{\vec{E}^{[2]}(\vec{r},t)\}$. It is substantial,
that both the  vector
spaces cannot be considered to be subspaces of any three-dimensional or six-dimensional
vector spaces, consisting of uniform objects. Moreover, it is evident, that
the affine space, defined in that way, cannot be metrizable, when
considering it to be a single whole. It means in its turn, that the set of objects, given by (\ref{eq5abcdef})
are not vectors in usual algebraic meaning. Even Pythagorean theorem,
for instance, cannot be used. 

It can be shown, that the system, analogous to (\ref{eq5abc}), (\ref{eq6bcd}), (\ref{eq7abcd}), (\ref{eq8abcd}) can be obtained for the second pair   of 
EM-fields (photon fields in quantum case), which differ by parities of vector and scalar quantities, entering in equations, under time reversal.  Really it is easily to see, that Maxwell equations along with dual transformation symmetry, established by Rainich, given by relations (\ref{eq1b}) - (\ref{eq1c}), are symmetric relatively the dual transformations of another kind of all the vector and scalar quantities, characterizing EM-field, which, for instance, for electric and magnetic field strengh vector-functions can be presented in the following matrix form
\begin{equation}
\label{eq1bca}
 \left[\begin{array} {*{20}c}  \vec {E{''}} \\ \vec {H{''}} \end{array}\right] = \left[\begin{array} {*{20}c} \cosh\vartheta& i\sinh\vartheta  \\ -i\sinh\vartheta&\cosh\vartheta \end{array}\right]\left[\begin{array} {*{20}c}  \vec {E} \\ \vec {H} \end{array}\right],
\end{equation}
where  $\vartheta$ is arbitrary continuous parameter,
$\vartheta \in [0,2\pi]$. The relation (\ref{eq1bca}) can be rewritten in the form
\begin{equation}
\label{eq1bcae}
 \left[\begin{array} {*{20}c}  \vec {E{''}} \\ \vec {H{''}} \end{array}\right] = \left[\begin{array} {*{20}c} \cos i\vartheta& \sin i\vartheta  \\ -\sin i\vartheta&\cos i\vartheta \end{array}\right]\left[\begin{array} {*{20}c}  \vec {E} \\ \vec {H} \end{array}\right]. 
\end{equation}
In particular, if $\vartheta$ is polar angle of coordinate system in the plane, determined by $\vec {E}$ and $\vec {H}$, the transformations (\ref{eq1bca}) represent themselves hyperbolic rotations in $(\vec {E}, \vec {H})$-plane. Let us call the transformations (\ref{eq1bca}) by hyperbolic dual transformations. It represents the interest to consider the following particular case of hyperbolic dual transformations. We can define parameter $\vartheta$   according to relation
\begin{equation}
\label{eq2bcae}
\tanh\vartheta = \frac{V}{c} = \beta,
\end{equation}
where $V$ is  the velocity of the frame of reference, moving along $x$-axis in 3D  real subspace of ${^1}R_4$  Minkowski space. We can  also to set up in conformity to the plane $(x_2, x_3)$ in  Minkowski space, the plane $ (\vec {E}, \vec {H})$ , in which $\vec {E}, \vec {H}$  are orthogonal and are directed along abscissa and ordinate axes correspondingly (or vice versa). Then we obtain
\begin{equation}
\label{eq3bcae}
\begin{split}
\raisetag{40pt}
&|\vec {E{''}}| = \frac {|\vec {E}| + \beta|\vec {H}|}{\sqrt{1-\beta^2}}\\ 
&|\vec {H{''}}| = \frac {|\vec {H}| - \beta|\vec {E}|}{\sqrt{1-\beta^2}}, 
\end{split}
\end{equation}
(or similar relations, in which $\vec {E},  \vec {H}$ are interchanged by places).
In vector form given transformations are 
\begin{equation}
\label{eq4bcae}
\begin{split}
\raisetag{40pt}
&\vec {E{''}} = \frac {\vec {E} + \frac{1}{c}[\vec {H} \times \vec{V} ]}{\sqrt{1-\beta^2}}\\ 
&\vec {H{''}} = \frac {\vec {H} - \frac{1}{c}[\vec {E} \times \vec{V} ]}{\sqrt{1-\beta^2}}, 
\end{split}
\end{equation}
Therefore, it is seen, that $\vec {E}$,  $\vec {H}$ are transformed like to $x_0$ and $x_1$ coordinates (or vice versa) of the space ${^1}R_4$. It follows from here, that both the vectors $\vec {E{''}}$,  $\vec {H{''}}$ have $t$-even and $t$-uneven components in general case. We see also that Lorentz-invariance of Maxwell equations is particular case of more general hyperbolic dual symmetry. It means, that restriction to only Lorentz-invariance in consideration of Maxwell equations' symmetry, which is usually used, constricts the concept on the EM-field itself and it is thereby constricting the possibilities for completeness of its practical usage.

Taking into account 
(\ref{eq1abcd}) we obtain the relations, which are similar to (\ref{eq1bcd}), which can be rewritten in the similar to (\ref{eq1c}) form, that is, we have
 \begin{equation} 
\label{eq1c'}
\begin{split}
\raisetag{40pt}                                                    
&\vec {E{''}} = \vec {E} \cos i\vartheta - i\vec {E} \sin i\vartheta\\
&\vec {H{''}} = \vec {H} \cos i\vartheta - i\vec {H}  \sin i\vartheta.
\end{split}
\end{equation}
It is proof in general case, that each of two independent  Maxwellian field components with even and uneven parities under space inversion is also compound and it consists of two independent   components with even and uneven parities under time reversal. Then imposing designations 
 \begin{equation} 
\label{eq10cc}
\begin{split}
\raisetag{40pt}                                                    
& \vec {E} \cos i\vartheta = \vec{E}^{[3]},  \vec {E} \sin i\vartheta = \vec{E}^{[4]} \\
&\vec {H} \cos i\vartheta = \vec{H}^{[3]}, \vec {H} \sin i\vartheta =  \vec{H}^{[4]}
\end{split}
\end{equation}
and considering the vector-functions $(\vec{E}^{[1]}(\vec{r},t)$, $\vec{E}^{[2]}(\vec{r},t))$ and $(\vec{H}^{[1]}(\vec{r},t)$, $\vec{H}^{[2]}(\vec{r},t))$ to be definitional domain for the vector-functions $ \vec {E{''}}(\vec{r},t)$, $\vec {H{''}}(\vec{r},t)$ correspondingly,
the Maxwell equations for the components of the field $(\vec{E}^{[1]}$, $\vec{H}^{[1]})$ and $(\vec{E}^{[2]}$, $\vec{H}^{[2]})$ have the same form and they are 
\begin{equation}
\label{eq5abcc}
\begin{split}
\raisetag{40pt}
&\left[\nabla\times(\vec{E}^{[3]}(\vec{r},t) - i\vec{E}^{[4]}(\vec{r},t)) \right] = \\
&- \mu_0 \left[\frac{\partial \vec{H}^{[3]}(\vec{r},t)}{\partial t} - i \frac{\partial \vec{H}^{[4]}(\vec{r},t)}{\partial t}\right]\\ 
&- \vec{j_g}^{[3]}(\vec{r},t) + i \vec{j_g}^{[4]}(\vec{r},t), 
\end{split}
\end{equation}
\begin{equation}
\label{eq6bcdd}
\begin{split}
\raisetag{40pt}
&\left[\nabla\times(\vec{H}^{[3]}(\vec{r},t) - i\vec{H}^{[4]}(\vec{r},t)) \right] = \\
&\epsilon_0 \left[\frac{\partial \vec{E}^{[3]}(\vec{r},t)}{\partial t} - i \frac{\partial \vec{E}^{[4]}(\vec{r},t)}{\partial t}\right]\\ 
&+ \vec{j_e}^{[3]}(\vec{r},t) - i \vec{j_e}^{[4]}(\vec{r},t), 
\end{split}
\end{equation}
\begin{equation}
\label{eq7abccd}
(\nabla \cdot (\vec{E}^{[3]}(\vec{r},t) - i\vec{E}^{[4]}(\vec{r},t))) = \rho^{[3]}_e(\vec{r},t) - i\rho^{[4]}_e(\vec{r},t),
\end{equation}
\begin{equation}
\label{eq8abccd}
(\nabla \cdot (\vec{H}^{[3]}(\vec{r},t) - i\vec{H}^{[4]}(\vec{r},t))) = \rho^{[3]}_g(\vec{r},t) - i\rho^{[4]}_g(\vec{r},t),
\end{equation}
where $\vec{j_e}^{31]}(\vec{r},t)$, $\vec{j_e}^{[4]}(\vec{r},t)$, $\vec{j_g}^{[3]}(\vec{r},t)$, $\vec{j_g}^{[4]}(\vec{r},t)$ are, correspondingly, electric  and magnetic current densities,
$\rho^{[3]}_e(\vec{r},t)$, $\rho^{[4]}_e(\vec{r},t)$, $\rho^{[3]}_g(\vec{r},t)$,  $\rho^{[4]}_g(\vec{r},t)$ are, correspondingly, electric  and magnetic charge densities, which transformed and designated like to field strengths and currents. In fact  the system of equations (\ref{eq5abcc}), (\ref{eq6bcdd}), (\ref{eq7abccd}), (\ref{eq8abccd}) represent itself corredtly integrated Maxwell equations for two kinds of 
EM-fields (photon fields in quantum case), which differ by parities of vector and scalar quantities, entering in equations, under time reversal. So, the components $\vec{E}^{[3]}(\vec{r},t)$, $\vec{H}^{[4]}(\vec{r},t)$, $\vec{j_e}^{[3]}(\vec{r},t)$ have uneven parity, $\vec{E}^{[4]}(\vec{r},t)$, $\vec{H}^{[3]}(\vec{r},t)$, $\vec{j_e}^{[4]}(\vec{r},t)$ have even parity, $\rho^{[3]}_e(\vec{r},t)$, $\rho^{[4]}_g(\vec{r},t)$ are scalars, $\rho^{[4]}_e(\vec{r},t)$, $\rho^{[3]}_g(\vec{r},t)$ are pseudoscalars. In the case, when $\vec{j{''}_g}(\vec{r},t) = 0$, $\rho{''}_g(\vec{r},t) = 0$ we obtain the equations of usual singly charge electrodynamics for two  components of EM-field with various parities under space inversion, at that either of the two consist also of two  components of EM-field with various parity under time reversal.

It is easily to see,  that invariants
for EM-field, consisting of two  hyperbolic dually symmetric parts, that is at  $\vartheta \neq 0$ have the form, analogous to (\ref{eq1bcde}) and they can be obtained, if parameter $\theta$ to replace by $i\vartheta$. They are \begin{equation}
\label{eq1bcdee}
 \left[ \vec {E}^2 - \vec {H}^2 + 2i(\vec {E}\vec {H})\right]  e^ {2\vartheta} = inv.
\end{equation}
Consequently, two real invariats at  $\vartheta \neq 0$ have the form
\begin{equation} 
\label{eq1cacde}
\begin{split}
\raisetag{40pt}                                                    
&(\vec {E}^2 - \vec {H}^2) e^{2\vartheta} = I_1{''} = inv, \\
&2(\vec {E}\vec {H}) e^{2\vartheta} = I_2{''}  = inv.
\end{split}
\end{equation}
  It follows  from  relation (\ref{eq1cacde}), that in  both the cases, that is,  at $\vartheta = 0$ and  at fixed $\vartheta \neq 0$, we obtain in fact well known EM-field invariants, since factor $e^ {2\vartheta}$ at fixed $\vartheta$ seems to be insufficient. At the same time at arbitrary $\vartheta$,which can be considered to be additional variable, the relation
\begin{equation} 
\label{eq1cacdeg}
\begin{split}
\raisetag{40pt}                                                    
 \frac {I_1{''}}{I_2{''}} =                        
 \frac {I_1}{I_2} = W = inv 
\end{split}
\end{equation}
is taking place. It is seen, that the value of $W$ is independent on $\vartheta$. It means physically, that the absolute values of both the  vector-functions $\vec {E}(\vec{r},t)$ and $\vec {H}(\vec{r},t)$ are changed synchronously by hyperbolic dual transformations.

So, the usage of complex number theory
allows to represent  correctly the electrodynamics for two photon fields,  which differs by parities under space inversion or time reversal by  the same single system of generalized  Maxwell equations. At the same time, we have two related sets, that is pairs of complex vector and scalar functions, which are ordered in their $P$- and $t$-parities. It corresponds to definition   of quaternions. 
Really, any quaternion number $x$ can be determined according to relation
\begin{equation} 
\label{eq1cacdegh}
\begin{split}
\raisetag{40pt}
&x = (a_1 + ia_2)e + (a_3 + ia_4)j,
\end{split}
\end{equation}
where $\{a_m\}\in R, m = \overline{1,4}$ and
 $e, i, j, k$ produce basis, elements of which are satisfying the conditions  
\begin{equation} 
\label{eq15cacdegh}
\begin{split}
\raisetag{40pt}\\
&(ij) = k, (ji) = -k, (ki) = j, (ik)= -j,\\
&(ei)= (ie) = i, (ej)= (je) = j, (ek) = (ke) = k.
\end{split}
\end{equation}

Let us designate the quantities
\begin{equation} 
\label{eq1cadegh}
\begin{split}
\raisetag{40pt} 
&(\vec{E}^{[1]}(\vec{r},t) - i\vec{E}^{[2]}(\vec{r},t)) +
(\vec{E}^{[3]}(\vec{r},t) - i\vec{E}^{[4]}(\vec{r},t)) j =\\ &\vec{\mathfrak{E}}(\vec{r},t)\\
&(\vec{H}^{[1]}(\vec{r},t) - i\vec{H}^{[2]}(\vec{r},t)) +
(\vec{H}^{[3]}(\vec{r},t) - i\vec{H}^{[4]}(\vec{r},t)) j =\\ &\vec{\mathfrak{H}}(\vec{r},t)\\
&(\vec{j_e}^{[1]}(\vec{r},t) - i \vec{j_e}^{[2]}(\vec{r},t)) +
(\vec{j_e}^{[3]}(\vec{r},t) - i \vec{j_e}^{[4]}(\vec{r},t)) j =\\ &\vec{\mathfrak{j_e}}(\vec{r},t)\\
&(- \vec{j_g}^{[1]}(\vec{r},t) + i \vec{j_g}^{[2]}(\vec{r},t)) +
(- \vec{j_g}^{[3]}(\vec{r},t) + i \vec{j_g}^{[4]}(\vec{r},t)) j =\\ &\vec{\mathfrak{j_g}}(\vec{r},t)\\
&(\rho^{[1]}_e(\vec{r},t) - i\rho^{[2]}_e(\vec{r},t)) +
(\rho^{[3]}_e(\vec{r},t) - i\rho^{[4]}_e(\vec{r},t)) j =\\ &\mathfrak{\rho_e}(\vec{r},t)\\
&(\rho^{[1]}_g(\vec{r},t) - i\rho^{[2]}_g(\vec{r},t)) +
(\rho^{[3]}_g(\vec{r},t) - i\rho^{[4]}_g(\vec{r},t)) j = \\ &\mathfrak{\rho_g}(\vec{r},t),
\end{split}
\end{equation}
where
 \begin{equation} 
\label{eq31cadegh}
\begin{split}
\raisetag{40pt} 
&\vec{E}^{[1]}(\vec{r},t), \vec{H}^{[2]}(\vec{r},t), \vec{j_e}^{[1]}(\vec{r},t),\\
&\vec{j_g}^{[2]}(\vec{r},t), \rho^{[1]}_e(\vec{r},t), \rho^{[2]}_g(\vec{r},t)
\end{split}
\end{equation} are $P$-uneven, $t$-even,
 \begin{equation} 
\label{eq32cadegh}
\begin{split}
\raisetag{40pt}
&\vec{E}^{[2]}(\vec{r},t), \vec{H}^{[1]}(\vec{r},t), \vec{j_e}^{[2]}(\vec{r},t),\\ &\vec{j_g}^{[1]}(\vec{r},t), \rho^{[2]}_e(\vec{r},t), \rho^{[1]}_g(\vec{r},t)
\end{split}
\end{equation}
 are $P$-uneven, $t$-uneven,
 \begin{equation} 
\label{eq33cadegh}
\begin{split}
\raisetag{40pt}
&\vec{E}^{[3]}(\vec{r},t), \vec{H}^{[4]}(\vec{r},t), \vec{j_e}^{[3]}(\vec{r},t),\\ &\vec{j_g}^{[4]}(\vec{r},t), \rho^{[3]}_e(\vec{r},t), \rho^{[4]}_g(\vec{r},t)
\end{split}
\end{equation}
 are $P$-even, $t$-even,
\begin{equation} 
\label{eq34cadegh}
\begin{split}
\raisetag{40pt}
&\vec{E}^{[4]}(\vec{r},t), \vec{H}^{[3]}(\vec{r},t), \vec{j_e}^{[4]}(\vec{r},t),\\ &\vec{j_g}^{[3]}(\vec{r},t), \rho^{[4]}_e(\vec{r},t), \rho^{[3]}_g(\vec{r},t)
\end{split}
\end{equation}
 are $P$-even, $t$-uneven.  According to definition   of quaternions $\vec{\mathfrak{E}}(\vec{r},t)$, $\vec{\mathfrak{H}}(\vec{r},t)$, $\vec{\mathfrak{j_e}}(\vec{r},t)$, $\vec{\mathfrak{j_g}}(\vec{r},t)$, $\mathfrak{\rho_e}(\vec{r},t)$,
$\mathfrak{\rho_g}(\vec{r},t)$ are quaternions. It means, that EM-field has quaternion structure and dual and hyperbolic dual symmetry of  Maxwell equations will take proper account, if all the  vector and scalar quantities to represent in quaternion form. 
Consequently, we have
\begin{equation}
\label{eq5abcce}
\begin{split}
\raisetag{40pt}
\left[\nabla\times(\vec{\mathfrak{E}}(\vec{r},t)) \right] = 
- \mu_0 \left[\frac{\partial \vec{\mathfrak{H}}(\vec{r},t)}{\partial t} \right] 
- \vec{\mathfrak{j_g}}(\vec{r},t), 
\end{split}
\end{equation}
\begin{equation}
\label{eq6bcdde}
\begin{split}
\raisetag{40pt}
\left[\nabla\times(\vec{\mathfrak{H}}(\vec{r},t)) \right] = 
\epsilon_0 \left[\frac{\partial \vec{\mathfrak{E}}(\vec{r},t)}{\partial t} \right] 
&+ \vec{\mathfrak{j_e}}(\vec{r},t), 
\end{split}
\end{equation}
\begin{equation}
\label{eq7abccde}
(\nabla \cdot (\vec{\mathfrak{E}}(\vec{r},t))) = \mathfrak{\rho_e}(\vec{r},t),
\end{equation}
\begin{equation}
\label{eq8abccde}
(\nabla \cdot (\vec{\mathfrak{H}}(\vec{r},t))) = \mathfrak{\rho_g}(\vec{r},t)
\end{equation}
Therefore, symmetry of Maxwell equations under  dual transformations of both the kinds allows along with  generalization of Maxwell equations themselves   to  extend the field of application of Maxwell equations. It means also, that dual electrodynamics, developed by Tomilchick and co-authors, see, for instance, \cite{Tomilchick}, obtains additional ground. Basic field equations in dual electrodynamics \cite{Tomilchick}, \cite{Berezin}, being to be written separately for two type of independent photon fields with various parities under space inversion or time reversal, will be isomorphic to Maxwell equations in complex form.  It was in fact shown partly  earlier in  \cite{Tolkachev}, \cite{Berezin}, where  complex charge was taken into consideration. At the same time, all aspect of dual symmetry, leading to four-component quaternion form of Maxwell equations seem to be representing in \cite{Dovlatova_Yerchuck} for the first time. 

\section{Symmetry of Mechanics and Electrodynamics Differential  Equations and Quantum Theory}

The results of algebraic symmetry studies can be
generalized, if to take into consideration,
that differential equations $(\ref{eq4a}, \ref{eq4b})$  are invariant
under the same transformation $(\ref{eq10})$, $(\ref{eq11})$. Therefore, it is
seen, that the correspondence between the symmetry of differential equations and
 the mathematical nature [in the concept of the number theory] of the
quantities, incoming in given equations seems to be taking place. It was done in \cite{A_Dovlatova_D_Yerchuck} and will be reviewed in the paper presented. The following 
statements were established:

9.\textit{Differential equations, which are invariant under transformations of groups,
which are symmetry groups of mathematical numbers (considered in the
frame of the number theory) determine the mathematical nature of the quantities,
incoming in given equations}.
In the case of invariance of  differential equations under transformation $(\ref{eq10})$, that is in the case of invariance under transformations of multiplicative group of complex numbers, the proof seems to be evident. Really,  the multiplication, for instance, of full set of
field function on complex numbers means,
that the functions themselves have to be
complex. 
It seems to be essential, that differential
equations for dynamics of nonrelativistic classical mechanics
 are invariant under
transformation $(\ref{eq10})$ too. Really, the dynamics of classical mechanics systems is described by Lagrange equations or by equivalent  canonical Hamilton equations. It is known, that, for instance, Hamilton equations are invariant under contact transformations of variables, that is under the transformation of linear elements - positions and directions, but not points. The transformation $(\ref{eq10})$ is referred to given class. It means, that by
quantization all physical
quantities, which determine  the dynamics of classical mechanics systems  have to be represented by quantum-mechanical description
 by one of the variants of the
representation of complex numbers, in
particular, taking into account (\ref{eq66}), by Hermitian matrices.
Therefore, we have the proof of the
statement, being to be the proof of the
main postulate of quantum mechanics.

10.\textit{To any mechanical quantity can be set
up in the correspondence the Hermitian
matrix by quantization.}
It is in fact the consequence of the
statement 9.
The choose of construction of mathemaical apparatus of quantum mechanics on the base of Hermitian matrices is convenient,
however, it is the only one variant from the
infinity of variants of the representations
of quantum mechanical quantities by complex numbers.
We have to remark, that the
description of the processes in classical
mechanics by means of complex number
is also correct. But in very many practical
cases, for instance, for mechanical tasks,
described by Newton equation, force and
impulse can be characterized by the same
phase factor (that is, it can be not taking
into consideration).
Let us consider the case of invariance of  differential equations under transformation of anticommutative group of quaternion numbers.
It is the case of electrodynamics.

11.\textit{To any electrodynamics quantity can
be set up in the correspondence the
Quaternion (that is twice-Hermitian)
matrix by quantization of EM-field}.
 
 The presence along with vector quaternion characteristics  the independent scalar quaternion characterics of EM-field allows to  describe EM-field instead of unobservable vector and scalar potentials by observable electric field 4-vector-function with the components $E_\alpha(\vec{r},t) = \{E_x(\vec{r},t), E_y(\vec{r},t), E_z(\vec{r},t), i \frac{c \rho_e(\vec{r},t)}{\lambda}\}$ and (or in the case of free EM-field) by means of magnetic field 4-vector-function $H_\mu(\vec{r},t) = \{H_x(\vec{r},t), H_y(\vec{r},t), H_z(\vec{r},t), i \frac{c \rho_m(\vec{r},t)}{\lambda}\}$, where  $i c \rho_e(\vec{r},t)$, $ic\rho_m(\vec{r},t)$ are the $j_4(\vec{r},t)$-component of 4-current density, corresponding to contribution of electric and magnetic component  of charge densities correspondingly, $\lambda$ is  conductivity, which for the case of EM-field propagation in vacuum is  $\lambda_v$ = $\frac{1}{120\pi}$ $(Ohm)^{-1}$. 
Then electric $\vec{j}_e(\vec{r},t)$
 (and magnetic $\vec{j}_g(\vec{r},t)$ in general case) current densities in
the right side of Maxwell equations can be represented
by well known relations $\vec{j}_e(\vec{r},t) = \rho_e(\vec{r},t) \vec{v}_e(\vec{r},t)$, where $\vec{v}_e(\vec{r},t)$is charge velocity in joint system \{EM-field + matter\} and analogous relation (in general
case) for magnetic $\vec{j}_g(\vec{r},t)$ current density. Now the invariance of Maxwell equations under transformations $(\ref{eq10})$ becomes to be evident. Therefore, take into account the
statement 11, we obtain the independent proof, that
all EM-feld quantities have to be considered minimum
being to be complex quantities (for correct description
of electromagnetic phenomena). Given picture is practically always used both in the theory and in practical
applications. However, even given representation in some
cases is insufficient, it  concerns, for instance, the dynamics of optical transitions \cite{D_Yearchuck_A_Dovlatova}. At the same time, it is easily to see, that Maxwell equations are invariant under transformations of quaternion non-abelian multiplicative group. In its turn, it leads to conclusion, that really to EM-field functions  can
be set up in the correspondence the
Quaternion (that is twice-Hermitian)
matrices by the quantization of EM-field. It is in fact the consequence of the presence along with symmetry of Maxwll equations under transformations, given by  $(\ref{eq10})$, the Rainich \cite{Rainich} dual symmetry and additional dual symmetry, established in \cite{Dovlatova_Yerchuck}. Let us remark for comparison, that the equations of the dynamics of mechanical systems are not invariant  under transformations of quaternion multiplicative group. It is the consequence of non-abelian character of given group.

Therefore, the symmetry study of main differential equations of mechanics and electrodynamics  has shown, that differential equations, which
are invariant under transformations of groups, which are symmetry groups of mathematical
numbers (considered in the frame of the number theory) determine the mathematical nature of the
quantities, incoming in given equations. The main postulate of quantum mechanics, consisting in that, that to any mechanical
quantity can be set up into the correspondence the Hermitian matrix by quantization was proved. High symmetry of Maxwell equations, consisting in the presence along with gauge symmetry  under transformations, given by  $(\ref{eq10})$, the Rainich dual symmetry and additional hyperbolic dual symmetry, established in \cite{Dovlatova_Yerchuck}, allowed to show, that to EM-field functions, incoming in given equations, can be set up into the correspondence the Quaternion (twice-Hermitian) matrices by their quantization.

\section{Cavity Dually Symmetric Electrodynamics}

Let us find the conserving quantities, which correspond to dual and hyperbolic dual symmetries of Maxwell equations. It seems to be interesing to realize given task on concrete practically essential example of cavity EM-field. At the same time, in order to built the Lagrangian, which is adequate to given task, it seems to be reasonable to solve the following concomitant task - to find dually symmetric  solutions of Maxwell equations. It seems to be  understandable, that the general solutions of differential equations can also possess by the same symmetry, which have starting differential equations, nevetheless dual symmetry of the solutions of Maxwell equations was earlier not found.

\subsection{\textbf{Classical Cavity EM-Field}}

Suppose EM-field in volume rectangular cavity without any matter inside it and made up of perfectly electrically conducting walls. Suppose also, that the field is linearly polarized and without restriction of commonness let us choose the one of two possible polarization of   EM-field electrical component $\vec{E}(\vec{r},t)$ along $x$-direction. Then the vector-function $E_x(z,t) \vec{e}_x$  can be represented in well known form of Fourier sine series
\begin{equation}
\label{eq1ab}
\vec{E}^{[1]}(\vec{r},t) = 
E_x(z,t) \vec{e}_x = \left[\sum_{\alpha=1}^{\infty}A^{E}_{\alpha}q_{\alpha}(t)\sin(k_{\alpha}z)\right]\vec{e}_x ,
\end{equation}
where $q_{\alpha}(t)$ is amplitude of $\alpha$-th normal mode of the cavity, $\alpha \in N$, $k_{\alpha} = \alpha\pi/L$, $A^{E}_{\alpha}=\sqrt{2 \omega_{\alpha}^2m_{\alpha}/V\epsilon_0}$, $\omega_{\alpha} = \alpha\pi c/L$, $L$ is cavity length along z-axis, $V$ is cavity volume, $m_{\alpha}$ is parameter, which is introduced to obtain the analogy with mechanical harmonic oscillator. Let us remember, that the expansion in Fourier series instead of  Fourier integral expansion is  determined by known  discontinuity of $\vec{k}$-space, which is the result of finiteness of cavity volume. Particular sine case of Fourier  series is consequence of boundary conditions 
\begin{equation}
\label{eq1abc}
[\vec{n} \times\vec{E}]|_S = 0, (\vec{n} \vec{H})|_S = 0,
\end{equation}
which are held true for the perfect cavity considered. Here $\vec{n}$ is the normal to the surface $S$ of the cavity. It is easily to show, that $E_x(z,t)$ represents itself  a standing wave along z-direction.

Let us analyse the solutions of Maxwell equations for EM-field in a cavity in comparison with known solutions from the literature and to draw the attention to some mathematical details, which have, however, substantial physical consequences,  allowing to extend our insight to EM-field nature. For given reasons, despite on analysis simplicity, we will produce the consideration in details.
 Using the equation
\begin{equation}
\label{eq2ab}
\epsilon_0 \frac{\partial\vec{E}(z,t)}{\partial t } = \left[\nabla\times\vec{H}(z,t)\right],
\end{equation}
 we obtain  the expression for magnetic field
 \begin{equation}
\label{eq3ab}
\vec{H}(\vec{r},t) =  \left[\sum_{\alpha=1}^{\infty}A^{E}_{\alpha}\frac{\epsilon_0}{k_{\alpha}}\frac{dq_{\alpha}(t)}{dt}\cos(k_{\alpha}z) + f_{\alpha}(t)\right]\vec{e}_y,
\end{equation}   
where  $\{f_{\alpha}(t)\}$, $ \alpha \in N $, is the set of arbitrary  functions of the time. It is evident, that the expression for $ \vec{H}(\vec{r},t)$ (\ref{eq3ab}) is satisfying to boundary conditions (\ref{eq1abc}). 
The partial solution, in which the functions $\{f_{\alpha}(t)\}$ are identically zero, that is,  $\vec{H}(\vec{r},t)$ is
\begin{equation}
\label{eq3abc}
\vec{H}^{[1]}(\vec{r},t) =  \left[\sum_{\alpha=1}^{\infty}A^{E}_{\alpha} \frac{\epsilon_0}{k_{\alpha}} \frac{dq_{\alpha}(t)}{dt}\cos(k_{\alpha}z)\right] \vec{e}_y,
\end{equation} 
 is always used in all the EM-field literature. 
However, even in given case it is evident, that the Maxwellian field is complex field. Really, 
 using the equation
\begin{equation}
\label{eq4ab}
\left[ \nabla\times\vec{E}\right] = -\frac{\partial \vec{B}}{\partial t} = -\mu_0 \frac{\partial \vec{H}}{\partial t}
\end{equation}
it is easily to find the class of field functions $\{q_{\alpha}(t)\}$. They will satisfy to  differential equations
\begin{equation}
\label{eq5ab}
\frac{d^2q_{\alpha}(t)}{dt^2} + \frac{k_{\alpha}^2}{\mu_0\epsilon_0} q_{\alpha}(t)=0, \alpha \in N.
\end{equation}
Consequently,  we have
\begin{equation}
\label{eq6ab}
q_{\alpha}(t) = C_{1\alpha} e^{i\omega_{\alpha}t} + C_{2\alpha} e^{-i\omega_{\alpha}t}, \alpha \in N,
\end{equation}
where  $C_{1\alpha}, C_{2\alpha}, \alpha \in N$ are arbitrary constants.
Thus, real-valued free Maxwell field equations are resulting in well known in the theory of differential equations  
situation - the solutions are complex-valued functions. It means, that generally the field functions for free Maxwellian field in the cavity produce complex space. So, we obtain additional independent argument, that the  known conception, on the only  real-quantity definiteness of EM-field, has to be corrected. On the other hand, the equation (\ref{eq5ab}) has also the only real-valued general solution, which can be represented in the form
\begin{equation}
\label{eq6abc}
q_{\alpha}(t) = B_{\alpha} \cos(\omega_{\alpha}t + \phi_{\alpha}),
\end{equation}
where  $B_{\alpha}, \phi_{\alpha}, \alpha \in N$ are arbitrary constants. It is substantial, that the functions in real-valued general solution have a definite t-parity. 

Thus, we come independently on the previous consideration in Sec.III and Sec.IV to the conclusion,  that classical Maxwellian EM-field can be both real-quantity defined and complex-quantity defined.
 
It is interesting, that
there is the second physically substantial solution of Maxwell equations. Really, from general expression (\ref{eq3ab}) for the  field $\vec{H}(\vec{r},t)$ 
it  is easily to obtain differential equations  for $\{f_{\alpha}(t)\}$, $ \alpha \in N $, 
\begin{equation}
\label{eq7ab}
\begin{split}
&\frac{d f_{\alpha}(t)}{dt} + A^{E}_{\alpha}\frac{\epsilon_0}{k_{\alpha}}\frac{\partial^2q_{\alpha}(t)}{\partial t^2}\cos(k_{\alpha}z) \\
&- \frac {1}{\mu_0} A^{E}_{\alpha}k_{\alpha}q_{\alpha}(t)\cos(k_{\alpha}z) = 0.
\end{split}
\end{equation}
The  formal solution of given equations 
in general case is
\begin{equation}
\label{eq8ab}
f_{\alpha}(t) =  A^{E}_{\alpha} \cos(k_{\alpha}z)\left[\frac{k_{\alpha}}{\mu_0} \int\limits _{0}^{t} q_{\alpha}(\tau)d\tau -\frac{dq_{\alpha}(t)}{dt}\frac{\epsilon_0}{k_{\alpha}}\right]
\end{equation}
Therefore, we have the second  solution  of Maxwell equations for $\vec{H}(\vec{r},t)$ in the form
\begin{equation}
\label{eq9ab}
\vec{H}^{[2]}(\vec{r},t) = -\left\{\sum_{\alpha=1}^{\infty} A^{H}_{\alpha} q_{\alpha}'(t)\cos(k_{\alpha}z) \right\}\vec{e}_y,
\end{equation}
where $A^{H}_{\alpha}=\sqrt{2 \omega_{\alpha}^2m_{\alpha}/V\mu_0}$.
Similar consideration gives the second  solution for $\vec{E}(\vec{r},t)$ 
\begin{equation}
\label{eq10ab}
\vec{E}^{[2]}(\vec{r},t) = \left\{\sum_{\alpha=1}^{\infty} A^{E}_{\alpha}q_{\alpha}''(t)\sin(k_{\alpha}z)\right\}\vec{e}_x,
\end{equation}
The functions $q_{\alpha}'(t)$ and $q_{\alpha}''(t)$ in relationships (\ref{eq9ab}) and (\ref{eq10ab})  are
\begin{equation}
\label{eq11ab}
\begin{split}
&q_{\alpha}'(t) = {\omega_{\alpha}}\int\limits _{0}^{t} q_{\alpha}(\tau)d\tau\\
&q_{\alpha}''(t) = {\omega_{\alpha}}\int\limits _{0}^{t} q_{\alpha}'(\tau')d\tau'
\end{split}
\end{equation}
correspondingly. Owing to the fact, that the solutions have simple form of harmonic trigonometrical functions, the 
second solution for electric field differs from the first solution the only by sign, that is substantial,  and by inessential integration constants.  Integration constants  can be taken into account by means of redefinition of factor $m_{\alpha}$ in field amplitudes.    It is also evident, that if vector-functions $\vec{E}(\vec{r},t)$ and $\vec{H}(\vec{r},t)$ are the solutions of Maxwell equations, then vector-functions $\hat{T}\vec{E}(\vec{r},t)$ and $\hat{T}\vec{H}(\vec{r},t)$, where $\hat{T}$ is time inversion operator, are also the solutions of Maxwell equations. Moreover, if starting vector-function, to which operator $\hat{T}$ is applied is $t$-even, then there is $t$-uneven  solution, for instance, for magnetic component in the form 
\begin{equation}
\label{eq11abc}
\frac{\hat{T}[t \vec{H}(\vec{r},t)]}{t}, 
\end{equation}
where $t$ is time. 
It can be shown in a similar way, that dually symmetric solutions, which are $P$-even and $P$-uneven are also existing.

Therefore, there are the solutions with various combinations of the signs for vector-functions $\vec{E}(\vec{r},t)$ and $\vec{H}(\vec{r},t)$, which are realized simultaneously, that is, their linear combination with coefficients from the field $C$ of complex numbers will represent  
 the solution of Cauchy problem for Maxwell equations in correspondence with  known theorem, that the solution of Cauchy problem  for any systems of homogeneous linear equations in partial derivatives exists and it is unique  in the vicinity of any point of the  initial surface  (in the case, when the point selected is not characteristic point and the function, which determines given hypersurface is continuously differentiable). In other words, we obtain again  the agreement with Maxwell equation symmetry consideration.   Given property of EM-field  seems to be essential, since it permits passing for the processes, which seemingly are forbidden by CPT-theorem. For example, let us consider the resonance system EM-field plus matter in the cavity, in particular, the so called dressed state of some quasiparticles' system. Suppose, that wave function can be factorized, matter part is $P$- and $t$-even under space  and time inversion transformations, while EM-field part is  $P$-uneven. CPT-invariance will be preserved, since EM-field has simultaneously with $t$-even the  $t$-uneven component, determined by expression  (\ref{eq11abc}) [that is we have the same conclusion, obtained in previous Section].  Therefore, $t$-parity of the function $q_{\alpha}'(t)$ can be various, and in the case, if we  choose $t$-parity to be identical to the parity of the function $q_{\alpha}(t)$, the solution will be different in the meaning, that the field vectors will have opposite  $t$-parity in comparison with the  first  solution. It is  also evident, that boundary conditions are fullfilled for all the cases considered. 

To build the Lagrangian, we can choose the following sets of EM-field functions $\{u^{s,\pm}_{\alpha}(x)\}, s = 1, 2, \alpha \in N$, 
\begin{equation}
\label{eq19abc}
\begin{split}
&\{u^{1,\pm}_{\alpha}(x)\} = \{\sqrt{\epsilon_0}A^{E}_\alpha\sin k_\alpha(x_3) [q_\alpha(x_4) \pm i q^{''}_\alpha(x_4)]\}\\
&\{u^{2,\pm}_{\alpha}(x) = \{\sqrt{\mu_0}A^{H}_\alpha\cos k_\alpha(x_3) [-q{'}_\alpha(x_4) \pm \frac {i}{\omega_\alpha}\frac{dq_\alpha(x_4)}{dx_4}]\}
\end{split}
\end{equation} 
The functions $\{u^{s,\pm}_{\alpha}(x)\}, s = 1, 2, \alpha \in N$ are built from the components of   
the expansion in Fourier series of the fields $\vec{E}^{[1]}(\vec{r},t), \vec{E}^{[2]}(\vec{r},t)$ and
$\vec{H}^{[2]}(\vec{r},t), \vec{H}^{[1]}(\vec{r},t)$ correspondingly. At the same time the sets $\{u^{s,\pm}_{\alpha}(x)\}, s = 1, 2, \alpha \in N$ produce at fixed $x$ two  orthogonal countable bases, corresponding to  $s = 1, 2$ in two Hilbert spaces, which are formed by vectors $\mathfrak U^{[s,\pm]}(u^{s,\pm}_{1}(x), u^{s,\pm}_{2}(x), ...)$ for variable $x \in {^1}R_4$.
Really scalar product of two arbitrary vectors $\mathfrak U_i^{[s,\pm]}(u^{s,\pm}_{1}(x_i), u^{s,\pm}_{2}(x_i), ...)$ and $\mathfrak U_j^{[s,\pm]}(u^{s,\pm}_{1}(x_j), u^{s,\pm}_{2}(x_j), ...)$, that is
\begin{equation}
\label{eq19bcd}
\langle \mathfrak U_i^{[s,\pm]}(x_i)\mid \mathfrak U_j^{[s,\pm]}(x_j)\rangle
\end{equation}
is equal to 
\begin{equation}
\label{eq19cde}
\sum_{\alpha = 1}^{\infty}\int\limits _{0}^{L}u^{*s,\pm}_{\alpha}(x_{4,i}, z)u^{s,\pm}_{\alpha}(x_{4,j}, z) dz, s = 1, 2, 
\end{equation}
that means, that it is restricted, since the sum over $s$ represents the energy of the field in restricted volume.  Consequently, the norm of vectors can be defined by the relationship
\begin{equation}
\label{eq19def}
\begin{split}
&\|\mathfrak U^{[s,\pm]}(x)\| = \sqrt{\langle \mathfrak U^{[s,\pm]}(x)\mid \mathfrak U^{[s,\pm]}(x)\rangle} = \\
&\sqrt{\sum_{\alpha = 1}^{\infty}\int\limits _{0}^{L}u^{*s,\pm}_{\alpha}(x_{4,i}, z)u^{s,\pm}_{\alpha}(x_{4,j}, z) dz}, s = 1, 2. 
\end{split}
\end{equation}
Then vector distance is
\begin{equation}
\label{eq19efg}
d(\mathfrak U^{[s,\pm]}(x_i), \mathfrak U^{[s,\pm]}(x_j)) = \|\mathfrak U^{[s,\pm]}(x_i) - \mathfrak U^{[s,\pm]}(x_j)\|.
\end{equation}
So we obtain, that the vectors $\{\mathfrak U^{[s,\pm]}(x)\}$, $x \in {^1}R_4$ produce the space $L_2$ and taking into account the Riss-Fisher theorem, it means, that given vector space is complete, that in its turn means, that the spaces of vectors $\{\mathfrak U^{[s,\pm]}(x)\}$, $x \in {^1}R_4, s = 1, 2$, are Hilbert spaces.
Consequently, Lagrangian $L(x)$ can be represented in the following form
\begin{equation}
\label{eq20abcd}
\begin{split}
&L(x) = \sum_{s=1}^{2}\sum_{\mu=1}^{4}\sum_{\alpha=1}^{\infty} \frac{\partial u^{s,\pm}_{\alpha}(x)}{\partial x_\mu}\frac{\partial u^{*s,\pm}_{\alpha}(x)}{\partial x_\mu} \\
&- \sum_{s=1}^{2}\sum_{\mu=1}^{4}\sum_{\alpha=1}^{\infty} K(x) u^{s,\pm}_{\alpha}(x) u^{*s,\pm}_{\alpha}(x),
\end{split}
\end{equation}
where $ K(x)$  is factor, depending on the set of variables $ x = \{x_\mu\}, \mu =  \overline {1,4}$. 

Let us find the conserving quantity, corresponding to dual symmetry of Maxwell equations.   Dual transformation, determined by relation (\ref{eq1bc})
is the  transformation the only in the space of field three-dimensional vector-functions $\vec{E}, \vec{H}$, (let us designate it by  $(\vec{E}, \vec{H})$-space) and it does not touch upon the coordinates. It seems to be conveniet to define in given space the reference frame, then the  transformation, given by (\ref{eq1bc}) is the rotation of two component matrix vector-function
\begin{equation}
\label{eq21abcd}
 \|F\| = \left[\begin{array} {*{20}c}  \vec {E} \\ \vec {H} \end{array}\right].
\end{equation}
 Instead of two Hilbert space for two sets of vectors $\{\mathfrak U^{[s,\pm]}(x)\}$, $x \in {^1}R_4, s = 1, 2$ we can also define one Hilbert space for block row matrix vector function set
\begin{equation}
\label{eq22abcd}
\|\mathfrak U^{}(x)\| = \left[\mathfrak U^{[1,\pm]}(x) \mathfrak U^{[2,\pm]}(x)\right]
\end{equation}
with the set of  components, being to be row two-component matrices
\begin{equation}
\label{eq22abcde}
\{\|U_{\alpha}(x)\|\} = \{\left[u^{1,\pm}_{\alpha}(x) u^{2,\pm}_{\alpha}(x)\right]\},
\end{equation}
where $ \alpha \in N$. 
In general case, instead of parameter $\theta$ we can define rotation angles $\theta_{ik}$, $i, k = \overline {1,3}$ in 2D-planes of $(\vec{E}, \vec{H})$ functional space. It is evident, that $\theta_{ik}$ are antisymmetric under the indices $i, k = \overline {1,3}$.
According to N\"{o}ther theorem, the conserving quantity, corresponding to parameters $\theta_{ik}$ in dual transformations (\ref{eq1bc}), that is at $\theta_{ik}$ = $\theta_{12}$ is determined by 
relations like to (\ref{eq19a}) and (\ref{eq19b}). So,  we obtain
\begin{equation}
\label{eq19bac}
S^{\mu}_{12} = -[\sum_{\alpha=1}^{\infty}\frac{\partial{L}}{\partial(\partial_{\mu}\|U^*_{\alpha}\|)}  \|Y_{\alpha}\|] + c.c.,
\end{equation}
 where $ \mu = \overline {1,4}$  and it was taken into account, that $\|X_{\alpha}\|$ in matrix relation (\ref{eq19bac}), which is like to (\ref{eq19a}), is equal to zero. 
The factor $\frac{\partial{L}}{\partial(\partial_{\mu}\|U^*_{\alpha}\|)}$ in  (\ref{eq19bac})  is row matrix
\begin{equation}
\label{eq21cabde}
\frac{\partial{L}}{\partial(\partial_{\mu}\|U*_{\alpha}\|)} = \left[\frac{\partial{L}}{\partial(\frac{\partial u^{*,1\pm}_\alpha}{\partial x_\mu})} \frac{\partial{L}}{\partial(\frac{\partial u^{*,2\pm}_\alpha}{\partial x_\mu})}\right],
\end{equation}
matrix $\|Y_{\alpha}\|$ is product of matrices $ \|I_{\alpha}\|$  and $ \widetilde{\|U_{\alpha}(x)\|}$, that is 
\begin{equation}
\label{eq23cabde}
\|Y_{\alpha}\| = \|I_{\alpha}\|\left[\begin{array} {*{20}c} u_{\alpha }^{1\pm}  \\ u_{\alpha}^{2\pm}  \end{array}\right],
\end{equation}
where $\|I_{\alpha}\|$ is the matrix, which corresponds to infinitesimal operator of dual  or hyperbolic dual  transformations of $\alpha$-th mode of cavity EM-field. It represents in general case the product of three matrices, corresponding to rotation along three mutually perpendicular axes in 3D functional space above defined. So $\|I_{\alpha}\| =
\|I^1_{\alpha}\| \|I^2_{\alpha}\|\|I^3_{\alpha}\|$. The transformations in the form, which is given by (\ref{eq1bc}) correspond to $\theta_{23} = \theta$, $\theta_{12} = 0, \theta_{31} = 0$, that is,  $\|I^2_{\alpha}\| = \|I^3_{\alpha}\| = E$, where $E$ is unit $[2\times 2]$-matrix.  In the absence of  dispersive medium in the cavity $\|I_\alpha\|$ will be independent on $\alpha$. Moreover, it is easily to see, that   
infinitesimal operator with matrix $\|I_\alpha\|$ is the same for dual transformations, determined by   (\ref{eq1bc}) and hyperbolic dual  transformations, determined by   (\ref{eq1bca}). Really $\|I_\alpha\|$ in both the cases is
\begin{equation}
\label{eq62b}
\|I_\alpha\| = \left[\begin{array} {*{20}c}0&1 \\-1&0   \end{array}\right], 
\end{equation}
 for any $\alpha \in N$.

Conserving quantity is
\begin{equation}
\label{eq20cab}
\begin{split}
S^{4}_{12} = 
-\frac{i}{c} \int\{[\sum_{\alpha=1}^{\infty}\frac{\partial{L}}{\partial(\partial_{\mu}\|U^*_{\alpha}\|)}  \|Y_{\alpha}\|] + c.c.\}d^3x
\end{split}
\end{equation}
The structure of (\ref{eq20cab}) unambiguously indicates, that it is the component of  spin tensor \cite{Bogush}, \cite{Shirkov}, to which dual vector component can be set in the correspondence according to relation  
\begin{equation}
\label{eq24cab}
\begin{split}
&S^{4}_i = \varepsilon_{ijk}S^{4}_{jk}  =\\
&-\varepsilon_{ijk}\frac{i}{c} \int\{[\sum_{\alpha=1}^{\infty}\frac{\partial{L}}{\partial(\partial_{\mu}\|U^*_{\alpha}\|)}  \|Y_{\alpha}\|]_{jk} + c.c.\}d^3x, 
\end{split}
\end{equation}
where $\varepsilon_{ijk}$ is completely antisymmetric Levi-Civita 3-tensor.

Therefore, we obtain, that  the same physical conserving quantity corresponds to
dual and hyperbolic dual symmetry of Maxwell equations.  Taking into account the expressions for Lagrangian (\ref{eq20abcd}) and for  infinitesimal operator (\ref{eq62b}), in the geometry choosed, when vector $\vec{E}$ is directed along abscissa axis, vector $\vec{H}$ is directed along ordinate axis in  $(\vec{E}, \vec{H})$ functional space, we have
\begin{equation}
\label{eq63ac}
S^{\mu}_{12} = \sum_{\alpha=1}^{\infty}[\frac{\partial u^{*,1\pm}_\alpha}{\partial x_\mu}u^{2\pm}_\alpha - \frac{\partial u^{*,2\pm}_\alpha}{\partial x_\mu}u^{1\pm}_\alpha]  + c.c.
\end{equation}
and
\begin{equation}
\label{eq25cab}
S^{4}_3 = \varepsilon_{312}S^{4}_{12}  = -\frac{i}{c} \int\{[\sum_{\alpha=1}^{\infty}\frac{\partial{L}}{\partial(\partial_{\mu}\|U^*_{\alpha}\|)}  \|Y_{\alpha}\|] + c.c.\}d^3x, 
\end{equation}
It is projection of spin on the propagation direction. Therefore, we have in given case right away physically significant quantity - spirality. 

The relations (\ref{eq19bac}), (\ref{eq24cab}), (\ref{eq63ac}), (\ref{eq25cab}) show, that  spin of classical relativistic EM-field in the cavity and, correspondingly, spirality are additive
quantities and they represent the sum of cavity spin  [spirality]  modes. On the connection  of the conserving quantity, which is  invariant of dual symmetry, with spin was indicated in \cite{Tomilchick}, where free EM-field was considered with traditional Lagrangian, which uses vector potentials to be field functions.
The result obtained together with aforecited result in \cite{Tomilchick} lift 
dilemma on the necessity of using of given quantity by consideration of classical 
EM-field. Really, the situation was to some extent paradoxical, and it can be displayed by the following conversation between two disputant
physicists. "Spin  exists" - has insisted the first, referring on the appearance of additional tensor component in total tensor of moment - intrinsic moment - to be consequence of Minkowsky space symmetry under Lorentz transformations, "Spin  does not exists" - has insisted the second, referring on the metrized tensor of the moment, in which spin part is equal to zero \cite{Bogush} in distinction from canonical tensor. In other words, both disputants were in one's own way right. Dual symmetry leads to unambiguous conclusion "Spin  exists" and has to be taken into consideration by the solution of tasks, concerning both classical and quantum electrodynamics. Moreover, spin takes on special leading significance among the physical characteristics of EM-field, since the only spin (spirality in the simplest case above considered) combine two subsystems of photon fields, that is the subsystem of two fields, which have definite $P$-parity  (even and uneven) with the subsystem of two fields, which have definite $t$-parity (also even and uneven) into one system. In fact, we obtain the proof for four component structure of EM-field to be a single whole, that is confirmation along with the possibility of the representation of EM-field in four component quaternion form, given by (\ref{eq5abcce}), (\ref{eq6bcdde}), (\ref{eq7abccde}), (\ref{eq8abccde}), [sufficient condition]
the necessity of given representation. It
extends the overview on the nature of EM-field itself. It seems to be remarkable, that given result on the special leading significance of spin is in agreement with result in \cite{D_Yearchuck_A_Dovlatova}, where was shown,  that spin is quaternion vector of the state in Hilbert space, defined under ring of quaternions, of any quantum system (in the
frame of the chain model considered) interacting with EM-field.

It is  interesting, that the charge und current, being to be the components of 4-vector, which are transformed by corresponding representation of Lorentz group, are invariants of hyperbolic dual transformations, that is, they are also Lorentz invariants in the case, when both charge und current  are taken separately by $|\vec {E}| = |\vec {H}|$. The proof is evident, if to take into account, that  Lorentz transformations are particular case of hyperbolic dual transformations. It is seen immediately from the expressions for 4-current  and it means, that observers in various inertial frames will register the same value of the charge in correspondence with conclusion in \cite{Tomilchick}. It is connected with invariance of Lagrange equations and   expressions for 4-current by multiplication of field functions on arbitrary complex number, established in Section 1, since   hyperbolic dual transformations like to Rainich dual transformations are equivalent to the multiplication of field functions on some  complex number.

\subsection{Connection between Gauge Invariance of  EM-field and Analicity of its  Vector-Functions}

The methods of theory of function of complex variable seem to be also useful along with algebraic methods for  the study of complex fields. The first example is the  conclusion, that Maxwell equations for free EM-field, for which electric and magnetic vector-functions are suggested to be real vector-functions, represent themselves the analicity condition for the complex-valued vector-function 
\begin{equation}
\label{eq53q}
\vec{F}(\vec{r},t) \equiv \vec{H}(\vec{r},t) - i \vec{E}(\vec{r},t)
\end{equation}
 of two variables $\vec{r}$ and $t$, where $\vec{r}$ is variable, which belong to any spacelike hypersurface $V$ $\subset {^{1}R_4}$, that is $\vec{r} \in R^3$, $t\in (0,\infty)$ is time. The proof is evident, it is sufficient to write down Cauchy-Riemann conditions for complex-valued vector-function (\ref{eq53q}).

It seems to be interesting to ascertain, whether is there the connection between symmetry of dynamical systems, in particular, between gauge symmetry, and analytical properties of quantities, which are invariant under corresponding symmetry qroups. Let us consider  
 the complex-valued function $Q(\vec{r},t)$ = $Q_1(\vec{r},t) + iQ_2(\vec{r},t)$, which are defined by (\ref{eq22}, \ref{eq23}), if integration limits  in (\ref{eq22}, \ref{eq23}) are  variable, that is, we have then  the function of the same two variables $\vec{r}$ and $t$,  $\vec{r} \in R^3$, $t\in (0,\infty)$. At the same time it is also function of field functions $\|u(x)\|$, which satisfy  to Lagrange equations, that is, differential equations of the second order in partial derivatives. It follows from definition of differential equation solutions, that the field functions $\|u(x)\|$ are continuously differentiable functions, their first partial derivatives are also continuously differentiable functions and the second partial derivatives are continuous functions. So integrands in (\ref{eq22}, \ref{eq23}) are continuous functions of variables $\vec{r}$ and $t$. It is sufficient for variable integration limit  differentiation of integrals in relationships (\ref{eq22}, \ref{eq23}). Let us introduce complex vector-scalar variable $z = \vec{r} + ict$. 
Then the following statement takes place

8.\textit{Gauge-invariant complex-valued quantity of any complex relativistic classical field, that is, complex charge field function represents itself analytical function in complex "plane", determined by variable} $z = \vec{r} + ict$. 

To prove the statement, it is sufficient to show, that $ReQ(\vec{r},t)$ and $Im Q(\vec{r},t)$ of the function $Q(\vec{r},t) = Q_1(\vec{r},t) + iQ_2(\vec{r},t)$ are satysfying to Cauchy-Riemann conditions, that is, the following relationships take place
\begin{equation}
\label{eq6q}
\frac{\partial Q_1(\vec{r},t)}{\partial \vec{r}} = \frac{\partial Q_2(\vec{r},t)}{\partial t}, 
\end{equation}
\begin{equation}
\label{eq7q}
\frac{\partial Q_1(\vec{r},t)}{\partial t} = - \frac{\partial Q_2(\vec{r},t)}{\partial \vec{r}}.
\end{equation}
Let us solve (\ref{eq6q}) and (\ref{eq7q}) regarding to $Q_2(\vec{r},t)$ (the quantity $Q_1(\vec{r},t)$ is considered to be given). It is apparent, that

\begin{equation}
\label{eq54a}
\begin{split}
\raisetag{40pt}
&\frac{\partial Q_1(\vec{r},t)}{\partial \vec{r}} = \left.-\left[\frac{\partial{L(t,\vec{r}')}}{\partial \left(\frac{\partial u_i}{\partial x_4}\right)} u_{i} - \frac{\partial{L(t,\vec{r}')}}{\partial \left(\frac{\partial u_i}{\partial x_4}\right)^{*}} u_{i}^*\right]\right|_{\vec{r}} = \\ 
&\frac{\partial Q_2(\vec{r},t)}{\partial t}. 
\end{split}
\end{equation}
Consequently $Q_2(\vec{r},t)$ is
\begin{equation}
\label{eq55a}
\begin{split}
\raisetag{40pt}
&Q_2(\vec{r},t) = -\int\limits_{(t)}\frac{\partial{L(\vec{r},t')}}{\partial \left(\frac{\partial u_i}{\partial x_4}\right)} u_{i}dt' - \\
&\int\limits_{(t)}\frac{\partial{L(\vec{r},t')}}{\partial \left(\frac{\partial u_{i}^*}{\partial x_4}\right)} u_{i}^*dt' + f(\vec{r}).
\end{split}
\end{equation}
The equation for determination of $f(\vec{r})$ is
\begin{equation}
\label{eq56a}
\begin{split}
\raisetag{40pt}
&\frac{d f(\vec{r})}{d \vec{r}} = \frac{\partial}{\partial \vec{r}}\int\limits_{(t)}\left[\frac{\partial{L(\vec{r},t')}}{\partial \left(\frac{\partial u_i}{\partial x_4}\right)} u_{i} - \frac{\partial{L(\vec{r},t')}}{\partial \left(\frac{\partial u_{i}^*}{\partial x_4}\right)} u_{i}^*\right]dt' \\
&+ \frac{\partial}{\partial t} \int\limits_{(\vec{r})}\left[\frac{\partial{L(t,\vec{r}')}}{\partial \left(\frac{\partial u_i}{\partial x_4}\right)} u_{i} - \frac{\partial{L(t,\vec{r}')}}{\partial \left(\frac{\partial u_{i}^*}{\partial x_4}\right)} u_{i}^*\right]d\vec{r}'. 
\end{split}
\end{equation} 
So we have 
\begin{equation}
\label{eq57a}
\begin{split}
\raisetag{40pt}
&Q_2(\vec{r},t) =\\
&\int\limits_{(\vec{r})}\left\{\frac{\partial}{\partial t} \int\limits_{(\vec{r}'')}\left[\frac{\partial{L(\vec{r}',t)}}{\partial \left(\frac{\partial u_i}{\partial x_4}\right)}u_{i} - \frac{\partial{L(\vec{r}',t)}}{\partial \left(\frac{\partial u_{i}^*}{\partial x_4}\right)}u_{i}^*\right]d\vec{r}'\right\}d\vec{r}''.
\end{split}
\end{equation} 
Then, in suggestion, that dynamic system studied is autonomous, that is
$L(\vec{r},t) = L(\vec{r})$,  occupies volume $ v \subset R_3$ and taking into account the coincidence of integration ranges $(\vec{r}) = (\vec{r}'')$, we will have
\begin{equation}
\label{eq58a}
\begin{split}
\raisetag{40pt}
&Q_2(\vec{r},t) = \\
&v \int\limits_{(\vec{r})}\left[\frac{\partial{L(\vec{r}')}}{\partial \left(\frac{\partial u_i}{\partial x_4}\right)} \frac{\partial u_{i}}{\partial t} - \frac{\partial{L(\vec{r}')}}{\partial \left(\frac{\partial u_{i}^*}{\partial x_4}\right)}\frac{\partial u_{i}^*}{\partial t}\right]d\vec{r}'.
\end{split}
\end{equation} 
Further, taking into consideration, that general solution of general relativistic equation is superposition of monochromatic plane waves, which have the view 
$ u_i(t) \sim e^{-i \frac{\mathcal E}{\hbar} t}$, and making a transformation of variable $t \rightarrow ict = x_4$ in the simplest case of one plane wave we  obtain 
\begin{equation}
\label{eq59q}
\begin{split}
\raisetag{40pt}
&Q_2(\vec{r},x_4) = \\
&\frac{v \mathcal E }{\hbar c} \int\limits_{(\vec{r})}\left[\frac{\partial {L(\vec{r}')}}{\partial \left(\frac{\partial u_i}{\partial x_4}\right)} u_{i}(\vec{r}',x_4)\right. 
+ \left.\frac{\partial {L(\vec{r}')}}{\partial \left(\frac{\partial u_{i}^*}{\partial x_4}\right)} u_{i}^*(\vec{r}',x_4)\right] d\vec{r}'. 
\end{split}
\end{equation} 
We see, that relationships (\ref{eq22}) and (\ref{eq59q}) are coinciding to scaling factor. They will coincide fully, if to make a transformation of parameter $\beta \rightarrow \beta ^{'} = \beta \frac{v \mathcal E}{\hbar c}$. The statement is proved. 

The converse can also be proven and can be employed for independent establishing of existence of some physical quantities in complex fields. For the example studied the suggestion on analicity of charge function of EM-field leads to the existence of two quantities - real (electric) and imaginary (magnetic) components of the charge, which are invariant under gauge transformations.

\subsection{\textbf{Quantized Cavity EM-Field}}

The quantization of EM-field  was proposed for the first time still at the earliest stage of quantum physics in the works \cite{Born},\cite{Born_Heisenberg}, where quantum theory of dipole radiation was considered and the energy fluctuations in radiation field of blackbody have been calculated. The idea of Born-Jordan EM-field quantization is regarding of EM-field components to be matrices. At the same time quite another idea - to set up in the 
correspondence to each mode of radiation field the quantized harmonic oscillator, was proposed for the first time by Dirac  \cite{P.Dirac} and it is widely used in quantum electrodynamics (QED) including quantum optics \cite{Scully}, it is canonical quantization. Nevetheless at present in EM-field theory the first idea of quantization is also used. For instance, matrix representation of Maxwell equations in quantum optics \cite{Scully} corresponds to given idea.

EM-field potentials are used to be field functions by canonical quantization. At the same time to describe free  EM-field   it is sufficient to choose immediately the observable quantities - vector-functions $\vec{E}(\vec{r},t)$ and $\vec{H}(\vec{r},t)$ - to be field functions.
We use further given idea by EM-field quantization.

\subsubsection{\textbf{Time-Local Quantization of Cavity EM-Field}}

 We can start like to canonical quantization, from classical Hamiltonian, which for the first partial classical solution of Maxwell equations is

\begin{equation}
\label{eq12ab}
\begin{split}
&\mathcal{H}^{[1]}(t) = \frac{1}{2}\iiint\limits_{(V)}\left[\epsilon_0E_x^2(z,t)+\mu_0H_y^2(z,t)\right]dxdydz\\
&= \frac{1}{2}\sum_{\alpha=1}^{\infty}\left[m_{\alpha}\nu_{\alpha}^2q_{\alpha}^2(t) + \frac{p_{\alpha}^2(t)}{m_{\alpha}} \right],
\end{split}
\end{equation}
where
\begin{equation}
\label{eq42abc}
p_{\alpha} = m_{\alpha} \frac{dq_{\alpha}(t)}{dt}.
\end{equation}

 So, taking into consideration the relationship for Hamiltonian $\mathcal{H}^{[1]}(t)$ we set in correspondence to canonical variables ${q}_{\alpha}(t), {p}_{\alpha}(t)$, determined by the first partial solution of Maxwell equations,  the operators by usual way
\begin{equation}
\label{eq25ab}
\begin{split}
&\left[\hat {p}_{\alpha}(t) , \hat {q}_{\beta}(t)\right] = i\hbar\delta_{{\alpha}\beta}\\
&\left[\hat {q}_{\alpha}(t) , \hat {q}_{\beta}(t)\right] = \left[\hat {p}_{\alpha}(t) , \hat {p}_{\beta}(t)\right] = 0,
\end{split}
\end{equation}
where
$\alpha, \beta \in N$.
Introducing the operator functions  of time $\hat{a}_{\alpha}(t)$ and $ \hat{a}^{+}_{\alpha}(t)$
\begin{equation}
\label{eq26ab}
\begin{split}
&\hat{a}_{\alpha}(t) = \frac{1}{ \sqrt{ 2 \hbar  m_{\alpha} \omega_{\alpha}}} \left[ m_{\alpha} \omega_{\alpha}\hat {q}_{\alpha}(t) + i \hat {p}_{\alpha}(t)\right]\\
&\hat{a}^{+}_{\alpha}(t) = \frac{1}{ \sqrt{ 2 \hbar  m_{\alpha} \omega_{\alpha}}} \left[ m_{\alpha} \omega_{\alpha}\hat {q}_{\alpha}(t) - i \hat {p}_{\alpha}(t)\right],
\end{split}
\end{equation}
we obtain the operator functions of canonical variables in the form
\begin{equation}
\label{eq27ab}
\begin{split}
&\hat {q}_{\alpha}(t) = \sqrt{\frac{\hbar}{2 m_{\alpha} \omega_{\alpha}}} \left[\hat{a}^{+}_{\alpha}(t) + \hat{a}_{\alpha}(t)\right]\\
&\hat {p}_{\alpha}(t) = i \sqrt{\frac{\hbar m_{\alpha} \omega_{\alpha}}{2}} \left[\hat{a}^{+}_{\alpha}(t) - \hat{a}_{\alpha}(t)\right]. 
\end{split}
\end{equation}
Then EM-field  operator functions are obtained right away and they are
\begin{equation}
\label{eq28ab}
\hat{\vec{E}}(\vec{r},t) = \{\sum_{\alpha=1}^{\infty} \sqrt{\frac{\hbar \omega_{\alpha}}{V\epsilon_0}} \left[\hat{a}^{+}_{\alpha}(t) + \hat{a}_{\alpha}(t)\right] \sin(k_{\alpha} z)\} \vec{e}_x,
\end{equation}

\begin{equation}
\label{eq29ab}
\hat{\vec{H}}(\vec{r},t) = i \{\sum_{\alpha=1}^{\infty} \sqrt{\frac{\hbar \omega_{\alpha}}{V\mu_0}} \left[\hat{a}^{+}_{\alpha}(t) - \hat{a}_{\alpha}(t)\right] \cos(k_{\alpha} z)\} \vec{e}_y,
\end{equation} 
Taking into account the relationships (\ref{eq28ab}), (\ref{eq29ab}) and Maxwell equations, it is easily to find  an explicit form for the dependencies of operator functions   $\hat{a}_{\alpha}(t)$ and $ \hat{a}^{+}_{\alpha}(t)$ on the time. They are
\begin{equation}
\label{eq30ab}
\begin{split}
&\hat{a}^{+}_{\alpha}(t) = \hat{a}^{+}_{\alpha}(t = 0) e^{i\omega_{\alpha}t},\\
&\hat{a}_{\alpha}(t) = \hat{a}_{\alpha}(t = 0) e^{-i\omega_{\alpha}t},
\end{split}
\end{equation}
where $\hat{a}^{+}_{\alpha}(t = 0), \hat{a}_{\alpha}(t = 0)$ are constant, complex-valued in general case, operators. 

Physical sense of operator time dependent functions $\hat{a}^{+}_{\alpha}(t)$ and $\hat{a}_{\alpha}(t)$ is well known. They are creation  and annihilation operator of the $\alpha$-mode photon.
They are continuously differentiable operator functions of time. It means, that the time of  photon creation (annihilation) can be determined strictly, at the same time operator  functions $\hat{a}^{+}_{\alpha}(t)$ and $\hat{a}_{\alpha}(t)$  do not curry any information on the place, that is on space coordinates of given event.

It seems to be essential, that complex exponential dependencies in (\ref{eq30ab}) cannot be replaced by the real-valued harmonic trigonometrical functions. Really, if to suggest, that 
\begin{equation}
\label{eq31ab}
\hat{a}^{+}_{\alpha}(t) = \hat{a}^{+}_{\alpha}(t = 0)\cos\omega_{\alpha}t,
\end{equation}
then we obtain, that the following relation has to be taking place
\begin{equation}
\label{eq32ab}
\begin{split}
&[\hat{a}^{+}_{\alpha}(t = 0) - \hat{a}_{\alpha}(t = 0)]^{-1}[\hat{a}^{+}_{\alpha}(t = 0)\\
& + \hat{a}_{\alpha}(t = 0)] = \tan\omega_{\alpha}t.
\end{split}
\end{equation}
We see, that left-hand side in relation (\ref{eq32ab})  does not depend on time, right-hand side is depending. The contradiction obtained establishes an assertion.
Therefore, the quantized Maxwellian EM-field is  complex-valued  field in full correspondence with pure algebraic conclusion in Sec.III. 

Consequently, there is difference between classical and  quantized EM-fields, since classical EM-field can be determined by both complex-valued and real-valued functions. 
The fields $\vec{E}^{[2]}(\vec{r},t), \vec{H}^{[2]}(\vec{r},t)$ can be quantized in  the same way.
The operators $\hat{a}{''}_{\alpha}(t)$, $\hat{a}{''}^{+}_{\alpha}(t)$ are introduced analogously to (\ref{eq26ab}).
\begin{equation}
\label{eq33ab}
\begin{split}
&\hat{a}{''}_{\alpha}(t) = \frac{1}{ \sqrt{ 2 \hbar  m_{\alpha} \omega_{\alpha}}} \left[ m_{\alpha} \omega_{\alpha}\hat {q}{''}_{\alpha}(t) + i \hat {p}{''}_{\alpha}(t)\right]\\
&\hat{a}{''}^{+}_{\alpha}(t) = \frac{1}{ \sqrt{ 2 \hbar  m_{\alpha} \omega_{\alpha}}} \left[ m_{\alpha} \omega_{\alpha}\hat {q}{''}_{\alpha}(t) - i \hat {p}{''}_{\alpha}(t)\right]
\end{split}
\end{equation}
For the operators of field function we obtain
\begin{equation}
\label{eq34ab}
\begin{split}
&\hat{\vec{E}}^{[2]}(\vec{r},t) = \\
&\{\sum_{\alpha=1}^{\infty} \sqrt{\frac{\hbar \omega_{\alpha}}{V\epsilon_0}} \left[\hat{a}{''}^{+}_{\alpha}(t) + \hat{a}{''}_{\alpha}(t)\right] \sin(k_{\alpha} z)\} \vec{e}_1,
\end{split}
\end{equation}
\begin{equation}
\label{eq35ab}
\begin{split}
&\hat{\vec{H}}^{[2]}(\vec{r},t) = \\
&\{\sum_{\alpha=1}^{\infty} \sqrt{\frac{\hbar \omega_{\alpha}}{V\mu_0}} (-i) \left[\hat{a}{''}^{+}_{\alpha}(t) - \hat{a}{''}_{\alpha}(t)\right] \cos(k_{\alpha} z)\} \vec{e}_2.
\end{split}
\end{equation}
 In accordance with definition of complex quantities we  can built the following combination of solutions, satisfying to Maxwell equtions
\begin{equation}\label{eq36ab}
(\vec{E}^{[1]}(\vec{r},t), \vec{E}^{[2]}(\vec{r},t)) \rightarrow \vec{E}^{[1]}(\vec{r},t) +  i \vec{E}^{[2]}(\vec{r},t) = \vec{E}(\vec{r},t),
\end{equation}
\begin{equation}\label{eq37ab}
(\vec{H}^{[2]}(\vec{r},t), \vec{H}^{[1]}(\vec{r},t)) \rightarrow \vec{H}^{[2]}(\vec{r},t) +  i \vec{H}^{[1]}(\vec{r},t) = \vec{H}(\vec{r},t).
\end{equation}
Consequently, the electric and magnetic field operators for quantized EM-field, corresponding to general solution of Maxwell equations,  are
\begin{equation}\label{eq33}
\begin{split}
&\hat{\vec{E}}(\vec{r},t) =  \{\sum_{\alpha=1}^{\infty} \sqrt{\frac{\hbar \omega_{\alpha}}{V\epsilon_0}} \{\left[\hat{a}^{+}_{\alpha}(t) + \hat{a}_{\alpha}(t)\right]\\
& + i \left[\hat{a}{''}_{\alpha}(t) + \hat{a}{''}^{+}_{\alpha}(t)\right]\} \sin(k_{\alpha} z)\} \vec{e}_x,
\end{split}
\end{equation}
and
\begin{equation}\label{eq34}
\begin{split}
&\hat{\vec{H}}(\vec{r},t) =  \{\sum_{\alpha=1}^{\infty} \sqrt{\frac{\hbar \omega_{\alpha}}{V\mu_0}}\{ \left[\hat{a}^{}_{\alpha}(t) - \hat{a}^{+}_{\alpha}(t)\right] \\
& + i \left[\hat{a}{''}_{\alpha}(t) - \hat{a}{''}^{+}_{\alpha}(t)\right] \} \cos(k_{\alpha} z) \} \vec{e}_y,
\end{split}
\end{equation}
It is substantial, that both field operators $\hat{\vec{E}}(\vec{r},t)$ and $\hat{\vec{H}}(\vec{r},t)$ are Hermitian operators.

The method of EM-field quantization above considered is in fact the development of canonical quantization, proposed by Dirac. Further development can be made, if to 
take into account the independence and equal rights of all the coordinates $x_\mu,
\mu = \overline {1,4}$ in Minkowsky space $^{1}R_4$. Really, all physical events are taking place on finite segment of time. It leads in application to electrodynamics to  discontinuity of $\omega$- space of possible light frequences like to discontinuity of $\vec{k}$-space, which is the result of finiteness of cavity volume. It is interesting that, Dirac himself has in \cite{P.Dirac} written, that the theory proposed is not strictly relativistic, since the time everywhere is considered to be $c$-number instead of to consider it symmetrically with the space coordinates. From here follows unambiguously, that quantum electrodynamics, based on Dirac canonical EM-field quantization method is not fully relativistic and, correspondingly, it is not fully quantum theory. 
Nevetheless, despite the Dirac opinion, the time is  considered usually (in standard formulation of quantum theory) to be  not a dynamical observable, but a mere parameter marking the
evolution of a quantum system, that is the time is believed to be an external variable, which is independent on the dynamics of any given system.  It  is connected with Pauli's conclusion, based on well known his theorem, that the introduction of an time  operator $\hat{T}$ must fundamentally be abandoned and that the time $t$ in quantum mechanics has to be regarded to be an ordinary number \cite{Pauli}. In other words, consequence of Pauli's theorem is the nonequality in rights of time coordinate in comparison with space coordinates for description of quantum systems. Recently, given theorem was reconsidered (that became to be known for us from the results, reported by Galapon on the 7th International Conference "Quantum Theory and Symmetries", August 7-13, 2011, Prague \cite{Galapon}). 
 In particular, it has been proved \cite{Galapon}, \cite{Galapon_E}, that in quantum theory to classical variable "time", in contrast to conclusion of Pauli, can be put in the correspondence self-adjoint time  operator, like to space coordinates, energy, impulse et cetera.  Thus, the equality in rights of time coordinate and space coordinates was reestablished.

\subsubsection{\textbf{Space-Local  Quantization of Cavity EM-Field}} 

We will consider for the simplicity the dependence of EM-field vector-functions the only on $z$-space coordinate, which is choosed in propagation direction in $R_3$ $\in$ $^{1}R_4$. The generalization on 3D-case is simple and will be not considered. 
Taking into account the independence and equal rights of all the coordinates $x_\mu,
\mu = \overline {1,4}$ in Minkowsky space $^{1}R_4$ we can also make Fourier transform on the segment [0, T], where $T$ is fixed time value, that is to represent the EM-field vector-functions in the form
\begin{equation}\label{eq35}
E^{[1]}_x(z,t)\vec{e}_x = \left[ \sum_{\alpha=1}^{\infty}A'_{\alpha}q_{\alpha}(z)\sin(\omega_{\alpha}t)\right] \vec{e}_x,
\end{equation}
\begin{equation}\label{eq36}
\begin{split}
& {H}^{[1]}_y(z,t)\vec{e}_y = \\
& \left\{ - \epsilon_0 \sum_{\alpha=1}^{\infty}\left[ A'_{\alpha}\omega_{\alpha}\cos(\omega_{\alpha}t)\int\limits_{0}^{z} q_{\alpha}(z')dz' + H_{\alpha 0}(t)\right]\right\}\vec{e}_y,
\end{split}
\end{equation}
where $q_{\alpha}(z)$,$\alpha \in N$,  is  $\alpha$-th normal mode of the 4-dimensional cavity, which include time coordinate along with space coordinates, 
\begin{equation}\label{eq37}
k_{\alpha} = \frac{\alpha\pi}{cT},  A'_{\alpha} = \sqrt{\frac{2 \omega_{\alpha}^2 m_{\alpha}}{T \epsilon_0}}, \omega_{\alpha} = \frac{\alpha \pi}{T},
\end{equation}
  $\{H_{\alpha 0}(t)\}$, $\alpha \in N$, is the set of arbitrary  functions of the time. Then the Hamiltonian can be obtained taking into account the expressions for $ E^{[1]}_x(z,t)$, ${H}^{[1]}_y(z,t)$ and
integrating.  So we will have
\begin{equation}\label{eq38}
G^{[1]}(z) = \frac{1}{2}\sum_{\alpha=1}^{\infty}\{m_{\alpha}\omega^{2}_{\alpha}\left[\frac{dq_{\alpha}'(z)}{dz}\right]^{2} + \frac{1}{c^2} \omega^{4}_{\alpha}m_{\alpha}\left[q_{\alpha}'(z)\right]^{2}\},
\end{equation}
where the case with  $\{H_{\alpha 0}(t)\} \equiv 0$, $t \in [0, T]$ for all $\alpha \in N$  is choosed and
\begin{equation}\label{eq39}
 q_{\alpha}'(z) = \int\limits_{0}^{z} q_{\alpha}(z')dz'.
\end{equation}
By redefition of  the variables in accordance with relations
\begin{equation}\label{eq40a}
\begin{split}
&q_{\alpha}''(z) = \frac{1}{c}\omega_{\alpha} q_{\alpha}'(z),\\
&p_{\alpha}''(z) = m_{\alpha}\omega_{\alpha} \frac{dq_{\alpha}'(z)}{dz},
\end{split}
\end{equation}
the  Hamiltonian $G^{[1]}(z)$ will have the canonical form
\begin{equation}\label{eq41}
G^{[1]}(z) = \frac{1}{2}\sum_{\alpha=1}^{\infty}\{\frac{[p_{\alpha}''(z)]^2}{m_{\alpha}} + m_{\alpha}\omega^{2}_{\alpha}[q_{\alpha}''(z)]^{2}\}.
\end{equation}
It means, that
space coordinates' dependent quantization of cavity EM-field can be  realized in a similar manner with above described time dependent quantization. 
 So, we can define quite analogously the quantization rules by the relationships
\begin{equation}\label{eq42}
\begin{split}
&\left[\hat{p}{''}_{\alpha}(z) , \hat {q}{''}_{\beta}(z)\right] = i\lambda_{0}\delta_{{\alpha}\beta}\\
&\left[\hat {q}{''}_{\alpha}(z) , \hat {q}{''}_{\beta}(z)\right] = \left[\hat {p}{''}_{\alpha}(z) , \hat {p}{''}_{\beta}(z)\right] = 0,
\end{split}
\end{equation}
where $\alpha,\beta \in N$, $\lambda_{0}$ is analogue of Planck constant.  It is evident from $\lambda_{0}$-definition by (\ref{eq42}), that $\lambda_{0}$ and Planck constant have the same dimension, however their numerical coincidence seems to be unobvious, since Planck constant characterizes the "seizure" of the time by propagating of EM-field, while $\lambda_{0}$ characterises the "seizure" of the space.

The operators $\hat{a}{''}_{\alpha}(z)$, $\hat{a}{''}^{+}_{\alpha}(z)$ are defined also analogously to operators $\hat{a}{}_{\alpha}(t)$, $\hat{a}{}^{+}_{\alpha}(t)$ and they are
\begin{equation}\label{eq43}
\begin{split}
&\hat{a}{''}_{\alpha}(z) = \frac{1}{ \sqrt{ 2 m_{\alpha} \lambda_{0}  \omega_{\alpha}}} \left[m_{\alpha} \omega_{\alpha}\hat {q}{''}_{\alpha}(z) + i \hat {p}{''}_{\alpha}(z)\right]\\
&\hat{a}{''}^{+}_{\alpha}(z) = \frac{1}{ \sqrt{ 2 m_{\alpha} \lambda_{0} \omega_{\alpha}}} \left[m_{\alpha} \omega_{\alpha}\hat {q}{''}_{\alpha}(z) - i \hat {p}{''}_{\alpha}(z)\right].
\end{split}
\end{equation}
The dependencies of given scalar operator functions on coordinate $z$ in an explicit form for Maxwellian EM-field can be easily obtained by means of solutions of Maxwell equations and they are
\begin{equation}\label{eq44}
\begin{split}
&\hat{a}^{+}_\alpha(z) = \hat{a}^{+}_\alpha (0) e^{ik_\alpha z}\\
&\hat{a}_\alpha(z) = \hat{a}_\alpha(0) e^{-ik_\alpha z},
\end{split}
\end{equation}
where $\hat{a}^{+}_{\alpha}(0), \hat{a}_{\alpha}(0)$ are constant, complex-valued in general case, operators.
Let us remark in passing, that the dependencies (\ref{eq44}) on $z$-coordinate are similar to  dependencies $\hat{a}^{+}_{\alpha}(t), \hat{a}_{\alpha}(t)$ on time, which are given by  (\ref{eq30ab}).

From relationships  (\ref{eq43}) we obtain the expressions for operators of canonical variables $\hat {q}{''}_{\alpha}(z)$ and $\hat {p}{''}_{\alpha}(z)$ in the form
\begin{equation} \label{eq45}
\begin{split}
&\hat {q}{''}_{\alpha}(z) = \sqrt{\frac{\lambda_{0}}{2 m_{\alpha}\omega_{\alpha}}} \left[\hat{a}{''}^{+}_{\alpha}(z) + \hat{a}{''}_{\alpha}(z)\right]\\
&\hat {p}{''}_{\alpha}(z) = i \sqrt{\frac{m_{\alpha}\lambda_{0} \omega_{\alpha}}{2}} \left[\hat{a}{''}^{+}_{\alpha}(z) - \hat{a}{''}_{\alpha}(z)\right]. 
\end{split}
\end{equation}
Then it is easily to show, that Hamilton operator $\hat{G}^{[1]}(z)$ can be represented in the simple form
\begin{equation}\label{eq46}
\hat{G}^{[1]}(z) = \sum_{\alpha=1}^{\infty}\lambda_{0} \omega_{\alpha}\left[\hat{a}{''}^{+}_{\alpha}(z)\hat{a}{''}_{\alpha}(z) + \frac{1}{2}\right],
\end{equation}
which determines physical meaning of the operators
$\hat{a}{''}^{+}_{\alpha}(z)$ and $\hat{a}{''}_{\alpha}(z)$. It is evident, that they are operators of creation and annihilation of the photon at space coordinate $z$. So, we see, that it is possible by space coordinates' dependent quantization   to determine the place of photon creation (annihilation), however it is impossible to determine the time of photon creation (annihilation). Therefore we have reverse picture to the case of  the time dependent quantization, where (see previous Subsection)  it is possible to determine the time of photon creation (annihilation) and it is impossible to determine the place of photon creation (annihilation). The view of (\ref{eq46}), which is coinciding with view of known expressions for canonical  quantization, if $\lambda_{0}$ to replace by $\hbar$, confirms the conclusion, that dimension of constant of space coordinates' dependent quantization and dimension of Planck  constant are identical, that is $[\lambda_{0}]$ = $[\hbar]$.

From relationships  (\ref{eq43}) and (\ref{eq42}) we can obtain the expressions for commutation relations of the creation and annihilation operators $\hat{a}{''}^{+}_{\alpha}(z)$ and $\hat{a}{''}_{\alpha}(z)$. They are
\begin{equation}\label{eq47}
[\hat{a}{''}_{\alpha}(z),\hat{a}{''}^{+}_{\beta}(z)] = \hat{e}\delta_{\alpha\beta},
\end{equation}
where $\hat{e}$ is unit operator, $\alpha,\beta \in N$.

It seems to be evident, that the second case of EM-field quantization, that is   space coordinates' dependent quantization is acceptable for the quantization of any Coulomb field, which has nonzeroth curl, that takes place in  1D and in 2D systems. It was passed earlier for impossible  to quantize any Coulomb field, see for example \cite{Dutra}. The quantization of Coulomb field in lowdimensional aforesaid systems corresponds to the presence of own life of radiation Coulomb field in given systems, that is Coulomb field in lowdimensional  systems has the character of radiation field and it can exist without the sources, which have created given field.  Given conclusion seems to be substantial to gain a better understanding, for instance, of the properties of organic conductors, perfect nanowires and nanotubes, graphene and the systems like them, including 1D and 2D biological subsystems. 

The expressions for 
the operators of vector-functions of EM-field are similar in their structure to expressions, given by (\ref{eq33}), (\ref{eq34}) and they are
\begin{equation}\label{eq48}
\hat{\vec{E}}^{[1]}(\vec{r},t) = \{i\sum_{\alpha=1}^{\infty}  \sqrt{\frac{\lambda_{0}\omega_{\alpha}}{T\epsilon_0}} \sin\omega_{\alpha}t \left[\hat{a}{''}^{+}_{\alpha}(z) - \hat{a}{''}_{\alpha}(z)\right]\}\vec{e}_x
\end{equation}
 and
\begin{equation}\label{eq49}\begin{split}
&\hat{\vec{H}}^{[1]}(\vec{r},t) =\\ 
&\{-\sum_{\alpha=1}^{\infty}\sqrt{\frac{\lambda_{0}\omega_{\alpha}}{T\mu_0}} \cos\omega_{\alpha}t\left[\hat{a}{''}^{+}_{\alpha}(z) + \hat{a}{''}_{\alpha}(z)\right]\}\vec{e}_y.\end{split}
\end{equation}  
We see, that the field operators $\hat{\vec{E}}^{[1]}(\vec{r},t)$ $\hat{\vec{H}}^{[1]}(\vec{r},t)$ are local operators in the space $R_3$, that allows to enter the photon wave function in coordinate representation, that is, to solve the problem, which was accepted to be unsolvable in the principle \cite{Scully}, \cite{Akhiezer}, \cite{Dutra}. 

\subsubsection{\textbf{Space-Time Local Quantization of Cavity EM-Field}} 

Let us consider general case, corresponding to discrete both  $\omega$-space of possible light frequencies  and  $\vec{k}$-space of light wave vectors, which are  result of finiteness of 4-cavity space volume and time segment. Let us find the relations for EM-field vector-functions.  In the case of cavity electrodynamics considered  we have two $1D$ ranges of variables $t$  and $z$, which belong to segments $t \in [0, T]$, $z \in [0, L]$, that is, there is in fact  to be given 2D-range D(t,z), which can be considered to be definitional domain of vector-functions $\vec{E}(\vec{r},t)$ and $\vec{H}(\vec{r},t)$ of two  variables $t$  and $z$. In the case, when given functions are absolutely integrable over both the variables $t$  and $z$, they can be represented in the form of  multiple series, given by the relations
\begin{equation}\label{eq50}
\vec{E}(\vec{r},t) = \{\sum_{\alpha=1}^{\infty}\sum_{\beta=1}^{\infty} A^{'E}_{\alpha \beta} q_{\alpha}(t) q_{\beta}(z)\}\vec{e}_x,
\end{equation}
and 
\begin{equation}\label{eq51}
\vec{H}(\vec{r},t) = \{-\sum_{\alpha=1}^{\infty}\sum_{\beta=1}^{\infty} A^{'H}_{\alpha \beta} \frac{dq_{\alpha}(t)}{dt}\int\limits _{0}^{z} q_{\beta}(z')dz'\}\vec{e}_y, 
\end{equation}
where $\{q_{\alpha}(t)\}$, $\{q_{\beta}(z)\}$, $\alpha$, $\beta$ $\in$ $N$ are two systems of orthogonal functions, $A^{'E}_{\alpha \beta}$, $A^{'H}_{\alpha \beta}$ are coefficients in given expansions, which depend on both the indices $\alpha$ and $\beta$. Both two-fold series will be two-fold Fourier series, if the sets  $\{q_{\alpha}(t)\}$, $\{q_{\beta}(z)\}$ are two orthogonal systems of harmonical trigonometric functions. It is evident, that the sets $\{q_{\alpha}(t)\}$, $\{q_{\beta}(z)\}$ are independent from each other and produce bases with  $\aleph_0$ dimension
in the metrizable complete spaces $L_2$, which are, therefore, Hilbert spaces.  It follows from physical meaning in the case of definite direction of propagation, that between the bases $\{q_{\alpha}(t)\}$, $\{q_{\beta}(z)\}$ and, correspondingly, between both Hilbert spaces the mapping
\begin{equation}\label{eq52}
\Gamma: \{q_{\alpha}(t)\} \rightarrow \{q_{\beta}(z)\}
\end{equation}
is isomorphism, at that, if there is preferential (propagation) direction in $^1R_4$-space, both the sets have to be ordered in correspondence with running numbers. 
It means, that
\begin{equation}\label{eq53}
\begin{split}
&\vec{E}(\vec{r},t) = \{\sum_{\alpha=1}^{\infty}\sum_{\beta=1}^{\infty} A^{'E}_{\alpha \beta} q_{\alpha}(t) q_{\beta}(z)\}\vec{e}_x =\\ &\{\sum_{\alpha=1}^{\infty} A^{''E}_{\alpha} q_{\alpha}(t) q_{\alpha}(z)\}\vec{e}_x
\end{split}
\end{equation}
and
\begin{equation}
\label{eq54}
\begin{split}
&\vec{H}(\vec{r},t) = \{\sum_{\alpha=1}^{\infty}\sum_{\beta=1}^{\infty} A^{'H}_{\alpha \beta}\frac{dq_{\alpha}(t)}{dt}\int\limits_{0}^{z}  q_{\beta}(z')dz' \}\vec{e}_y = \\
&\{\sum_{\alpha=1}^{\infty} A^{'H}_{\alpha}\frac{dq_{\alpha}(t)}{dt} \int\limits_{0}^{z}q_{\alpha}(z')dz'\}\vec{e}_y,
\end{split}
\end{equation}
where $A^{''E}_{\alpha}$, $A^{'H}_{\alpha}$ are coefficients in given expansions, which depend now the only on index $\alpha$. We have considered the mathematical aspect. Physically the insert of Kronecker symbol $\delta_{\alpha\beta}$ in double sum in (\ref{eq50}), (\ref{eq51}) corresponds to renumbering of the massive $\{\beta\}$ in that way, in order to $\alpha$ and $\beta$ were running  the sets $\{\alpha\}$ and $\{\beta\}$ synchronously one after another with number growth. It is additional requirement, since, although both the sets $\{\alpha\}$ and $\{\beta\}$ have the same cardinal number $\aleph_0$ and although the mapping (\ref{eq52}) in view of its biectivity gives one-to-one relation
between both the sets, it can be realized along with synchronous running above indicated by infinite number of asynchronous running. The choice of synchronous running is determined by causality principle - the photons by  their propagation synchronously "lock on" the space and the time.
It means in its turn, that complete local quantization of EM-field becomes to be possible. 

It can be shown, that along with expression for $\vec{H}(\vec{r},t)$, given by (\ref{eq54}),  $z$-coordinate part can be represented in more symmetrical  form like to $t$-coordinate part in (\ref{eq3abc}), that is
\begin{equation}
\label{eq55}
\begin{split}
\vec{H}(\vec{r},t) = 
\{\sum_{\alpha=1}^{\infty} A^{''H}_{\alpha}\frac{1}{\omega_{\alpha}}\frac{dq_{\alpha}(t)}{dt} \left(\frac{1}{k_{\alpha}}\frac{dq_{\alpha}(z)}{dz}\right)\}\vec{e}_y,
\end{split}
\end{equation}
where  $t \in [0, T]$, $z \in [0, L]$ and $A^{''H}_{\alpha}$ is
\begin{equation}
\label{eq56}
A^{''H} = \sqrt{\frac{2\omega^2_{\alpha} m_{\alpha}}{\mu_{0} V T}}.
\end{equation}
 For 
$\vec{E}(\vec{r},t)$ we retain the relation, given by (\ref{eq53}), in which  $t \in [0, T]$, $z \in [0, L]$ and $A^{''E}_{\alpha}$ is
 \begin{equation}
\label{eq57}
A^{''E}_{\alpha} = \sqrt{\frac{2\omega^2_{\alpha} m_{\alpha}}{\epsilon_{0} V T}}.
\end{equation}

It seems to be essential, that the segment $[0, T]$ is not arbitrary, $T$ has to be equal to $\frac{L}{c}$, that ensures the synchronization above discussed of the EM-field propagation in the space and in the time. Then discrete behavior of EM-field in $\omega$-space will correspond to its discrete behavior in $\vec{k}$-space.

Let us designate
\begin{equation}
\label{eq58}
\begin{split}
&q_{\alpha}(z)q_{\alpha}(t) = q_{\alpha}(z,t),\\ 
&m_{\alpha}\frac{dq_{\alpha}(t)}{dt} = p_{\alpha}(t),\\
&\frac{1}{k_{\alpha}}\frac{dq_{\alpha}(z)}{dz} = p_{\alpha}(z),\\
&p_{\alpha}(z)p_{\alpha}(t) = p_{\alpha}(z,t).
\end{split}
\end{equation}
Then classical Hamiltonian density is 
\begin{equation}
\label{eq59}
\begin{split}
&\mathfrak{W}(z,t) = \frac{1}{2} \{ \epsilon_0 \left[\sum_{\alpha=1}^{\infty} \sqrt{\frac{2 \omega^2_{\alpha} m_{\alpha}}{\epsilon_{0} V T}} q_{\alpha}(z,t)\right]^2 + \\
&\mu_0 \left[\sum_{\alpha=1}^{\infty} \sqrt{\frac{2 \omega^2_{\alpha} m_{\alpha}}{\mu_{0} V T}} \frac{1}{\omega_{\alpha} m_{\alpha}} p_{\alpha}(z,t) \right]^2 \} = \\
&\frac{1}{2} \{\sum_{\alpha=1}^{\infty} \frac
{2 \omega^2_{\alpha} m_{\alpha}}{V T} q^2_{\alpha}(z,t)  +  \\
&\sum_{\alpha\neq\beta}\sum_{\beta=1}^{\infty}\frac{2 \omega_{\alpha} \omega_{\beta}}{V T} \sqrt{m_{\alpha} m_{\beta}}  q_{\alpha}(z,t)
q_{\beta}(z,t)  +  \\
&\sum_{\alpha=1}^{\infty} \frac
{2}{V T} \frac{1}{m_{\alpha}} p^2_{\alpha}(z,t)  +  \\
&\sum_{\alpha\neq\beta}^{}\sum_{\beta=1}^{\infty}\frac{2}{V T \sqrt{m_{\alpha} m_{\beta}}} p_{\alpha}(z,t) p_{\beta}(z,t)\}
\end{split}
\end{equation}  

It seems to be evident, that by integration over 4-volume both  the items with double sum will give contribution, which is equal to zero.  It is consequence of 
orthogonality of the functions $\{q_{\alpha}(t)\}$, $\{q_{\beta}(z)\}$. 
It means, that the Hamiltonian density can be choosed in the canonical form
\begin{equation}
\label{eq60q}
\begin{split}
\mathfrak{W}^{[1]}(z,t) = \frac{1}{V T}\sum_{\alpha=1}^{\infty} \left[ 
m_{\alpha} \omega^2_{\alpha} q^2_{\alpha}(z,t)  +    
 \frac{p^2_{\alpha}(z,t)}{m_{\alpha}} \right]    
\end{split}
\end{equation}
Then following to  Dirac canonical quantization method, we have 
\begin{equation}
\label{eq61a}
\begin{split}
&\left[\hat {p}_{\alpha}(z, t) , \hat {q}_{\beta}(z, t)\right] = i\hat{g}^{}(z, t) \delta_{{\alpha}\beta} \equiv \hat{g}^{(1)}(z, t) \\
&\left[\hat {q}_{\alpha}(z, t) , \hat {q}_{\beta}(z, t)\right] = \left[\hat {p}_{\alpha}(z, t) , \hat {p}_{\beta}(z, t)\right] = 0,
\end{split}
\end{equation}
It is substantial, that instead of scalar value we have $\hat{g}^{(1)}(z, t)$, that is, operator function of the variables $z$ and $t$.  Really, taking into account
(\ref{eq58}), we obtain
\begin{equation}
\label{eq64a}
\begin{split}
&[\hat{p}_{\alpha}(z,t),\hat{q}_{\beta}(z,t)] = [\hat{p}_{\alpha}(z) \hat{p}_{\alpha}(t),
\hat{q}_{\beta}(z)\hat{q}_{\beta}(t)] = \\ &i\hbar\delta_{\alpha\beta}\hat{p}_{\alpha}(z)\hat{q}_{\beta}(z) + i \lambda_0 \delta_{\alpha\beta} \hat{p}_{\alpha}(t) \hat{q}_{\beta}(t)
\end{split}
\end{equation}
Therefore, $\hat{g}^{(1)}(z, t)$ is
\begin{equation}
\label{eq65a}
\hat{g}^{(1)}(z, t) = i \delta_{\alpha\beta} [\hbar \hat{p}_{\alpha}(z) \hat{q}_{\beta}(z) +  \lambda_0  \hat{p}_{\alpha}(t) \hat{q}_{\beta}(t)]
\end{equation}
It is seen, that $\hat{g}^{(1)}(z, t)$ is dependent on both the sequence of indices $\alpha$, $\beta$ (in distinction from usual case) and on the sequence of operator functions in (\ref{eq65a}). In other words, there are else three operator functions of analogous structure. They are
\begin{equation}
\label{eq66a}
\hat{g}^{(2)}(z, t) = -i \delta_{\alpha\beta} [\hbar \hat{q}_{\beta}(z)\hat{p}_{\alpha}(z)  +  \lambda_0 \hat{q}_{\beta}(t) \hat{p}_{\alpha}(t)]\end{equation}
\begin{equation}
\label{eq62a}
\hat{g}^{(3)}(z, t) = i \delta_{\alpha\beta} [\hbar \hat{p}_{\beta}(z) \hat{q}_{\alpha}(z) +  \lambda_0  \hat{p}_{\beta}(t) \hat{q}_{\alpha}(t)]\end{equation}
\begin{equation}
\label{eq73}
\hat{g}^{(4)}(z, t) = -i \delta_{\alpha\beta} [\hat{q}_{\alpha}(z)\hbar \hat{p}_{\beta}(z)) +  \lambda_0  \hat{q}_{\alpha}(t)\hat{p}_{\beta}(t)]
\end{equation}
It seems to be convenient to define symmetrized operator $\hat{g}$ by all the four $\hat{g}^{(j)}(z, t)$ , $j = \overline {1,4}$,  functions,
that is 
\begin{equation}\label{eq67a}
\hat{g} = \frac{1}{4}\sum_{j=1}^{4}\hat{g}^{(j)}(z, t) = -\hbar \lambda_0 \hat{e}, 
\end{equation}
which is scalar, multiplied on unit operator. So,  
\begin{equation}\label{eq68a}
g = -\hbar \lambda_0
\end{equation}
The operator functions   $\hat{a}_{\alpha}(z,t)$ and $ \hat{a}^{+}_{\alpha}(z,t)$ are
\begin{equation}
\label{eq63a}
\begin{split}
&\hat{a}_{\alpha}(z, t) = \frac{1}{ \sqrt{ 2 \hbar \lambda_0  m_{\alpha} \omega_{\alpha}}} \left[ m_{\alpha} \omega_{\alpha}\hat {q}_{\alpha}(z, t) + i \hat {p}_{\alpha}(z, t)\right]\end{split}
\end{equation}
\begin{equation}
\label{eq69a}
\begin{split}
\hat{a}^{+}_{\alpha}(z, t) = \frac{1}{ \sqrt{ 2 \hbar \lambda_0  m_{\alpha} \omega_{\alpha}}} \left[ m_{\alpha} \omega_{\alpha}\hat {q}_{\alpha}(z, t) - i \hat {p}_{\alpha}(z, t)\right].
\end{split}
\end{equation}
Then the operator functions of canonical variables have the form
\begin{equation}
\label{eq70a}
\begin{split}
&\hat {q}_{\alpha}(z, t) = \sqrt{\frac{\hbar \lambda_0 }{2 m_{\alpha} \omega_{\alpha}}} \left[\hat{a}^{+}_{\alpha}(z, t) + \hat{a}_{\alpha}(z, t)\right]
\end{split}
\end{equation}
\begin{equation}
\label{eq71}
\begin{split}
&\hat {p}_{\alpha}(z, t) = i \sqrt{\frac{\hbar \lambda_0  m_{\alpha} \omega_{\alpha}}{2}} \left[\hat{a}^{+}_{\alpha}(z, t) - \hat{a}_{\alpha}(z, t)\right]. 
\end{split}
\end{equation}
It is easily to show, that operator functions $\hat{a}_{\alpha}(z,t)$ and $ \hat{a}^{+}_{\beta}(z,t)$ satisfy the following relation
\begin{equation}
\label{eq72}
[\hat{a}_{\alpha}(z, t), \hat{a}^{+}_{\beta}(z, t)] = -i \delta_{\alpha\beta} \hat{e}
\end{equation}
Taking into account the expressions for $\vec{E}(\vec{r},t)$ and $\vec{H}(\vec{r},t)$, given by (\ref{eq53}), (\ref{eq55}), we have for operators of EM-field vector functions
\begin{equation}\label{eq74}
\begin{split}
&\hat{\vec{E}}(\vec{r},t) = \{\sum_{\alpha=1}^{\infty} A^{''E}_{\alpha} \hat{q}_{\alpha}(t) \hat{q}_{\alpha}(z)\}\vec{e}_x = \\ &\{\sum_{\alpha=1}^{\infty} A^{''E}_{\alpha} \hat{q}_{\alpha}(z, t) \}\vec{e}_x = \\
&\{\sum_{\alpha=1}^{\infty} A^{''E}_{\alpha}\sqrt{\frac{\hbar \lambda_0 }{2 m_{\alpha} \omega_{\alpha}}} [\hat{a}^{+}_{\alpha}(z, t) + \hat{a}_{\alpha}(z, t)]\}\vec{e}_x
\end{split}
\end{equation}
and
\begin{equation}
\label{eq75}
\begin{split}
&\hat{\vec{H}}(\vec{r},t) = 
\{\sum_{\alpha=1}^{\infty} A^{''H}_{\alpha}\frac{1}{\omega_{\alpha}}\frac{d\hat{q}_{\alpha}(t)}{dt} (\frac{1}{k_{\alpha}}\frac{d\hat{q}_{\alpha}(z)}{dz})\}\vec{e}_y = \\
&\{\sum_{\alpha=1}^{\infty} A^{''H}_{\alpha}
\frac{1}{m_{\alpha}\omega_{\alpha}}\hat{p}_{\alpha}(t)
\hat{p}_{\alpha}(z) )\}\vec{e}_y,
\end{split}
\end{equation}
which, using the relations (\ref{eq58}) (\ref{eq71}), can be rewritten in the form
\begin{equation}
\label{eq76}
\begin{split}
&\hat{\vec{H}}(\vec{r},t) = \{\sum_{\alpha=1}^{\infty} A^{''H}_{\alpha}
\frac{1}{m_{\alpha}\omega_{\alpha}}
\hat{p}_{\alpha}(z, t)\}\vec{e}_y = \\
&i \{\sum_{\alpha=1}^{\infty} A^{''H}_{\alpha}
 \sqrt{\frac{\hbar \lambda_0}{2 m_{\alpha} \omega_{\alpha}}} [\hat{a}^{+}_{\alpha}(z, t) -\\
& \hat{a}_{\alpha}(z, t)]\}\vec{e}_y
\end{split}
\end{equation}
Therefore, by means of operator functions $\hat{a}^{+}_{\alpha}(z, t)$, $\hat{a}_{\alpha}(z, t)$ the local quantization of EM-field is realized, which allows to determine simultaneously along with time of creation (annihilation) of photons the space coordinate of given process. It allows to solve the problem of  the wave function of photons, which was discussed rather long time, see, for example, \cite{Kramers}, \cite{Power}, \cite{Bohm}, \cite{Scully}, but is was remained to be unresolved upto now. It can be built like to wave function for the other particles or quasiparticles, for example, like to wave function for electron.

\subsection{\textbf{Cavity 4-Currents}}

It represents the interest to calculate the 4-currents for given task. 
Let us  place all the vector-functions in pairs in accordance with their parity. Then we have the following pairs
\begin{equation}
\label{eq15abc}
(\vec{E}^{[1]}(\vec{r},t), \vec{E}^{[2]}(\vec{r},t)),
(\vec{H}^{[2]}(\vec{r},t), \vec{H}^{[1]}(\vec{r},t))
\end{equation}
in which both the $\vec{E}$-vectors and $\vec{H}$-vectors have the same space parity (polar and axial correspondingly) and differ from each other by $t$-parity, $t$-even and $t$-uneven  in accordance with their numbers in pairs. It means, that they trasform like to $x_4$ and $x_1$ coordinates in $^1R_4$. 
In a similar manner can be set the vector-functions with opposite to the vector-functions in (\ref{eq15abc}) space parity
\begin{equation}
\label{eq15bcd}
(\vec{E}^{[3]}(\vec{r},t), \vec{E}^{[4]}(\vec{r},t)),
(\vec{H}^{[4]}(\vec{r},t), \vec{H}^{[3]}(\vec{r},t)).
\end{equation}
Then, taking into account the definition of complex quantities to be pair of real defined quantities, taken in fixed order, we come in a natural way once again  to concept of  complex vector-functions, which describe Maxwellian EM-field.  In other words, we have in fact the quantities
\begin{equation}
\label{eq16ab}
\begin{split}
&\vec{E}^{[1]}(\vec{r},t) +  i \vec{E}^{[2]}(\vec{r},t) = \vec{E}_{1,2}(\vec{r},t),\\
&\vec{H}^{[2]}(\vec{r},t) +  i \vec{H}^{[1]}(\vec{r},t) = \vec{H}_{1,2}(\vec{r},t),
\end{split}
\end{equation}
and
\begin{equation}
\label{eq17ab}
\begin{split}
&\vec{E}^{[3]}(\vec{r},t) +  i \vec{E}^{[4]}(\vec{r},t) = \vec{E}_{3,4}(\vec{r},t),\\
&\vec{H}^{[4]}(\vec{r},t) +  i \vec{H}^{[3]}(\vec{r},t) = \vec{H}_{3,4}(\vec{r},t),
\end{split}
\end{equation}
where complex plane put in  the correspondence to $(y, z)$ real plane. In fact, by means of given process we have obtained the complex vector-functions, consisting of the components with different symmetry under space inversion. Initial P-symmetry of the imagine vector-components was changed into opposite.
 It seems to be convenient to determine  the space of EM-field vector-functions  under the ring of quaternions with another basis in comparison with basis, given by (\ref{eq15cacdegh}). We can use also the quaternion basis $\{e_i\}, i = \overline{0,3}$ with algebraic operations between elements, satisfying to relationships 
\begin{equation}
\label{eq40}
e_i e_j = \varepsilon_{ijk}e_k + \delta_{ij} e_{_0}, e_{_0} e_i = e_i, {e_{_0}}^2 = e_{_0}, i,j,k =\overline{1,3},
\end{equation}
where $\varepsilon_{ijk}$ is completely antisymmetric Levi-Civita 3-tensor. 

Let us define the vector biquaternion 
\begin{equation}
\label{eq17abc}
\vec{\Phi} = (\vec{E}^{[1]} + \vec{H}^{[2]}) + i (\vec{H}^{[1]} + \vec{E}^{[2]}),
\end{equation}
 which can be represented to be the sum of the biquaternions
\begin{equation}
\label{eq17bcd} 
\vec{\Phi} = \vec{F} + \vec{\tilde{F}},
\end{equation}
where $\vec{F} =\vec{E}^{[1]} + i (\vec{H}^{[1]}, \vec{\tilde{F}} = \vec{H}^{[2]} + i \vec{E}^{[2]}$.
 Then Maxwell equations for instance for two free photon fields with different $t$-parity are
\begin{equation}
\label{eq17cde} 
\nabla\vec{\Phi} = 0.
\end{equation}

  The generalized Maxwell equations in quaternion form with quaternion basis, given by (\ref{eq15cacdegh}), can also be rewritten in united quaternion-biquaternion form, which is more compact, if to use both the bases. It  is mathematically possible, since both the bases are independent on each other.

\subsubsection{\textbf{Classical Cavity 4-Currents}}

It is evident, that  
\begin{equation}
\label{eq18ab}
{j_{\mu,\pm}}(x) = j^{(1)}_{\mu,\pm}(x) + i j^{(2)}_{\mu,\pm}(x),
\end{equation}
where subscript $\pm$ corresponds to two possibilities for definition of complex vector-functions. Along with relationships (\ref{eq16ab}), (\ref{eq17ab}) they can be defined by the change of addition sign   in (\ref{eq16ab}), (\ref{eq17ab}) into opposite.
The quantity $j^{(1)}_{\mu,\pm}(x)$ is well known quantity, and it is determined by
\begin{equation}
\label{eq19ab}
\begin{split}
&j^{(1)}_{\mu,\pm}(x) = -\frac{i e}{\hbar c}\sum_{\alpha = 1}^{\infty}\sum_{s=1}^{2}\left[ \frac{\partial{L(x)}}{\partial(\partial_{\mu}u^{s,\pm}_\alpha(x))} u^{s,\pm}_{\alpha}(x)\right] \\
&+ \frac{i e}{\hbar c}\sum_{\alpha = 1}^{\infty}\sum_{s=1}^{2}\left[ \frac{\partial{L(x)}}{\partial(\partial_{\mu}u^{*s,\pm}_\alpha(x))} u^{*s,\pm}_{\alpha}(x)\right],
\end{split}
\end{equation}
where $L(x)$ is Lagrange function and  $u^{s,\pm}_{\alpha}(x), s = 1,2$ are
\begin{equation}
\label{eq19abcd}
\begin{split}
&u^{1,\pm}_{\alpha}(x) = \sqrt{\epsilon_0}A^{E}_\alpha \sin k_\alpha(x_3) [q_\alpha(x_4) \pm i q^{''}_\alpha(x_4)]\\
&u^{2,\pm}_{\alpha}(x) = \sqrt{\mu_0}A^{H}_\alpha \cos k_\alpha(x_3) [-q{'}_\alpha(x_4) \pm i \frac {1}{\omega_\alpha}\frac{dq_\alpha(x_4)}{dx_4}]
\end{split}
\end{equation} 
The functions $u^{s,\pm}_{\alpha}(x), s = 1,2, \alpha \in N$ are built from the components of   
the expansion in Fourier series of the fields $\vec{E}^{[1]}(\vec{r},t), \vec{E}^{[2]}(\vec{r},t)$ and
$\vec{H}^{[2]}(\vec{r},t), \vec{H}^{[1]}(\vec{r},t)$ correspondingly.   

To determine the current density $j^{(2)}_{\mu,\pm}(x)$ we have to take into consideration, that gauge symmetry group of EM-field is two-parametric group $\Gamma(\alpha,\beta) = U_{1}(\alpha) \otimes \mathfrak R(\beta)$, where $\mathfrak R(\beta)$ is abelian multiplicative group of real numbers (excluding zero). It leads also to existence for EM-field of complex 4-current densities including complex charge density component. 
 
The current density $j^{(2)}_{\mu,\pm}(x)$ is given by the expression
\begin{equation}
\label{eq21ab}
\begin{split}
&j^{(2)}_{\mu,\pm}(x) = -\frac{ie}{\hbar c}\sum_{\alpha = 1}^{\infty}\sum_{s=1}^{2}\left[ \frac{\partial{L(x)}}{\partial(\partial_{\mu}u^{s,\pm}_\alpha(x))} u^{s,\pm}_{\alpha}(x)\right] \\
&- \frac{ie}{\hbar c}\sum_{\alpha = 1}^{\infty}\sum_{s=1}^{2}\left[ \frac{\partial{L(x)}}{\partial\partial_{\mu}u^{*s,\pm}_\alpha(x)} u^{*s,\pm}_{\alpha}(x)\right].
\end{split}
\end{equation}

It can be easily shown, that $j_{3}^{1,\pm}(\vec{r},t)$ is always equal to zero for any set of twice continuously differentiable functions  $\{q_{\alpha}(t)\}, \alpha \in N$. The expression for arbitrary set of twice continuously differentiable functions  $\{q_{\alpha}(t)\}, \alpha \in N$, for $j_{3}^{2,\pm}(\vec{r},t)$ is
\begin{equation}
\label{eq21abc}
\begin{split}
&j_{3}^{2,\pm}(\vec{r},t) = -\frac{2ie}{\hbar c^2V}\sum_{\alpha = 1}^{\infty}m_{\alpha}\omega_{\alpha}^3\sin 2k_{\alpha}z \times\\
&\{\left[i|q_{\alpha}(t)\pm
i\omega_{\alpha}^2 \int\limits_{0}^{t} \int\limits_{0}^{t{''}}q_{\alpha}(t')dt'dt{''}|^2\right]\\ - &\left[|\omega_{\alpha}\int\limits _{0}^{t}q_{\alpha}(t')dt' \mp \frac{i}{\omega_{\alpha}}\frac{dq_{\alpha}(t)}{dt}|^2\right]\}.
\end{split}
\end{equation}
The relationship (\ref{eq21abc}) is true  for both the variants in superposition
\begin{equation}
\label{eq21bcd}
\vec{H}^{[i]}(\vec{r},t) +  i \vec{H}^{[j]}(\vec{r},t) = \vec{H}^{[ij]}(\vec{r},t), i \neq j, i,j = 1,2.
\end{equation}
Taking into account relationship (\ref{eq6ab}), that is the set $\{q_{\alpha}(t)\}, \alpha \in N$, which satisfy the Maxwell equations   we  will have
\begin{equation}
\label{eq22ab}
\begin{split}
&j_{3}^{2,\pm}(\vec{r},t)= -\frac{8ie}{\hbar c^2V}\sum_{\alpha = 1}^{\infty}m_{\alpha}\omega_{\alpha}^3\sin 2k_{\alpha}z \times\\
&[C_{1\alpha}C^*_{2\alpha}e^{2i\omega_{\alpha}t} + C^*_{1\alpha}C_{2\alpha}e^{-2i\omega_{\alpha}t}],
\end{split}
\end{equation}
the expression for arbitrary set of twice continuously differentiable functions  $\{q_{\alpha}(t)\}, \alpha \in N$, for $j_{4}^{1,\pm}(\vec{r},t)$ is
\begin{equation}
\label{eq23ab}
\begin{split}
&j_{4}^{1,\pm}(\vec{r},t) = -\frac{2e}{\hbar c^2V}\sum_{\alpha = 1}^{\infty}m_{\alpha}\omega_{\alpha}^2\{\sin^2k_{\alpha}z \times\\
&[\frac{dq^*_{\alpha}(t)}{dt}\mp i\frac{dq^{*''}_{\alpha}(t)}{dt}][q_{\alpha}(t) \pm q^{''}_{\alpha}(t)]\\ +
&[\frac{dq_{\alpha}(t)}{dt} \pm i\frac{dq^{''}_{\alpha}(t)}{dt}][- q^*_{\alpha}(t) \pm q^{*''}_{\alpha}(t)]\\
&+ \cos^2k_{\alpha}z [\frac{1}{\omega_{\alpha}}\frac{d^2q^*_{\alpha}(t)}{dt^2} \pm i\omega_{\alpha} q^*_{\alpha}(t)] \times\\
&[\frac{1}{\omega_{\alpha}}\frac{dq_{\alpha}(t)}{dt}
\mp i\omega_{\alpha}\int\limits_{0}^{t}q_{\alpha}(t{'}dt{'}] \\ + &[\frac{-1}{\omega_{\alpha}}\frac{d^2q_{\alpha}(t)}{dt^2} \pm i\omega_{\alpha}q_{\alpha}(t)] \times\\
&[\frac{1}{\omega_{\alpha}}\frac{dq^*_{\alpha}(t)}{dt}
\mp i\omega_{\alpha}\int\limits_{0}^{t}q^*_{\alpha}(t{'})dt{'}]\},
\end{split}
\end{equation}
where $q^{''}_{\alpha}(t) = \omega_{\alpha}^2 \int\limits_{0}^{t} \int\limits_{0}^{t{''}}q_{\alpha}(t')dt'dt{''}$.
It is evident from relationship (\ref{eq23ab}), that in the case  of real-valued sets of twice continuously differentiable functions $\{q_{\alpha}(t)\}, \alpha \in N$, $j_{4}^{1,\pm}(\vec{r},t)$ is equal to zero. For complex-valued functions, determined by (\ref{eq6ab}), we will have
\begin{equation}
\label{eq23abc}
j_{4}^{1,\pm}(\vec{r},t) = \frac{8ie}{\hbar c^2V}\sum_{\alpha = 1}^{\infty}m_{\alpha}\omega_{\alpha}^3 (|C_{1\alpha}|^2 - |C_{2\alpha}|^2).
\end{equation}
It is seen from (\ref{eq23abc}), that $j_{4}^{1,\pm}(\vec{r},t)$ in the  case of Maxwellian EM-field is constant, which is equal to zero at $|C_{1\alpha}| = |C_{2\alpha}|$, that is for all real-valued functions and for complex-valued functions $\{q_{\alpha}(t)\}, \alpha \in N$, which differ each other by arguments of constants $C_{1\alpha}$ and $C_{2\alpha}$. Further, for  the current density $j^{(4)}_{2,\pm}(x)$ we have
\begin{equation}
\label{eq23bcd}
\begin{split}
&j_{4}^{2,\pm}(\vec{r},t) = -\frac{2e}{\hbar c^2V}\sum_{\alpha = 1}^{\infty}\{m_{\alpha}\omega_{\alpha}^2\sin^2k_{\alpha}z \frac{d}{dt}(|q_{\alpha}(t)|^2)  + \\
&\omega_{\alpha}^4\frac{d}{dt}(|\int\limits_{0}^{t} \int\limits_{0}^{t{''}}q_{\alpha}(t')dt'dt{''}|^2) \mp  \frac{d}{dt}[q_{\alpha}(t)\int\limits_{0}^{t} \int\limits_{0}^{t{''}}q^*_{\alpha}(t')dt'dt{''}]\\
& \times i\omega_{\alpha}^2\pm
\frac{d}{dt}[q^*_{\alpha}(t)\int\limits_{0}^{t} \int\limits_{0}^{t{''}}q_{\alpha}(t')dt'dt{''}]i\omega_{\alpha}^2
+ m_{\alpha}\omega_{\alpha}^2\cos^2k_{\alpha}z \\
&\times [\frac{1}{\omega_{\alpha}^2}\frac{d}{dt}(|\frac{dq_{\alpha}(t)}{dt}|^2)                                                                                                \pm i \frac{d}{dt}(\frac{dq^*_{\alpha}(t)}{dt}\int\limits_{0}^{t}q_{\alpha}(t')dt')\\
&+ \omega_{\alpha}^2 \frac{d}{dt}(|\int\limits_{0}^{t}q_{\alpha}(t')dt'|^2) \mp i \frac{d}{dt}(\frac{dq_{\alpha}(t)}{dt}\int\limits_{0}^{t}q^*_{\alpha}(t')dt')]\}.
\end{split}
\end{equation}
For complex-valued functions, determined by (\ref{eq6ab}), we obtain
\begin{equation}
\begin{split}
\label{eq23ccd}
&j_{4}^{2,\pm}(\vec{r},t) = \frac{8ie}{\hbar c^2V}\sum_{\alpha = 1}^{\infty}m_{\alpha}\omega_{\alpha}^3 \cos 2k_{\alpha}z \times\\
&[C_{1\alpha}C^*_{2\alpha}e^{2i\omega_{\alpha}t} - C^*_{1\alpha}C_{2\alpha}e^{-2i\omega_{\alpha}t}].
\end{split}
\end{equation}

It can be shown, that continuity equation
\begin{equation}
\label{eq24ab}
\frac{\partial j^{\pm}_{\mu} (x)}{\partial x_\mu} = 0 
 \end{equation}
is fulfilled  for both general  case and for Maxwellian EM-field functions considered. 

\subsection{\textbf{Quantized Cavity 4-Currents}}

Let us calculate the 4-current densities, which correspond to quantized dually symmetric EM-field, that is to the field, which consist of two components with even and  uneven parities under time reversal or space inversion of both the EM-field vector functions $\vec{E}(\vec{r},t)$ and $\vec{H}(\vec{r},t)$. Let us consider for distinctness the case of two-component EM-field, in which $\vec{E}(\vec{r},t)$- components and $\vec{H}(\vec{r},t)$- components have the same $P$-parity (uneven and even corresondingly) and differ each other by $t$-parity. Given choose corresponds to classical consideration in previous subsection, that allows to compare the results for classical and quantized dually symmetric EM-field. Consequently, we can use the set of  EM-field
vector functions, analogous to (\ref{eq19abcd}), in which the operator functions are set up in conformity to canonical variables. 
 
$\hat{\vec{u}}^{s,\pm}_{\alpha}(x), s = 1, 2$ are
\begin{equation}
\label{eq29abcd}
\begin{split}
&\hat{\vec{u}}^{1,\pm}_{\alpha}(x) = \sqrt{\epsilon_0}A^{E}_\alpha \sin k_\alpha(x_3) [\hat{q}_\alpha(x_4) \pm i \hat{q}^{''}_\alpha(x_4)]\vec{e}_1\\
&\hat{\vec{u}}^{2,\pm}_{\alpha}(x) = \sqrt{\mu_0}A^{H}_\alpha \cos k_\alpha(x_3)\times \\
&[-\hat{q}{'}_\alpha(x_4) \pm i \frac {1}{\omega_\alpha}\frac{d\hat{q}_\alpha(x_4)}{dx_4}]\vec{e}_2,
\end{split}
\end{equation}
where $\vec{e}_1 \equiv \vec{e}_x$, $\vec{e}_2 \equiv \vec{e}_y$,    $\hat{q}{'}_\alpha(x_4)$,  $\hat{q}^{''}_\alpha(x_4)$ are operator functions, which are setting up in the conformity  to classical variables ${q}{'}_\alpha(x_4)$,  ${q}^{''}_\alpha(x_4)$, defined by (\ref{eq11ab}). They are
\begin{equation}
\label{eq41ab}
\begin{split}
&\hat{q}_{\alpha}'(t) = {\omega_{\alpha}}\int\limits _{0}^{t} \hat{q}_{\alpha}(\tau)d\tau\\
&\hat{q}_{\alpha}''(t) = {\omega_{\alpha}}\int\limits _{0}^{t} \hat{q}_{\alpha}'(\tau')d\tau'
\end{split}
\end{equation}
correspondingly.
The functions $\hat{u}^{s,\pm}_{\alpha}(x), s = 1,2, \alpha \in N$ can be built from the components of   
the expansion in Fourier series of quantized dually symmetric EM-field, which consist of two components with even and  uneven parities under time reversal, that is, of    $\hat{\vec{E}}^{[1]}(\vec{r},t), \hat{\vec{E}}^{[2]}(\vec{r},t)$ and
$\hat{\vec{H}}^{[2]}(\vec{r},t), \hat{\vec{H}}^{[1]}(\vec{r},t)$, given by (\ref{eq34}).   Therefore, we have
\begin{equation}
\label{eq42ab}
\begin{split}
&\hat{\vec{u}}^{1,\pm}_{\alpha}(\vec{r},t) = \sqrt{\frac{\hbar \omega_{\alpha}}{V}} \{\left[\hat{a}_{\alpha}(t) + \hat{a}^{+}_{\alpha}(t)\right]\\
& + i \left[\hat{a}{''}_{\alpha}(t) + \hat{a}{''}^{+}_{\alpha}(t)\right]\} \sin(k_{\alpha} z)\} \vec{e}_x,
\end{split}
\end{equation}
\begin{equation}
\label{eq43ab}
\begin{split}
&\hat{\vec{u}}^{2,\pm}_{\alpha}(\vec{r},t) =\sqrt{\frac{\hbar \omega_{\alpha}}{V}}\{ \left[\hat{a}^{}_{\alpha}(t) - \hat{a}^{+}_{\alpha}(t)\right] \\
& + i \left[\hat{a}{''}_{\alpha}(t) - \hat{a}{''}^{+}_{\alpha}(t)\right] \} \cos(k_{\alpha} z) \} \vec{e}_y,
\end{split}
\end{equation}
where superscript $\pm$ means, that in (\ref{eq36ab}) and (\ref{eq37ab}) by definition of complex   EM-field vector function operators $\hat{\vec{E}}(\vec{r},t)$ and $\hat{\vec{H}}(\vec{r},t)$ along with the sign plus, the sign minus   can  be used. We also consider the case of $\hat{\vec{H}}(\vec{r},t)$ formation along with given by (\ref{eq37ab}) (with both the signs in the sum) the following case
\begin{equation}\label{eq44ab}
(\hat{\vec{H}}^{[1]}(\vec{r},t), \hat{\vec{H}}^{[2]}(\vec{r},t)) \rightarrow \hat{\vec{H}}^{[1]}(\vec{r},t) \pm i \hat{\vec{H}}^{[2]}(\vec{r},t) = \hat{\vec{H}}^{\pm}(\vec{r},t).
\end{equation}
It seems to be evident, that 4-current density operator can be determined by the expressions, 
coinciding with classical relations (\ref{eq18ab}), (\ref{eq19ab}), (\ref{eq21ab}), in which  all the physical quantities  are operators.  
For the operator $\hat{j}_{3}^{1,\pm}(\vec{r},t)$ we have
\begin{equation}
\label{eq45ab}
\begin{split}
&\hat{j}_{3}^{1,\pm}(\vec{r},t) = Re\hat{j}_{3}^{\pm}(\vec{r},t) = \frac{ie}{2cV}\sum_{\alpha = 1}^{\infty}k_{\alpha}\omega_{\alpha}\sin{2k_{\alpha}z}\times\\
&\{|i\left[\hat{a}{''}_{\alpha}(t) - \hat{a}{''}^{+}_{\alpha}(t)\right] \mp 
 \left[\hat{a}{}_{\alpha}(t) - \hat{a}{}^{+}_{\alpha}(t)\right]|^2 +\\ &i^2|i\left[\hat{a}{''}_{\alpha}(t) - \hat{a}{''}^{+}_{\alpha}(t)\right] \mp 
 \left[\hat{a}{}_{\alpha}(t) - \hat{a}{}^{+}_{\alpha}(t)\right]|^2\},
\end{split} 
\end{equation}
which is equaled to zero. The same result is obtained in the case of magnetic field operator, determined by (\ref{eq44ab}). For the operator $\hat{j}_{3}^{2,\pm}(\vec{r},t)$ we obtain the relation
\begin{equation}
\label{eq46ab}
\begin{split}
&\hat{j}_{3}^{2,\pm}(\vec{r},t) = Im\hat{j}_{3}^{\pm}(\vec{r},t) = -\frac{2ie}{cV}\sum_{\alpha = 1}^{\infty}k_{\alpha}\omega_{\alpha}\sin{2k_{\alpha}z}\times\\
&\left\{[\hat{a}{}_{\alpha}(t)]^2 + [\hat{a}{}^{+}_{\alpha}(t)]^2 + [\hat{a}{''}_{\alpha}(t)]^2 + [\hat{a}{''}^{+}_{\alpha}(t)]^2\right\}, 
\end{split} 
\end{equation}
which is the same for  magnetic field operator, determined by (\ref{eq44ab}).
Therefore the operator of current density $\hat{j}_{3}^{\pm}(\vec{r},t)$ is independent on sign in expressions for the field operators, based on (\ref{eq34})
and it is independent on  the sequence of $\hat{\vec{H}}^{[i]}(\vec{r},t)$, $i = 1, 2$, in
\begin{equation}
\label{eq47ab}
\hat{\vec{H}}^{ij\pm}(\vec{r},t) = \hat{\vec{H}}^{[i]}(\vec{r},t) \pm i \hat{\vec{H}}^{[j]}(\vec{r},t),
\end{equation}
where $i, j = 1, 2$, $i \neq j$.

It seems to be essential, that EM-field quantization is not binded to Maxwell equations in general case. It means, that the relations (\ref{eq45ab}), (\ref{eq46ab}) are true for more general fields. In the case of
Maxwellian EM-field, using explicit expressions for operator scalar functions given by (\ref{eq30ab}) for 
$\hat{a}{}_{\alpha}(t)$, $\hat{a}{}^{+}_{\alpha}(t)$ and similar relations for $\hat{a}{''}_{\alpha}(t)$, $\hat{a}{''}^{+}_{\alpha}(t)$ the expression (\ref{eq46ab})
has the form
\begin{equation}
\label{eq48ab}
\begin{split}
&\hat{j}_{3}^{2,\pm}(\vec{r},t) = Im\hat{j}_{3}^{\pm}(\vec{r},t) = -\frac{2ie}{cV}\sum_{\alpha = 1}^{\infty}k_{\alpha}\omega_{\alpha}\sin{2k_{\alpha}z}\times\\
&\{[\hat{a}{}_{\alpha}(t = 0)]^2 e^{-i\omega_{\alpha}t} + [\hat{a}{}^{+}_{\alpha}(t = 0)]^2 e^{i\omega_{\alpha}t} + \\ &[\hat{a}{''}_{\alpha}(t = 0)]^2 e^{-i\omega_{\alpha}t} + [\hat{a}{''}^{+}_{\alpha}(t = 0)]^2 e^{i\omega_{\alpha}t}\}. 
\end{split}
\end{equation}
Let us find now the fourth component of 4-vector operator of current density $\hat{j}_{4}^{\pm}(\vec{r},t)$, which determines the charge density. For real part
$\hat{j}_{4}^{1,\pm}(\vec{r},t)$ we have
\begin{equation}
\label{eq49ab}
\begin{split}
&\hat{j}_{4}^{1,\pm}(\vec{r},t) = Re\hat{j}_{4}^{\pm}(\vec{r},t) = \\ &\pm\frac{2e}{c^2V}\sum_{\alpha =1}^{\infty}k_{\alpha}\omega^2_{\alpha} [\{\hat{a}{''}_{\alpha}(t),\hat{a}{}^{+}_{\alpha}(t)\} - \{\hat{a}{}_{\alpha}(t),\hat{a}{''}^{+}_{\alpha}(t)\}],
\end{split}
\end{equation}
where the expressions in braces are anticommutators.
In the case of Maxwellian EM-field there is the connection between $\hat{a}{}_{\alpha}(t)$, $\hat{a}{}^{+}_{\alpha}(t)$ and  $\hat{a}{''}_{\alpha}(t)$, $\hat{a}{''}^{+}_{\alpha}(t)$, since, although they correspond to different particular solutions of Maxwell equations, the solutions are related and the connection between them can be found. It leads to connection between corresponding creation and annihilation operators for two related EM-fields with different $t$-parity.  It can be shown, that the  following relations take place
\begin{equation}
\label{eq50ab}
\begin{split}
&\hat{a}{''}_{\alpha}(t) = \omega^2_{\alpha}\int\limits_{0}^{t}[\int\limits_{0}^{t^{''}}\hat{a}{}_{\alpha}(t')dt']dt^{''}\\
&\hat{a}^{+''}_{\alpha}(t) = \omega^2_{\alpha}\int\limits_{0}^{t}[\int\limits_{0}^{t^{''}}\hat{a}^{+}_{\alpha}(t')dt']dt^{''}
\end{split}
\end{equation}
Then, taking into account the expressions for operator scalar functions, given by (\ref{eq30ab}) for 
$\hat{a}{}_{\alpha}(t)$, $\hat{a}{}^{+}_{\alpha}(t)$ and similar relations for $\hat{a}{''}_{\alpha}(t)$, $\hat{a}{''}^{+}_{\alpha}(t)$, we obtain from (\ref{eq49ab}), that  for Maxwellian EM-field $Re\hat{j}_{4}^{\pm}(\vec{r},t)$ is equal to zero. Therefore we see, that all real part of 4-vector $\hat{j}_{\mu}^{\pm}(\vec{r},t)$ is equal to zero. It corresponds to well known case of nondual single charge electrodynamics. 

For imaginary part $\hat{j}_{4}^{2,\pm}(\vec{r},t)$ we have the relation
\begin{equation}
\label{eq51ab}
\begin{split}
&\hat{j}_{4}^{2,\pm}(\vec{r},t) = Im\hat{j}_{4}^{\pm}(\vec{r},t) = \\ &\frac{2ie}{c^2V}\sum_{\alpha =1}^{\infty}[k_{\alpha}\omega^2_{\alpha} \{[\hat{a}{}^{+}_{\alpha}(t)]^2 - [\hat{a}{}_{\alpha}(t)]^2 + \\ &[\hat{a}{''}^{+}_{\alpha}(t)]^2 - \hat{a}{''}_{\alpha}(t)]^2\}\cos{2k_{\alpha}z} - 2\omega^2_{\alpha}\hat{e}],
\end{split}
\end{equation}
from which the relation for Maxwellian EM-field can be obtained in the manner, analogous to obtaining of expression (\ref{eq48ab}).

Let us verify the implementation of differential conservation law
\begin{equation}
\label{eq52ab}
\frac{\partial \hat{j}^{\pm}_{\mu} (x)}{\partial x_\mu} = 0. 
 \end{equation}

Taking into account (\ref{eq46ab}) and (\ref{eq51ab}) in the case of $Im\hat{j}_{\mu,\pm} (x)$ we have
\begin{equation}
\label{eq53ab}
\begin{split}
&\frac{\partial [Im \hat{j}^{\pm}_{\mu} (x)]}{\partial x_\mu} = \frac{\partial\hat{j}_{3}^{2,\pm}(\vec{r},t)}{\partial x_3} + 
\frac{\partial\hat{j}_{4}^{2,\pm}(\vec{r},t)}{\partial x_4} = \\
&-\frac{4ie}{cV}\sum_{\alpha =1}^{\infty}k^2_{\alpha}\omega_{\alpha}\cos{2k_{\alpha}z} \{[[\hat{a}_{\alpha}(t)]^2 + [\hat{a}^{+}_{\alpha}(t)]^2 + \\
&[\hat{a}{''}_{\alpha}(t)]^2 + [\hat{a}{''}^{+}_{\alpha}(t)]^2] - [[\hat{a}_{\alpha}(t)]^2 + [\hat{a}^{+}_{\alpha}(t)]^2 + \\
&[\hat{a}{''}_{\alpha}(t)]^2 + [\hat{a}{''}^{+}_{\alpha}(t)]^2]\} = 0.
\end{split} 
\end{equation}
Here  the commutation relations 
\begin{equation}
\label{eq54ab}
[\hat{a}_{\alpha}(t), \hat{a}^{+}_{\alpha}(t)] = \hat{e}, \alpha \in N, \end{equation} and the  relations
\begin{equation}
\label{eq55ab}
\frac{d\hat{a}_{\alpha}(t)}{dt} = \frac{1}{i\hbar}[\hat{a}_{\alpha}(t), \hat{\mathcal{H}}_{\alpha}(t)], \alpha \in N,
\end{equation}
where $\hat{\mathcal{H}}_{\alpha}(t)$ is the Hamiltonian, corresponding to cavity $\alpha$-mode, were used. The Hamiltonian $\hat{\mathcal{H}}_{\alpha}(t)$ is given by relation
\begin{equation}
\label{eq56ab}
 \hat{\mathcal{H}}_{\alpha}(t) = \hbar \omega_{\alpha}\left[\hat{a}{}^{+}_{\alpha}(t)\hat{a}{}_{\alpha}(t) + \frac{1}{2}\right]
\end{equation}
 For calculation of derivatives of operators $\hat{a}^{+}_{\alpha}(t)$, $\hat{a}{''}_{\alpha}(t)$, $\hat{a}{''}^{+}_{\alpha}(t)$ the relations,
analogous to (\ref{eq55ab}), were used.
Taking into account (\ref{eq49ab}), (\ref{eq54ab}), (\ref{eq55ab}), (\ref{eq56ab}) in the case of $Re\hat{j}_{\mu,\pm} (x)$ we have
\begin{equation}
\label{eq57ab}
\begin{split}
&\frac{\partial [Re \hat{j}^{\pm}_{\mu} (x)]}{\partial x_\mu} =  
\frac{\partial\hat{j}_{4}^{1,\pm}(\vec{r},t)}{\partial x_4} = \\
&\pm\frac{2e}{c^3V}\sum_{\alpha =1}^{\infty}\omega^3_{\alpha} 
\{[\hat{a}^{+}_{\alpha}(t), \hat{a}{''}_{\alpha}(t)] + [\hat{a}{''}_{\alpha}(t), \hat{a}^{+}_{\alpha}(t)] - \\
&[\hat{a}{''}^{+}_{\alpha}(t), \hat{a}_{\alpha}(t)] - [\hat{a}_{\alpha}(t), \hat{a}{''}^{+}_{\alpha}(t)]\} = 0. 
\end{split} 
\end{equation}
Therefore differential conservation law, given by (\ref{eq52ab}), is fulfilled both for Maxwellian EM-field and in general case. It is seen also, that direct calculation gives really nonzero imaginary part for current densities  for free cavity EM-field. It  can be considered to be direct confirmation of above obtained conclusion on the existence of imaginary component of charge in free EM-field. It gives the possibility to define the set of EM-field functions in general case, that is for  EM-field with and without sources. EM-field functions can be defined to be the components of 4-current density, or in renormalized form, if to divide each component into complex conductivity of the medium $\lambda$. It means in its turn, that the propagation of  EM-field in vacuum, that is free EM-field is also characterized by the value of vacuum conductivity $\lambda_v$, like to dielectric $\epsilon_0$ and magnetic $\mu_0$ vacum permittivities. Therefore, instead unobservable vector and scalar potentials, EM-field can be characterized by, for instance, electric field 4-vector-function with the components $E_\alpha(\vec{r},t) = \{E_x(\vec{r},t), E_y(\vec{r},t), E_z(\vec{r},t), i \frac{c \rho_e(\vec{r},t)}{\lambda}\}$, where  $i {c \rho_e}(\vec{r},t)$ is the $j_4(\vec{r},t)$-component of 4-current density, corresponding to contribution of electric component of EM-field. For the case of EM-field propagation in vacuum $\lambda = \lambda_v$. Alternative characterization by means of magnetic field 4-vector-function $H_\mu(\vec{r},t) = \{H_x(\vec{r},t), H_y(\vec{r},t), H_z(\vec{r},t), i \frac{c \rho_m(\vec{r},t)}{\lambda}\}$ seems to be equivalent for free EM-field in vacuum, if to take into account, that for Maxwellian  free EM-field the components of electric field 3-vector-functions and magnetic field 3-vector-functions are bounded up between themselves by Cauchy-Riemann analicity condition.  Here  $i {c \rho_m}(\vec{r},t)$ is the $j_4(\vec{r},t)$-component of 4-current density, corresponding to contribution of magnetic component of EM-field. However the characterization by means of the only single 4-vector-function, for instance by  $H_\mu(\vec{r},t)$ becomes to be nonequivalent in general case, in particular, in the case of single-charge ED, for which $\rho_m(\vec{r},t) = 0$. It means, that in general case both 4-vector-functions $E_\alpha(\vec{r},t)$ and $H_\mu(\vec{r},t)$ have to be used.

 It is remarkable, that the notion of characteristic medium resistance, including characteristic vacuum resistance $Z_0 = \sqrt{\frac{\mu_0}{\epsilon_0}} = 120 \pi$ Ohm is widely used in technique, connected with EM-wave propagation, in particular, in radiospectroscopy technique \cite{Poole}. Therefore, $\lambda_v = \frac{1} {Z_0} = \frac{1} {120\pi} (Ohm)^{-1}$.

The proof of the existence  for EM-field along with vector force characteristics $\vec{E}(\vec{r},t)$-field strength vector-function and $\vec{H}(\vec{r},t)$-field strength vector-function the additional scalar force characteristic - charge, allows  to be nearing to understanding of, in Dirac characterization, "strange pequliarity of light quantum" \cite{P.Dirac}, consisting in that, that according to Dirac,  light quantum, apparently, discontinues its existence, when it is in one of its stationary states - in zero state - in which its impulse and  energy are equal to zero.  Dirac consideres the absorption process to be jump of  light quantum in zero state and emission process to be jump from given state in the state, where its  existence is physically evident, so it seems that it was recreated. From the absence of restriction on number of light quanta, which can be emerged by given way, Dirac suggested \cite{P.Dirac}, that there is infinitely many of light quanta in zero state, that is, in vacuum state in modern terminology.  

\section{Fermi-Liquid Model of  Quantized EM-Field Structure}

The conclusion on quaternion structure  of EM-field and the proof of existence of the charge, being to be inherent characteristic of EM-field allow to understand the nature of photons and to explain their two kinds' behavior. On the one hand the photons seem to be charge neutral particles, that is confirmed
 in a number of  observations, including astronomic observations, and experiments on light absorption, transmission, reflection, Rayleigh and Raman scatterings and so on. On the other hand, the photons seem to be charged particles, that is confirmed, for example, by observations of zigzag-like light propagation between adjacent rainclouds during storm, since zigzag-like light propagation in the same conditions (that is by the absence of lightning  between the same rainclouds does not takes place) and it is characteristic for charge particles, that was confirmed  long ago.  It allows to suggest, that EM-field is characterized by spin-charge separation effect, leading to appearance of the photons of two kinds.
We have reviewed in details the known mechanisms of spin-charge separation and  have chosen the  mechanism, which seems to be the most appropriate for the description of given effect in EM-field.  It is the model of 1D quantum Fermi  liquid, which was obtained by  generalization of  well known Su-Schrieffer-Heeger (SSH)-model \cite{SSH}, \cite{SSH_PRB} of formally 1D  Fermi gas model, which was very succesfull in the description of electronic properties of quasi-1D organic conjugated conductors. The model of 1D quantum Fermi  liquid will be developed and expanded in its application from matter systems to the description of EM-field. Let us remark, that the idea to describe EM-field in terms of liquid is coming from experimental observations of the  EM-field  propagation in microwave range. There exists among the experts, dealing with microwave techniques, winged (stock) phrase, that microwave field in waveguides leaks like to water. So, 1D quantum Fermi  liquid model will be further considered in details  in applicability to EM-field description. 

It seems to be substantial for modelling the result, obtained by Dirac, that dynamical system, which consists of the ensemble of identical bosons is equivalent to dynamical system, which consists of the ensemble of oscillators. In other words, quantized EM-field can be represented instead of oscillator system by equivalent many-particle
system of bosons, interacting with each other.  It seems to be evident, that each single boson will possess by spin with value S = 1. The simplest analogue in the physics of condensed matter of the system of interacting S = 1 bosons is carbon. So, we come to the model of linearly polarized EM-field to be the chain of bosons, which is like in its mathematical description to the chain of carbon atoms in \textit{trans}-polyacetylene (t-PA), at that both in "atomic" and "electronic" structure. One-dimensionality of the task can be argued in the following way. It is shown above, that for description of EM-field instead of unobservable vector and scalar potentials the 4-vector of electrical and/or magnetic field strength can be used. Consequently, to describe linearly polarized EM-field in Euclidian space $R_3$ it is sufficient to specify the propagation direction, that is vector $\vec k$  and to define $\vec E$. Given vectors determine the plane, in which a  frame of reference with $z$-axis along  the propagation direction and orthogonal to it $x$-axis can be set. Taking into account the homogeneity of Minkowski space $R_4$ and homogeneity of free EM-field in it, free EM-field can be modelled by the set of noninteracting (or weak interacting) between themselves "boson-atomic" chains, similarly to many carbon-based, that is, also spin-1 boson-based, conjugated polymer structures located along propagation direction. What concerned the "atomic" structure, we have to include the contribution of vacuum fluctuations, which presents in oscillator task and which is absent in the case of boson set \cite{Scully}. The presence of charge, being to be scalar characteristic of 
EM-field gives the possibility to  model "electronic" structure of equivalent boson chain like to t-PA  electronic structure, that is consisting of "$\sigma$-subsystem" and "$\pi$-subsystem". It becomes to be understandable, if to take into account, that charge space distribution is directly connected with  $\vec E$ space distribution. In other words, the presence of $E_z$-component will determine the appearance of EM-field charge "$\sigma$-subsystem", while $E_x$-component will determine the appearance of EM-field charge "$\pi$-subsystem", at that, like to  t-PA, its distribution in space $R_3$  will be twice degenerated. Given conclusion can be argued on the basis above discussed quaternion structure of EM-field in the following way. $E_x$-polar component by EM-field propagation every other half-period alters its sign, at the same time $E_x$-axial component does not alters its sign, which is equivalent to appearance of alternating single-double interbosonic  "$\pi$-bonds" in EM-field charge "$\pi$-subsystem", at that two configurations - single-double and double-single are topologically equivalent. Consequently, we come to conclusion, that the interaction between equivalent to oscillators "bosonic atoms" can be described within the frames of Fermi gas model in zero-th order approximation or in the frames of Fermi liquid model in the first order approximation. Mathematical description in zero-th order approximation will be similar to well known above mentioned SSH-model with some corrections, concerning two branch of quasiparticles, given in \cite{Yerchuck_Dovlatova}. 

The concept of quantum Fermi liquid for description of 1D correlated electronic systems was  recovered in \cite{Dovlatova_Yerchuck_Borovik}.  Analogous model will be developed in given paper for description of propagation of quantized EM-field. Mathematically it is similar with 1D quantum Fermi liquid model, developed in \cite{Dovlatova_Yerchuck_Borovik}, however the physics is quite other. 

We will start from Hamiltonian
\begin{equation}
\label{Eq1m}
\hat{\mathcal{H}}(u) = \hat{\mathcal{H}}_{0}(u) + \hat{\mathcal{H}}_{\pi,t}(u) + \hat{\mathcal{H}}_{\pi,u}(u).
\end{equation}
Like to works  \cite{SSH}, \cite{SSH_PRB} we will consider Born-Oppenheimer approximation.
 The first term in  in 
(\ref{Eq1m}) is 
\begin{equation}
\label{Eq2m}
\begin{split}
\hat{\mathcal{H}}_{0}(u) = \sum_{m}\sum_{s}({E_m^k}\hat{a}^+_{m,s} \hat{a}_{m,s} + K u_m^2 \hat{a}^+_{m,s} \hat{a}_{m,s})
\end{split}
\end{equation}
is the  operator of kinetic energy of "boson atomic" motion (the first item), and the  operator of the $\sigma$-bonding energy
(the second item), $K$ is effective $\sigma$-bonds' spring constant, $u_m$ is configuration coordinate for $m$-th "boson atom",  which corresponds to translation of $m$-th "boson atom"  along the symmetry axis $z$ of the chain, $m = \overline{1,N}$, $N$ is the number of "boson atoms"  in the chain,  $m = \overline{1,N}$, $\hat{a}^+_{m,s}$, $\hat{a}_{m,s}$ are creation and annihilation operators of creation or annihilation of quasipartile with spin projection $s$ on the $m$-th chain site in  "$\sigma$-subsystem".

The second term in 
(\ref{Eq1m}) can be represented in the form of two components and it is 
\begin{equation}\begin{split}
\label{Eq3m}
&\hat{\mathcal{H}}_{\pi,t}(u) = \hat{\mathcal{H}}_{\pi,t_0}(u) + \hat{\mathcal{H}}_{\pi,\alpha_1}(u) =\\
&\sum_{m}\sum_{s}[(t_0 (\hat{c}^+_{m+1,s} \hat{c}_{m,s} + \hat{c}^+_{m,s} \hat{c}_{m+1,s})]  +\\
& (-1)^m 2 \alpha_1 u )(\hat{c}^+_{m+1,s} \hat{c}_{m,s} + \hat{c}^+_{m,s} \hat{c}_{m+1,s})],
\end{split}
\end{equation}
where $\hat{c}^+_{m,s}$, $\hat{c}_{m,s}$ are creation and annihilation operators of creation or annihilation of quasiparticle with spin projection $s$ on the $m$-th chain site in formally introduced "$\pi$-subsystem". 
It is the resonance interaction (hopping interaction in rest massless atomic lattice in tight-binding model approximation) of
 quasiparticles in "$\pi$-subsystem" of the whole  "electronic" system, which is considered to be Fermi liquid, and
in which the only constant and linear terms in Taylor series expansion of resonance integral about the dimerized state are  taking into account.

Operator $\hat{\mathcal{H}}(u)$ is invariant under spatial translations with period $2a$, where $a$ is projection of spacing between two adjacent "boson atoms" in undimerized lattice on chain axis direction. It means, that all various wave vectors $\vec{k}$  in $\vec{k}$-space will be in reduced zone with module of $\vec{k}$ in the range $-\frac{\pi}{2a} \leq k \leq \frac{\pi}{2a}$ \cite{SSH_PRB}. It can be considered like to usual semiconductors to be consisting of two subzones - conduction $(c)$ band and valence $(v)$ band. Then, it seems to be convenient to represent the operators $\{\hat{c}^+_{m,s}\}$,   $\{\hat{c}_{m,s}\}$, $m = \overline{1,N}$, in the form
\begin{equation}
\begin{split}
\label{Eq5m}
&\{\hat{c}_{m,s}\} = \{\hat{c}^{(c)}_{m,s}\} + \{\hat{c}^{(v)}_{m,s}\},\\
&\{\hat{c}^+_{m,s}\} = \{\hat{c}^{+(c)}_{m,s}\} + \{\hat{c}^{+(v)}_{m,s}\},
\end{split}
\end{equation} related to $\pi-c$- and $\pi-v$-band correspondingly,
and to define $\vec{k}$-space operators
\begin{equation}
\begin{split}
\label{Eq6m}
&\{\hat{c}^{(c)}_{k,s}\} = \{\frac{i}{\sqrt{N}}\sum_{m}\sum_{s}(-1)^{m+1}\exp(-ikma)\hat{c}^{(c)}_{m,s}\},\\
&\{\hat{c}^{(v)}_{k,s}\} = \{\frac{1}{\sqrt{N}}\sum_{m}\sum_{s}\exp(-ikma)\hat{c}^{(v)}_{m,s}\},
\end{split}
\end{equation}
$m = \overline{1,N}$. The principle, like to  MO LCAO is used in fact to build the operators $\{\hat{c}^{(c)}_{k,s}\}$ and $\{\hat{c}^{(v)}_{k,s}\}$, at that the antibonding character of $c$-band orbitals is taken into account by means of factor $i(-1)^{m+1}$.  Inverse to (\ref{Eq6m}) transform is
\begin{equation}
\begin{split}
\label{Eq7m}
&\{\hat{c}^{(c)}_{m,s}\} = \{\frac{1}{\sqrt{N}}\sum_{k}\exp{i[m(ka + \pi) - \frac{\pi}{2}]}\hat{c}^{(c)}_{k,s}\},\\
&\{\hat{c}^{(v)}_{m,s}\} = \{\frac{1}{\sqrt{N}}\sum_{k}\exp(ikma)\hat{c}^{(v)}_{k,s}\},
\end{split}
\end{equation}
$m = \overline{1,N}$. 

The $\sigma$-operators $\{\hat{a}^+_{m,s}\}$ and $\{\hat{a}_{m,s}\}$, $m = \overline{1,N}$ can also be represented in the form like to (\ref{Eq5m}) for $\pi$-operators and analogous to (\ref{Eq6m}),  transforms can be defined. Then the expression for the operator $\hat{\mathcal{H}}_{0}(u)$ can be rewritten
\begin{equation}
\begin{split}
\label{Eq8m}
&\hat{\mathcal{H}}_{0}(u) = \hat{\mathcal{H}}^{\sigma,c}_{0}(u) + \hat{\mathcal{H}}^{\sigma,v}_{0}(u) =
\sum_{m}\sum_{s}(E_m^k + K u_m^2) \times \\ 
&\frac{1}{N}\sum_{k}(\hat{a}^{+\sigma,c}_{k,s} \hat{a}^{\sigma,c}_{k,s} + \hat{a}^{+\sigma,v}_{k,s} \hat{a}^{\sigma,v}_{k,s}),
\end{split}
\end{equation}
where $\hat{a}^{+\sigma,c}_{k,s}$,  $\hat{a}^{\sigma,c}_{k,s}$ and  $\hat{a}^{+\sigma,v}_{k,s}$, $\hat{a}^{\sigma,v}_{k,s}$ are $\sigma$-operators of creation and annihilation, related to $\sigma$-$c$-band and to $\sigma$-$v$-band correspondingly. The independence of $E_m^k$ and $|u_m|$ on $m$, $m = \overline{1,N}$, means, that the expression  
$(E_m^k + K u_m^2)$ is independent on $m$. Then we obtain
\begin{equation}
\begin{split}
\label{Eq9m}
\hat{\mathcal{H}}_{0}(u) = \sum_{k}\sum_{s}(E^k + K u^2)(\hat{n}^{\sigma,c}_{k,s} + 
\hat{n}^{\sigma,v}_{k,s}),
\end{split}
\end{equation}
where
$\hat{n}^{\sigma,c}_{k,s}$ and  
$\hat{n}^{\sigma,v}_{k,s}$ are operators of number of  $\sigma$-quasiparticles in $\sigma$-$c$-band and  $\sigma$-$v$-band correspondingly.

The expression for $\hat{\mathcal{H}}_{\pi,t_0}(u)$ in terms of $\{\hat{c}^{(c)}_{k,s}\}$ and $\{\hat{c}^{(v)}_{k,s}\}$ is coinciding with known corresponding expression in \cite{SSH}, \cite{SSH_PRB} and it is
\begin{equation}
\begin{split}
\label{Eq10m}
\hat{\mathcal{H}}_{\pi,t_0}(u) =  \sum_{k}\sum_{s} 2t_0 \cos ka (\hat{c}^{+(c)}_{k,s}\hat{c}^{(c)}_{k,s} - \hat{c}^{+(v)}_{k,s}\hat{c}^{(v)}_{k,s})
\end{split}
\end{equation}
The expression for the second  part of operator $\hat{\mathcal{H}}_{\pi,t}(u)$ in terms of $\{\hat{c}^{(c)}_{k,s}\}$ and $\{\hat{c}^{(v)}_{k,s}\}$ is also coinciding in its form with known corresponding expression in \cite{SSH}, \cite{SSH_PRB} and it is given by
\begin{equation}
\begin{split}
\label{Eq10ma}
\hat{\mathcal{H}}_{\pi,\alpha_1}(u) =  \sum_{k}\sum_{s} 4 \alpha_1 u \sin ka (\hat{c}^{+(v)}_{k,s}\hat{c}^{(c)}_{k,s} + \hat{c}^{+(c)}_{k,s}\hat{c}^{(v)}_{k,s}), 
\end{split}
\end{equation}
where subscript $\alpha_1$ in Hamiltonian designation indicates on the taking into account the part of interaction, which is analogue for rest massless lattice of electron-phonon interaction in condensed matter lattices, and which is connected with resonance interaction  (hopping) processes.
When concern of field analogue of  electron-electron interaction
The  constant  term in Taylor series expansion of potential energy of "electron-electron" interaction about the dimerization coordinate and the term proportional to dimerization coordinate derivative of potential energy of "electron-electron" interaction will be taken into account in given consideration, since they seem to be  the most essential.
The expression for the $\hat{\mathcal{H}}_{\pi,u}(u)$, which describes the part of electron-phonon interaction field analogue, being to be the consequence of the  interaction between quasiparticles themselves (see further) in Fermi liquid state of
$\pi$-subsystem analogue in terms of $\hat{c}^{(c)}_{k,s}$
and $\hat{c}^{(v)}_{k,s}$
 can be represented in the form
\begin{equation}
\begin{split}
\label{Eq11m}
\hat{\mathcal{H}}_{\pi,u}(u) = \sum_{k}\sum_{k'}\sum_{s}\alpha_2(k, k',s) \hat{c}^{+(c)}_{k',s}\hat{c}^{+(v)}_{k',s}  \hat{c}^{(v)}_{k,s}\hat{c}^{(c)}_{k,s}.
\end{split}
\end{equation}
The constant independent on $u$ static term, which is
determined by electron-electron interaction analogue on different
"atomic" sites in a chain, that is the constant terms in
Taylor series expansion of potential energy of quasiparticle-quasiparticle interaction, which is field analogue of electron-electron interaction in the matter (in t-PA chains, for example) about the dimerization coordinate
was omitted in its explicit form from Hamiltonian in
given work, in order to establish the role of "phonon"-assisted part. The independent on $u$ static term is taking,
however, into consideration by calculation of coefficient
$\alpha_2(k, k',s)$ in "phonon"-assisted term.
Physically the identification of linear on displacement
$u$ parts of both resonance interaction (hopping) and the
pairwise interaction of quasiparticles in "$\pi$-subsystem" between themselves with"electron-phonon" interaction is understandable, if to take into account, that by "atomic"  displacements, like to CH-group displacements in t-PA,and resulting of given interaction, the "phonons" are generated, which
in its turn can by release of the place on, for instance,
m-th "atomic" site, to deliver the energy and impulse, which
are necessary for transfer of the quasiparticle (like to electron transfer in the matter)
from adjacent (m - 1)- or (m + 1)-position in chain in
the case of resonance interaction (hopping). For the case
the pairwise interaction of quasiparticles, it means, that
its linear on displacement $u$ part is realized by means
of "phonon" field, which transfers the energy and impulse
from one quasiparticle to another (and which can be not inevitable adjacent). 

Mathematically  it can be proved in the following way. The processes of pairwise interaction in $c$  and $v$ bands can be considered to be independent on each other. It means, that transition probability from the $\langle k_{l,s}|$-state to $\langle k_{j,s}|$-state in $c$-band and from $\langle k'_{l,s}|$-state to  $\langle k'_{j,s}|$-state  in $v$-band, which is proportional to coefficient $\alpha_2(k, k',s)$, can be expressed in the form of product of real parts of corresponding matrix elements, that is in the form
\begin{equation}
\begin{split}
\label{Eq12m}
&\alpha_2(k, k',s) \sim  Re\langle k_{l,s}|\hat{V}^{(c)}|k_{j,s}\rangle Re\langle k'_{l,s}|\hat{V}^{(v)}|k'_{j,s}\rangle = \\
&\sum_{k_{ph}}Re\langle k_{l,s}|\hat{V}^{(c)}|k_{ph}\rangle \langle k_{ph}|k_{j,s}\rangle \times \\
&\sum_{k_{ph}}Re\langle k'_{r,s}|\hat{V}^{(v)}|k_{ph}\rangle\langle k_{ph}|k'_{n,s}\rangle,  
\end{split}
\end{equation} 
where $\hat{V}^{(v)}$ = $V_{0(v)}\hat{e}$ ($\hat{e}$ is unit operator) is the first term in Taylor expansion of pairwise interaction of quasiparticles, for instance, with wave vectors  $k'_{r}$,  $k'_{n}$ and spin projection $s$ in $v$-band, that is, in ground state, $\hat{V}^{(c)}$ = $V_{1(c)} u \hat{e}$ is the second term in Taylor expansion of pairwise interaction in excited state (in c-band), that is, it is   product of configuration coordinate $u$ and coordinate derivative at $u = 0$  of operator of pairwise interaction of quasiparticles with wave vectors $k_{l}$, $k_{j}$ and spin projection $s$   in $c$-band, $k_{ph}$ is phonon wave vector, and the summation is realized over all the field phonon analogue  spectrum. At that, since the linear density of pairwise interaction is independent on $k$, which is the consequence of translation invariance of the chain, $V_{0(v)}$, $V_{1(c)}$ are constants. Therefore, the  pairwise interaction is considered to be accompanying by process of "phonon" generation, when "electronic" quasiparticles are already in excited state, that is,   in $c$-band (retardation effect of  field phonon analogue subsystem is suggested to be taking place like to retardation effect of phonon  subsystem in condensed matter). Then we have  $\hat{V}^{(c)} = {V}_{0(c)}u \hat{e}$, $\hat{V}^{(v)} = V_{0(v)} \hat{e}$.  A number of variants are possible along with process of "phonon" generation, corresponding to states of "electronic" quasiparticles in $c$-band above described.  The result will mathematically be quite similar,  if to interchange the role of $c$ and $v$ bands for given process. There seems to be possible the realization of both the stages (that is "phonon" generation and absorption) resulting of "electonic" quasiparticle interaction, in single $c$ or $v$ band. Simultaneous realization of phonon generation and absorption can also take place in both the bands. Mathematical description will be for all possible variants  similar and for distinctness we will consider only the first variant. For the simplicity we consider also the processes, in which the spin projection is keeping to be the same. It is evident also,  that in $z$-direction the impulse distribution is quasi-continuous (since the chain has  the macroscopic length $L = N a$).
Consequently, standard way $\sum_{k_{ph}}\rightarrow \frac{L}{2\pi}\int_{k_{ph}}$ can be used. Further, "phonon" states can be described by wave functions $\langle k_{ph}| = v_0 exp(ik_{ph} z)$, where $z \in [0,L]$, $k_{ph} \in [-\frac{\pi}{2a}$, $\frac{\pi}{2a}$], $v_0$ is constant. Therefore, from (\ref{Eq12m}) we have the expression
\begin{equation}
\begin{split}
\label{Eq13m}
&\alpha_2(k, k',s) = b |v_{0v}|^2 |v_{0c}|^2  V_{0(c)} u V_{0(v)} |\phi_{0cs}|^2 |\phi_{0vs}|^2 \times\\ 
&\frac{N}{2\pi(q_l - q_j)(q_r - q_n)} Re\{\exp[{i(k_l m_l - k_j m_j)a}] \exp{ika}\} \times \\
&Re\{\exp[{i(k'_r m_r - k'_n m_n)a}] \exp {ik'a}\},
\end{split}
\end{equation}
where $|\phi_{0cs}|^2$, $|\phi_{0vs}|^2$ are squares of the modules of the wave functions $|k_{j,s}\rangle$ and $|k'_{j,s}\rangle$ respectively, $k = k_{ph}(q_l - q_j)$, $k' = k'_{ph}(q_r - q_n)$ $q_l, q_j, q_r, q_n \in N$ with additional conditions $(q_l - q_j)a \leq L$, $(q_r - q_n)a  \leq L$, $b$ - is aspect ratio, which in principle can be determined by comparison with experiment. Here
the values $(q_l - q_j)$, $(q_r - q_n)$ determine the steps  in pairwise interaction with "phonon" participation and they are considered to be fixed. We will consider the processes for which $k = k'$, consequently, $(q_r - q_n)$ =  $(q_l - q_j)$. 

The relation (\ref{Eq13m})
 by $k_{l }m_l = k_{j} m_j$ and by $k_r m_r = k_n m_n$ transforms into
\begin{equation}
\begin{split}
\label{Eq14m}
&\alpha_2(k, k',s) = b |v_{0v}|^2 |v_{0c}|^2  V_{0(c)} u V_{0(v)} |\phi_{0cs}|^2 |\phi_{0vs}|^2 \times \\
&\frac{N}{2\pi[(q_l - q_j)]^2} \sin ka  \sin k'a.
\end{split}
\end{equation}
Let us designate 
\begin{equation}
\begin{split}
\label{Eq15m}
&b |v_{0v}|^2 |v_{0c}|^2  V_{0(c)}  V_{0(v)} |\phi_{0cs}|^2 |\phi_{0vs}|^2 \times \\
&\frac{N}{2\pi[(q_l - q_j)]^2} = 4 \alpha_2(s)
\end{split}
\end{equation}
Then, taking into account that spin projection $s$ is fixed, the dependence on $s$ can be omitted,  consequently  $\alpha_2(s) =  \alpha_2$. So we have
\begin{equation}
\begin{split}
\label{Eq16m}
&\hat{\mathcal{H}}_{\pi,u}(u) = \\ 
&\sum_{k}\sum_{k'}\sum_{s}4 \alpha_2 u \sin ka  \sin k'a \hat{c}^{+(c)}_{k',s}\hat{c}^{+(v)}_{k',s}  \hat{c}^{(v)}_{k,s}\hat{c}^{(c)}_{k,s}.
\end{split}
\end{equation}

Something otherwise will be treated the participation of the "phonons" in linear on $u$ part of   pairwise interaction,
if phonon generation is accompanying process of quasiparticle transition from $v$-band into $c$-band, that is, when the retardation effect of "phonon" subsystem can be neglected.  In given case
the expression for density of the "electron-phonon" coupling parameter $\alpha_2(k, k',s)$, which is related to the part of "electron-phonon" interaction, determined by interaction  between quasiparticles in field analogue of $\pi$-subsystem, which produce electromagnetic 1D Fermi liquid, is the following
\begin{equation}
\begin{split}
\label{Eq17m}
&\alpha_2(k, k',s) \sim  Re\langle k_{l,s}|\hat{V}^{}|k'_{j,s}\rangle = |v_{0v}|^2 |v_{0c}|^2   u V_{1} |\phi_{0s}|^2  \times\\
&\frac{N^2}{[2\pi]^2} 
\int_{k_{ph}}\exp[i(k_{ph} q a - k_{l} m_l a)]  \times\\
&\{\int_{k'_{ph}}\exp[i(k'_{ph} - k_{ph})q'a]  \times\\
&\exp[-i(k'_{ph} q' a - k'_{j} m_n a)]dk'_{ph}\}dk_{ph},
\end{split}
\end{equation}
where $q = m_j - m_l$, $q' = m_r - m_n$ are integers, satisfying foregoing relations, subscrips in left part are omitted, since fixed step is considered.
Then,  taking into account, that in  continuous limit by integration  the modules $k$ and $k'$ of  wave vectors $\vec{k}$ and $\vec{k'}$ are running all the $k$- and $k'$-values in $k$- and $k'$-spaces, we can  designate $(k_{ph} q a - k_{l}m_l a) = ka$, $(k'_{ph} q' a - k'_{j} m_j a) = k'a$, omitting the subscrips. So, we obtain
\begin{equation}
\begin{split}
\label{Eq18m}
&\alpha_2(k, k',s) \sim  Re\langle k_{l,s}|\hat{V}^{}|k'_{j,s}\rangle = |v_{0v}|^2 |v_{0c}|^2   u V_{1} |\phi_{0s}|^2  \times\\
&\frac{N^2}{[2\pi]^2} (\sin ka  \sin k'a + \cos ka  \cos k'a). \end{split}
\end{equation}
It was taken into account, that by $v$-band $\rightarrow$ $c$-band  transition of  quasiparticle, the impulse of emitted "phonon" by "vibronic"  system with "electronic" quasiparticle in $v$-band is equal to the impulse of absorbed "phonon" by "vibronic"  system with "electronic" quasiparticle in  $c$-band.

System of operators $\hat{c}^{+(c)}_{k',s}$, $\hat{c}^{+(v)}_{k',s}$,  $\hat{c}^{(v)}_{k,s}$, $\hat{c}^{(c)}_{k,s}$ corresponds to noninteracting  quasiparticles, and it is understandable, that in the case of  interacting  quasiparticles their linear combination has to be used 
\begin{equation}
\begin{split}
\label{Eq19m}
\left[\begin{array} {*{20}c}  \hat{a}^{(v)}_{k,s} \\  \hat{a}^{(c)}_{k,s} \end{array}\right] = \left[\begin{array} {*{20}c} \alpha_{k,s} & -\beta_{k,s}  \\  \beta_{k,s} & \alpha_{k,s} \end{array}\right] \left[\begin{array} {*{20}c}  \hat{c}^{(v)}_{k,s} \\  \hat{c}^{(c)}_{k,s}  \end{array}\right], 
\end{split}
\end{equation}
where matrix of transformation coefficients $\|A\|$ is
\begin{equation}
\begin{split}
\label{Eq20m}
\|A\| = \left[\begin{array} {*{20}c} \alpha_{k,s} & -\beta_{k,s} \\  \beta_{k,s} & \alpha_{k,s} \end{array}\right],  
\end{split}
\end{equation}
that is it is unimodular matrix with determinant $det\|A\|= \alpha^2_{k,s} + \beta^2_{k,s} = 1$.  Since $det\|A\|| \neq 0$, it means, that inverse 
transformation exists and it is given by the matrix
\begin{equation}
\begin{split}
\label{Eq21m}
\|A\|^{-1} = \left[\begin{array} {*{20}c} \alpha_{k,s} & \beta_{k,s}  \\ - \beta_{k,s} & \alpha_{k,s} \end{array}\right].  
\end{split}
\end{equation}
Then we obtain for the Hamiltonian $\hat{\mathcal{H}}_{\pi,u,\alpha_1}(u)$, which corresponds to SSH one-electron treatment of electron-phonon coupling, the following relation
\begin{equation}
\begin{split}
\label{Eq22m}
&\hat{\mathcal{H}}_{\pi,\alpha_1}(u) = \\
&\sum_{k}\sum_{s}\Delta_k [\alpha^2_{k,s} \hat{a}^{+(v)}_{k,s} \hat{a}^{(c)}_{k,s} - 
\alpha_{k,s} \beta_{k,s} \hat{a}^{+(v)}_{k,s}\hat{a}^{(v)}_{k,s} \\
&+ \beta_{k,s} \alpha_{k,s} \hat{a}^{+(c)}_{k,s}\hat{a}^{(c)}_{k,s} - 
\beta^2_{k,s} \hat{a}^{+(c)}_{k,s} \hat{a}^{(v)}_{k,s} 
+ \alpha^2_{k,s} \hat{a}^{+(c)}_{k,s} \hat{a}^{(v)}_{k,s} \\
&+ \alpha_{k,s} \beta_{k,s} \hat{a}^{+(c)}_{k,s} \hat{a}^{(c)}_{k,s}  -
\beta_{k,s} \alpha_{k,s} \hat{a}^{+(v)}_{k,s}\hat{a}^{(v)}_{k,s} - \beta^2_{k,s} \hat{a}^{+(v)}_{k,s} \hat{a}^{(c)}_{k,s}], 
\end{split}
\end{equation}
where $\Delta_k = 4 \alpha_1 u \sin ka$.
The diagonal part $\hat{\mathcal{H}}^d_{\pi,\alpha_1}(u)$ of operator $\hat{\mathcal{H}}_{\pi,\alpha_1}(u)$  is
\begin{equation}
\begin{split}
\label{Eq23m}
&\hat{\mathcal{H}}^d_{\pi,\alpha_1}(u) = \sum_{k}\sum_{s}2 \Delta_k \alpha_{k,s} \beta_{k,s} (\hat{n}^{(c)}_{k,s} - \hat{n}^{(v)}_{k,s}), 
\end{split}
\end{equation}
where $\hat{n}^{(c)}_{k,s}$ is density of operator of quasiparticles' number in $c$-band,  $\hat{n}^{(v)}_{k,s}$ is density of operator of quasiparticles' number in $v$-band.
 The part of pairwise interaction, which is  linear in displacement coordinate $u$ and leads to participation of the phonons,  is given by 
the Hamiltonian 
\begin{equation}
\begin{split}
\label{Eq24m}
&\hat{\mathcal{H}}_{\pi,u}(u) = \sum_{k}\sum_{k'}\sum_{s} 4 \alpha_2 u \sin ka  \sin k'a \times\\ &(\alpha^2_{k',s} \hat{a}^{+(c)}_{k',s}\hat{a}^{(v)}_{k',s} - \beta^2_{k',s} \hat{a}^{(v)}_{k',s} \hat{a}^{+(c)}_{k',s}\\
&+ \alpha_{k',s} \beta_{k',s} \hat{a}^{(c)}_{k',s} \hat{a}^{+(c)}_{k',s} -
\beta_{k',s} \alpha_{k',s} \hat{a}^{(v)}_{k',s} \hat{a}^{+(v)}_{k',s}) \\
&\times (\alpha^2_{k,s} \hat{a}^{+(c)}_{k,s} \hat{a}^{(v)}_{k,s} - \beta^2_{k,s} \hat{a}^{+(v)}_{k,s} \hat{a}^{(c)}_{k,s}\\
&+ \alpha_{k,s} \beta_{k,s} \hat{a}^{+(c)}_{k,s} \hat{a}^{(c)}_{k,s} - \beta_{k,s} \alpha_{k,s} \hat{a}^{+(v)}_{k,s} \hat{a}^{(v)}_{k,s}).
\end{split}
\end{equation}
The diagonal part $\hat{\mathcal{H}}^d_{\pi,u}(u)$ of operator $\hat{\mathcal{H}}_{\pi,u,\alpha_2}(u)$  is
\begin{equation}
\begin{split}
\label{Eq25m}
&\hat{\mathcal{H}}^d_{\pi,u}(u) = 4 \alpha_2 u \sum_{k}\sum_{k'}\sum_{s} \alpha_{k'} \beta_{k'} (\hat{n}^{(v)}_{k',s} - \hat{n}^{(c)}_{k',s}) \\
&\times \alpha_{k,s} \beta_{k,s} (\hat{n}^{(v)}_{k,s} - \hat{n}^{(c)}_{k,s}) \sin k'a \sin ka 
\end{split}
\end{equation}

The Hamiltonian $\hat{\mathcal{H}}_{\pi,t_0}(u)$ in terms of operator variables $\hat{a}^{(c)}_{k,s}$ $\hat{a}^{(v)}_{k,s}$ is
\begin{equation}
\begin{split}
\label{Eq26m}
&\hat{\mathcal{H}}_{\pi,t_0}(u) =  \sum_{k}\sum_{s} 2t_0 \cos ka [(\alpha^2_{k,s}  - \beta^2_{k,s}) (\hat{a}^{+(c)}_{k,s} \hat{a}^{(c)}_{k,s} - \\
&\hat{a}^{+(v)}_{k,s} \hat{a}^{(v)}_{k,s}) - 2 \alpha_{k,s} \beta_{k,s} (\hat{a}^{+(v)}_{k,s} \hat{a}^{(c)}_{k,s} + \hat{a}^{+(c)}_{k,s} \hat{a}^{(v)}_{k,s})]
\end{split}
\end{equation}
The diagonal part $\hat{\mathcal{H}}^d_{\pi,t_0}(u)$ of operator $\hat{\mathcal{H}}_{\pi,t_0}(u)$  is given by the relation
\begin{equation}
\begin{split}
\label{Eq27m}
\hat{\mathcal{H}}^d_{\pi,t_0}(u) = \sum_{k}\sum_{s} \epsilon_k (\alpha^2_{k,s}  - \beta^2_{k,s}) (\hat{n}^{(c)}_{k,s} -
\hat{n}^{(v)}_{k,s}), 
\end{split}
\end{equation}
where $\epsilon_k = 2t_0 \cos ka$. 

The operator transformation for the field analogue of $\sigma$-subsystem, which is similar  to (\ref{Eq19m}), shows, that the Hamiltonian $\hat{\mathcal{H}}_{0}(u)$ is invariant under given transformation, that is, it can be represented in the form, given by (\ref{Eq9m}).

To find the values of elements of matrices $\|A\|$ and  $\|A\|^{-1}$, the Hamiltonian $\hat{\mathcal{H}}_{}(u)$
has to be tested for conditional extremum on the variables $\alpha_{k}$, $\beta_{k}$ with condition $\alpha^2_{k,s}  - \beta^2_{k,s} = 1$. The corresponding Lagrange function $\hat{\mathfrak{E}}^L_{}(u)$  
is
\begin{equation}
\begin{split}
\label{Eq28m}
&\hat{\mathfrak{E}}^L_{}(u) = \sum_{k}\sum_{s}(\frac{\hat{P}^2}{2M^*} + K u^2)(\hat{n}^{\sigma,c}_{k,s} + 
\hat{n}^{\sigma,v}_{k,s}) \\
&+ \sum_{k}\sum_{s} \epsilon_k (\alpha^2_{k,s}  - \beta^2_{k,s}) (\hat{n}^{(c)}_{k,s} -
\hat{n}^{(v)}_{k,s}) \\
&+ \sum_{k}\sum_{s} 2 \Delta_k \alpha_{k,s} \beta_{k,s} (\hat{n}^{(c)}_{k,s} - \hat{n}^{(v)}_{k,s}) \\
&+ 4 \alpha_2 u \sum_{k}\sum_{k'}\sum_{s} \alpha_{k',s} \beta_{k',s} (\hat{n}^{(c)}_{k',s} - \hat{n}^{(v)}_{k',s}) \alpha_{k,s} \beta_{k,s}\\
&\times(\hat{n}^{(c)}_{k,s} - \hat{n}^{(v)}_{k,s}) \sin k'a \sin ka  + \lambda_{k,s} (\alpha^2_{k,s}  - 1 + \beta^2_{k,s})
\end{split}
\end{equation}
Then, the necessary condition for extremum is determined by  Lagrange equations
\begin{equation}
\begin{split}
\label{Eq29m}
&\frac{\partial{\hat{\mathfrak{E}}^L_{}(u)}}{\partial\alpha_{k}} = 2 \alpha_{k,s}\epsilon_k (\hat{n}^{(c)}_{k,s} - \hat{n}^{(v)}_{k,s}) + 2 \Delta_k  \beta_{k,s} (\hat{n}^{(c)}_{k,s} - \hat{n}^{(v)}_{k,s}) \\
&\times [1 + \frac{\alpha_2}{\alpha_1} \sum_{k'}\sum_{s} \alpha_{k',s} \beta_{k',s} \sin k'a (\hat{n}^{(c)}_{k',s} - \hat{n}^{(v)}_{k',s})] \\
&+ 2 \lambda_{k,s} \alpha_{k,s} = 0,
\end{split}
\end{equation}
\begin{equation}
\begin{split}
\label{Eq30m}
&\frac{\partial{\hat{\mathfrak{E}}^L_{}(u)}}{\partial\beta_{k,s}} = 2 \beta_{k,s}\epsilon_k (\hat{n}^{(v)}_{k,s} - \hat{n}^{(c)}_{k,s}) + 2  \Delta_k  \alpha_{k,s} (\hat{n}^{(c)}_{k,s} - \hat{n}^{(v)}_{k,s}) \\
&\times [1 + \frac{\alpha_2}{\alpha_1} \sum_{k'}\sum_{s} \alpha_{k',s} \beta_{k',s} \sin k'a (\hat{n}^{(c)}_{k',s} - \hat{n}^{(v)}_{k',s})] \\
&+ 2 \lambda_{k,s} \beta_{k,s} = 0
\end{split}
\end{equation}
and
\begin{equation}
\begin{split}
\label{Eq31m}
\frac{\partial{\hat{\mathfrak{E}}^L_{}(u)}}{\partial\lambda_{k,s}} = \alpha^2_{k,s}  - 1 + \beta^2_{k,s} = 0.
\end{split}
\end{equation}
Let us designate
\begin{equation}
\begin{split}
\label{Eq32m}
[1 + \frac{\alpha_2}{\alpha_1} \sum_{k'}\sum_{s} \alpha_{k',s} \beta_{k',s} \sin k'a (\hat{n}^{(c)}_{k',s} - \hat{n}^{(v)}_{k',s})] = \hat{\mathcal{Q}},
\end{split}
\end{equation}
then, passing on to observables in the Lagrange equations (\ref{Eq29m}) - (\ref{Eq31m}), we obtain for $\beta^2_{k,s}$, $\alpha^2_{k,s}$ and for product $\alpha_{k,s} \beta_{k,s}$ the relations
\begin{equation}
\begin{split}
\label{Eq33m}
\beta^2_{k,s} = \frac{1}{2}(1 \pm \frac{\epsilon_k}{\sqrt{\epsilon^2_k + \mathcal{Q}^2 \Delta^2_k}}), 
\end{split}
\end{equation}
\begin{equation}
\begin{split}
\label{Eq34m}
\alpha^2_{k,s} = \frac{1}{2}(1 \mp \frac{\epsilon_k}{\sqrt{\epsilon^2_k + \mathcal{Q}^2 \Delta^2_k}}), 
\end{split}
\end{equation}
\begin{equation}
\begin{split}
\label{Eq35m}
\alpha_{k,s} \beta_{k,s} = \frac{1}{2}\frac{\mathcal{Q} \Delta_k}{\sqrt{\epsilon^2_k + \mathcal{Q}^2  \Delta^2_k}}, 
\end{split}
\end{equation}
where $\mathcal{Q}$ is eigenvalue of operator $\hat{\mathcal{Q}}$.
The equation for factor $\mathcal{Q}$ is
\begin{equation}
\begin{split}
\label{Eq36m}
[1 + \frac{\alpha_2}{2\alpha_1} \sum_{k}\sum_{s}\frac{\mathcal{Q} \Delta_k \sin ka }{\sqrt{\epsilon^2_k + \mathcal{Q}^2 \Delta^2_k}} ({n}^{(c)}_{k,s} - {n}^{(v)}_{k,s})] = \mathcal{Q},
\end{split}
\end{equation}
where superscript {'} is omitted and  ${n}^{(c)}_{k,s}$ is eigenvalue of density  operator of quasiparticles' number in $c$-band,  ${n}^{(v)}_{k,s}$ is eigenvalue of density operator of quasiparticles' number in $v$-band.
It is evident, that at $Q = 1$ in (\ref{Eq33m}) - (\ref{Eq35m}) we have the case of SSH-model. It corresponds to the case, if $\frac{\alpha_2}{\alpha_1} \sum_{k}\sum_{s} \frac{1}{2}\frac{\Delta_k}{\sqrt{\epsilon^2_k +  \Delta^2_k}} \sin ka ({n}^{(c)}_{k,s} - {n}^{(v)}_{k,s})] \rightarrow 0$, which is realized, if $\alpha_2  \rightarrow 0$. Consequently, it seems to be interesting to consider the opposite case, when $|\frac{\alpha_2}{\alpha_1} \sum_{k}\sum_{s} \frac{1}{2}\frac{\Delta_k}{\sqrt{\epsilon^2_k + \Delta^2_k}}\sin ka ({n}^{(c)}_{k,s} - {n}^{(v)}_{k,s})]| \gg 1$. Passing on to continuum limit, in which $\sum_{k}\sum_{s}
\rightarrow 2 \frac{Na}{\pi} \int\limits_0^{\frac{\pi}{2a}}$, and assuming ${n}^{(v)}_{k,s} = 1$, ${n}^{(c)}_{k,s} = 0$, we have integral equation
\begin{equation}
\begin{split}
\label{Eq37m}
\frac{2 N u a \alpha_2}{\alpha_1 \pi t_0} \int\limits_0^{\frac{\pi}{2a}}\frac{\sin^2 ka}{\sqrt{1 - \sin^2 ka [1-(\frac{2 u \mathcal{Q}}{t_0})^2] }} dk = 1.
\end{split}
\end{equation}
In the case $|\frac{2 u \mathcal{Q}}{t_0}| < 1$ the relation (\ref{Eq37m}) can be rewritten in the form
\begin{equation}
\begin{split}
\label{Eq38m} 
&K\left\{\sqrt{1-\left(\frac{2\alpha_1 u \mathcal{Q}}{t_0}\right)^2}  \right\} - E\left\{\sqrt{1-\left(\frac{2\alpha_1 u \mathcal{Q}}{t_0}\right)^2} \right\} = \\
&\frac{\pi [t^2_0 - (2 u \mathcal{Q})^2]}{2 N u \alpha_2},
\end{split}
\end{equation}
where $K\left\{\sqrt{1-(\frac{2\alpha_1 u \mathcal{Q}}{t_0})^2}  \right\}$ and $E\left\{\sqrt{1-(\frac{2 u \mathcal{Q}}{t_0})^2} \right\}$ are complete elliptic integrals of the first and  the second kind, respectively. Expanding given  integrals into the series and restricting by the terms of the second-order of smallness, we obtain
\begin{equation}
\begin{split}
\label{Eq39m} \mathcal{Q} \approx \frac{t_0}{6 u } \sqrt{25 - 32 \frac{t_0 \alpha_1}{N u \alpha_2}}.
\end{split}
\end{equation}
It is evident, that the condition $|\frac{2 u \mathcal{Q}}{t_0}| < 1$  is held true by $\frac{1}{3}\sqrt{25 - 32 \frac{t_0 \alpha_1}{N u \alpha_2}} < 1$.

In the case $|\frac{2 u \mathcal{Q}}{t_0}| > 1$ the relation (\ref{Eq37m}) can be represented in the form
\begin{equation}
\begin{split}
\label{Eq40m}
\int\limits_0^{\frac{\pi}{2}}\frac{\cos^2 y} {\sqrt{1 - \sin^2 y [1-(\frac{t_0}{2 u \mathcal{Q}})^2] }} dy = - \frac{\pi \mathcal{Q} \alpha_1}{\alpha_2 N},
\end{split}
\end{equation}
where $y = \frac{\pi}{2} - ka$. It is equivalent to the equation
\begin{equation}
\begin{split}
\label{Eq41m}
&\left(\frac{t_0}{2 u \mathcal{Q}}\right) F\left\{\frac{\pi}{2},\sqrt{1-\left(\frac{t_0}{2 u \mathcal{Q}}\right)^2}  \right\}\\
&- E\left\{\frac{\pi}{2},\sqrt{1-\left(\frac{t_0}{2 u \mathcal{Q}}\right)^2}  \right\} = \\
&\frac{\pi \mathcal{Q} \alpha_1}{\alpha_2 N}\left[1 - \left(\frac{t_0}{2 u \mathcal{Q}}\right)^2\right],
\end{split}
\end{equation}
where $F\left\{\frac{\pi}{2},\sqrt{1-\left(\frac{t_0}{2 u \mathcal{Q}}\right)^2}  \right\}$ is the complete elliptic integral of the first kind.
The  approximation of elliptic integrals, like to  the approximation, given by  (\ref{Eq39m}), leads to the relation
\begin{equation}
\begin{split}
\label{Eq42m} \mathcal{Q} \approx \frac{-3 \alpha_2 N}{16} \left[ 1 \pm \sqrt{1 + \frac{80 \alpha_1 t_0}{9 N u \alpha_2  }}\right].
\end{split}
\end{equation}

In the case $\frac{2 u \mathcal{Q}}{t_0} = 1$ the relation (\ref{Eq37m}) is 
\begin{equation}
\begin{split}
\label{Eq43m}
\int\limits_0^{\frac{\pi}{2}}\cos^2 y dy = - \frac{\pi \alpha_1 \mathcal{Q}}{\alpha_2 N},
\end{split}
\end{equation}
where $y = \frac{\pi}{2} - ka$.
It is seen, that  in given case the value of parameter $Q$ is calculated exactly, and it is
\begin{equation}
\begin{split}
\label{Eq44m}
\mathcal{Q} = \frac{\alpha_2 N}{ 4\alpha_1} 
\end{split}
\end{equation} 
The values of energy of $\pi$-quasiparticles in $c$-band $E_k^{(c)}(u)$ and in  $v$-band $E_k^{(v)}(u)$ can be obtained in the following way
\begin{equation}
\begin{split}
\label{Eq45m}
E_k^{(c)}(u) = \frac{\partial{\mathfrak{E}^L_{}(u)}}{\partial{n^{(c)}_{k,s}}},
E_k^{(v)}(u) = \frac{\partial{\mathfrak{E}^L_{}(u)}}{\partial{n^{(v)}_{k,s}}},
\end{split}
\end{equation}
 where $\mathfrak{E}^L_{}(u)$ is the energy of "$\pi$-subsystem" of chain, which is obtained from Lagrange function operator, (\ref{Eq28m}), by passing on to observables. Therefore, we have
\begin{equation}
\begin{split}
\label{Eq46m}
&E_k^{(c)}(u) = \epsilon_k (\alpha^2_{k,s}  - \beta^2_{k,s}) + 2 \Delta_k \alpha_{k,s} \beta_{k,s} 
+ 8 \alpha_2 u \sin ka \\
&\times \sum_{k'}\sum_{s} \alpha_{k',s} \beta_{k',s} (\hat{n}^{(c)}_{k',s} - \hat{n}^{(v)}_{k',s})  \sin k'a  \alpha_{k,s} \beta_{k,s} \\
&= \epsilon_k (\alpha^2_{k,s}  - \beta^2_{k,s}) + 2 \Delta_k \alpha_{k,s} \beta_{k,s} \mathcal{Q}
\end{split}
\end{equation}
and 
\begin{equation}
\begin{split}
\label{Eq47m}
&E_k^{(v)}(u) = - \epsilon_k (\alpha^2_{k,s}  - \beta^2_{k,s}) - 2 \Delta_k \alpha_{k,s} \beta_{k,s} 
 - 8 \frac{\alpha_2} u \sin ka \\
&\times \sum_{k'}\sum_{s} \alpha_{k',s} \beta_{k',s} (\hat{n}^{(c)}_{k',s} - \hat{n}^{(v)}_{k',s})  \sin k'a  \alpha_{k,s} \beta_{k,s} \\
&=- \epsilon_k (\alpha^2_{k,s}  - \beta^2_{k,s}) - 2 \Delta_k \alpha_{k,s} \beta_{k,s} \mathcal{Q}.
\end{split}
\end{equation}
It is seen from (\ref{Eq46m}) and (\ref{Eq47m}), that $E_k^{(v)}(u) = - E_k^{(c)}(u)$. Taking into account the relations (\ref{Eq33m}) - (\ref{Eq35m}), we obtain
\begin{equation}
\begin{split}
\label{Eq48m}
E_k^{(c)}(u) =  \mp \frac{\epsilon^2_k}{\sqrt{\epsilon^2_k + \mathcal{Q}^2 \Delta^2_k}} + \frac{\mathcal{Q}^2 \Delta^2_k}{\sqrt{\epsilon^2_k + \mathcal{Q}^2  \Delta^2_k}}, 
\end{split}
\end{equation}

\begin{equation}
\begin{split}
\label{Eq49m}
E_k^{(v)}(u) =  \pm \frac{\epsilon^2_k}{\sqrt{\epsilon^2_k + \mathcal{Q}^2 \Delta^2_k}} - \frac{\mathcal{Q}^2 \Delta^2_k}{\sqrt{\epsilon^2_k + \mathcal{Q}^2  \Delta^2_k}}. 
\end{split}
\end{equation}
Therefore, we have two values for the enegy of quasiparticles, indicating on the possibility of the formation of the quasiparticles of two kinds. Upper sign in the first terms in (\ref{Eq48m}),   (\ref{Eq49m}) corresponds to the quasiparticles with the energy
\begin{equation}
\begin{split}
\label{Eq50m}
&E_k^{(c)}(u) =   \frac{\mathcal{Q}^2 \Delta^2_k  - \epsilon^2_k}{\sqrt{\epsilon^2_k + \mathcal{Q}^2  \Delta^2_k}},\\ 
&E_k^{(v)}(u) =   \frac{\epsilon^2_k - \mathcal{Q}^2 \Delta^2_k}{\sqrt{\epsilon^2_k + \mathcal{Q}^2  \Delta^2_k}}
\end{split}
\end{equation}
in $c$-band and $v$-band respectively. Lower sign in the first terms in (\ref{Eq48m}),   (\ref{Eq49m}) corresponds to  the quasiparticles with the energy
\begin{equation}
\begin{split}
\label{Eq51m}
&E_k^{(c)}(u) =   \sqrt{\epsilon^2_k + \mathcal{Q}^2  \Delta^2_k},\\ 
&E_k^{(v)}(u) =   - \sqrt{\epsilon^2_k + \mathcal{Q}^2  \Delta^2_k}
\end{split}
\end{equation}
in $c$-band and $v$-band respectively.
The quasiparticles of the second kind  at  $\mathcal{Q} = 1$ are quite similar quasiparticles in its mathematical description, that those ones, which were obtained in \cite {SSH}, \cite{SSH_PRB}. 

We have used the only necessary condition for extremum of the functions $E(\alpha_{k,s} \beta_{k,s})$. It was shown in \cite{Yerchuck_Dovlatova}, that for the  SSH-model  
the  sufficient conditions for the minimum are substantial, they change the role of both solutions.  Sufficient conditions for the minimum of the functions $E(\alpha_{k,s} \beta_{k,s})$
 can be  obtained by standard way, which was used in \cite{Yerchuck_Dovlatova}. It consist in that, that the second differential of  the energy, being to be the function of  three variables  ${\alpha}_{k,s}$,  ${\beta}_{k,s}$ and  
 $\lambda_{k,s}$,  has to be positively defined quadratic form. From the condition of positiveness of three principal minors of quadratic form coefficients we obtain  the following three sufficient conditions for the energy minimum

\paragraph{The first condition}

The first condition is
\begin{equation}
\label{Eq52m}
\begin{split}
&\{ \epsilon_k (1 - \frac{\epsilon_k}{\sqrt{\epsilon^2_k + \mathcal{Q}^2  \Delta^2_k}}) < \frac{(\mathcal{Q}\Delta_k)^2}{\sqrt{\epsilon^2_k + \mathcal{Q}^2  \Delta^2_k}} | 
({n}^c_{k,s} - {n}^v_{k,s}) < 0\}, \\
&\{\epsilon_k (1 - \frac{\epsilon_k}{\sqrt{\epsilon^2_k + \mathcal{Q}^2  \Delta^2_k}}) > \frac{(\mathcal{Q}\Delta_k)^2}{\sqrt{\epsilon^2_k + \mathcal{Q}^2  \Delta^2_k}} | ({n}^c_{k,s} - {n}^v_{k,s}) > 0 \}
\end{split}
\end{equation}
for the solution which coincides with SSH-solution at the value $\mathcal{Q} = 1$ (SSH-like solution) and
\begin{equation}
\label{Eq53m}
\begin{split}
&\{\epsilon_k (1 + \frac{\epsilon_k}{\sqrt{\epsilon^2_k + \mathcal{Q}^2  \Delta^2_k}}) < \frac{(\mathcal{Q}\Delta_k)^2}{\sqrt{\epsilon^2_k + \mathcal{Q}^2  \Delta^2_k}} | ({n}^c_{ks} - {n}^v_{ks}) < 0 \},\\
&\{\epsilon_k (1 + \frac{\epsilon_k}{\sqrt{\epsilon^2_k + \mathcal{Q}^2  \Delta^2_k}}) > \frac{(\mathcal{Q}\Delta_k)^2}{\sqrt{\epsilon^2_k + \mathcal{Q}^2  \Delta^2_k}} | ({n}^c_{ks} - {n}^v_{ks}) > 0 \}
\end{split}
\end{equation}
 for the additional solution. It is seen, that the first condition is realizable for the quasiparticles of both the kinds, at that for both near equilibrium  state $({n}^c_{ks} - {n}^v_{ks} < 0)$  and  strongly nonequilibrium state $(n^c_{ks} - {n}^v_{ks} > 0$.  

\paragraph{The second condition}

The second condition is the same for both the solutions and it is
\begin{equation}
\label{Eq54m}
 (\frac{\epsilon^2_k}{\sqrt{\epsilon^2_k + \mathcal{Q}^2  \Delta^2_k}} - 2\frac{(\mathcal{Q}\Delta_k)^2}{\sqrt{\epsilon^2_k + \mathcal{Q}^2  \Delta^2_k}})^2 - \epsilon^2_k   + \frac{1}{4} (\mathcal{Q}\Delta_k)^2 > 0 
\end{equation}
It  is realizable for the quasiparticles of both the kinds.

\paragraph{The third condition}

For the SSH-like solution we have
\begin{equation}\label{Eq55m}
(3\frac{(\mathcal{Q}\Delta_k)^2}{\sqrt{\epsilon^2_k + \mathcal{Q}^2  \Delta^2_k}} + 4\frac{\epsilon^2_k}{\sqrt{\epsilon^2_k + \mathcal{Q}^2  \Delta^2_k}})({n}^c_{ks} - {n}^v_{ks}) > 0. 
\end{equation}
It  means,  that   SSH-like solution is unapplicable for the description of standard processes, passing near equilibrium state by any parameters. The quasiparticles, described by   SSH-like solution, can be created the only in strongly nonequilibrium state with inverse                                                                                                   
population of the levels in $c$- and $v$-bands. At the same time  the solution, which corresponds to upper signs in (\ref{Eq48m}), has to satisfy  to the following condition
\begin{equation}
\label{Eq56m}
(3\frac{(\mathcal{Q}\Delta_k)^2}{\sqrt{\epsilon^2_k + \mathcal{Q}^2  \Delta^2_k}} - 4\frac{\epsilon^2_k}{\sqrt{\epsilon^2_k + \mathcal{Q}^2  \Delta^2_k}})({n}^c_{ks} - {n}^v_{ks}) > 0, 
\end{equation}
which can be realized  both in near equilibrium and in strongly nonequilibrium states of the $\pi$-subsystem field analogue, which is considered to be EM-field analogue of quantum 1D Fermi liquid in condensed matter.

\paragraph{Ground State of  $t$-PA chain}

The continuum limit for the ground state energy of the "chain" with SSH-like quasiparticles will coincide mathematically with known solution \cite{SSH_PRB}, if to  replace $\Delta_k \mathcal{Q} \rightarrow \Delta_k$.  Let us calculate  the ground state energy $E^{[u]}_0(u)$ of the field "chain" with  quasiparticles' branch, which is stable near equilibrium. Taking into account, that in ground state ${n}^c_{k,s} = 0$, ${n}^v_{k,s} = 1$, in the continuum limit we have
\begin{equation}
\label{Eq57m}
E^{[u]}_0(u) = - \frac{2N a}{\pi}\int\limits_0^{\frac{\pi}{2a}} \frac{(\mathcal{Q}\Delta_k)^2 - 
\epsilon^2_k}{\sqrt{(\mathcal{Q}\Delta_k)^2 + 
\epsilon^2_k}}dk + 2NKu^2,
\end{equation}
then, calculating the integral and using the complete elliptic integral of the first kind $F(\frac{\pi}{2}, 1 - z^2)$ and the complete elliptic integral of the second kind
$E(\frac{\pi}{2}, 1 - z^2)$  we obtain
\begin{equation}
\label{Eq58m}
\begin{split}
&E^{[u]}_0(u) =  \frac{4Nt_0}{\pi}\{F(\frac{\pi}{2}, 1 - z^2) + \\
&\frac{1 + z^2}{1 - z^2}[E(\frac{\pi}{2}, 1 - z^2) - F(\frac{\pi}{2}, 1 - z^2)]\} + 2NKu^2, 
\end{split}
\end{equation}
where
$z^2 = \frac{2\mathcal{Q}\alpha_1 u}{t_0}$.
Approximation of ({\ref{Eq58m}}) at $z \ll 1$ gives
\begin{equation}
\label{Eq59m}
\begin{split}
&E^{[u]}_0(u) = N \{\frac{4t_0}{\pi} - \frac{6}{\pi}\ln\frac{2t_0}{\mathcal{Q}\alpha_1 u} \frac{4 (\mathcal{Q}\alpha_1)^2 u^2}{t_0} + \\
&\frac{28 (\mathcal{Q}\alpha_1)^2 u^2}{\pi t_0} + ...\} + 2NKu^2.
\end{split}
\end{equation}
It is seen from (\ref{Eq59m}), that the energy of quasiparticles, described by   solution, which corresponds to upper signs in (\ref{Eq48m}) has the form of Coleman-Weinberg potential with two minima at the values of dimerization coordinate $u_0$ and $-u_0$ like to the energy of quasiparticles, described by   SSH-solution \cite{SSH_PRB}. It is understandable, that subsequent consideration, including electrically neutral S=1/2 soliton formation and electrically charged spinless soliton formation, that is the appearence of the phenomenon of spin-charge separation,  by quantum Fermi liquid description of 1D systems  will be coinciding in its  mathematical form with starting  SSH-model.

Therefore, all qualitative conclusions of the model proposed in \cite{SSH_PRB} are holding in Fermi-liquid consideration of "$\pi$-subsystem" of the field "boson atomic chain" (instead of Fermi-gas consideration) for the quasiparticles, corresponding to the second-branch-solution. It seems to be also substantial, that Fermi-liquid treatment of electron-phonon interaction extends strongly the applicability limits of SSH-model along with its extension for description of  1D conjugated conductors  in the case of strong electron-phonon interaction \cite{Dovlatova_Yerchuck_Borovik} with its extension for description of quantized EM-field. It is evident, that the mechanism of the phenomenon
of spin-charge separation in 1D quantum Fermi-liquid is soliton mechanism, analogous to proposed by Jackiw and Rebbi \cite{Jackiw} on the basis of field theory positions and its applicability to physical fields, in particular, to EM-field, seems to be natural. It means, like to SSH-model [formally Fermi-gas model, see, however the remark in \cite{Dovlatova_Yerchuck_Borovik}], that when elementary spin and charge carrier, for instance, an electron,
is added to an  neutral  chain, it
can break up into two pieces, one of which carries the
electron’s charge and the other its spin. Given result
bears a clear family relation with the phenomenon
of antecedent spin-charge separation in the 1D electron gas theory of Luther and Emery \cite{Luther},
but it is quite different from Anderson spinon-holon mechanism \cite{Anderson}, \cite{Anderson_P_W}.   
Very strong argument in favour of the model proposed is given by experimentally observed  phenomenon of electron-positron annihilation. So it is well known, and theoretically described, see for example \cite{Berestecky}, that by direct annihilation  of electron-positron pair two photons are produced. At the same time the explanation, for instance,  for the case, when the relative velocity of annihilating particles tends to zero, how from two particles with spin value 1/2 also  two particles with spin value 1 are produced, is in fact absent. At the same time the  boson model of EM field explains given disagreement between electron-positron annihilation and theoretical viewpoint on the photon to be the spin-1 particle in a natural way - from two particles with spin value 1/2, that is from  electron-positron pair also  two particles  - two photons -  with spin value 1/2 are produced by electron-positron pair annihilation.

Thus, the results obtained allow to propose the reasonable explanation of the existence in the felds with
charges of chargeless particles - solitons with nonzero
spin value, which in the case of EM-field is equal to 1/2 instead  of prevalent viewpoint, that photons possess by spin
S = 1. The photons in quantized EM-field are main excitations in its oscillator structure, which is equivalent to
spin S = 1 "boson-atomic" structure. It has strong resemblance with  well known spin S = 1 boson matter structure, for example with 
 carbon atomic backbone structure in many conjugated
polymer chains. The photons have two kind nature. The
photons of the  first kind represent themselves neutral
EM-solitons of SSH-soliton family. They are main excitations in so-called "undoped"  structure of EM-fieeld,
including free EM-field in vacuum. Naturally, they have
nonzero size, that is they cannot be considered to be
point objects. It seems to be evident, that like to  SSH-model, the main excitations in "doped" "boson-atomic" structure of EM-field will be charged spinless
EM-solitons, which also can be referred to SSH-soliton
family. It seems to be  reasonable to suggest, that "doping" can
be effective  realized in the medium like to rain-clouds, although
detailed mechanism has to be additionally studied. 

The
representation of photons to be the result of spin-charge
separation effect in field analogue of quantum Fermi liquid and their assignment with main excitations in ground state of "$\pi$-subsystem" in rest masless
"boson-atomic" EM-field structure - chargeless spin 1/2
topological solitons - makes substantially more clear the
nature of corpuscular-wave dualism. Really, like to matter atomic structure, the quantized EM-field represents
itself the discrete "boson-atomic" structure, the individual
 bosons in which produce the lattice like to  genuine atomic lattice
structure in condensed matter. The main difference between EM-field "atomic" lattice structure and atomic  lattice structure
in condensed matter consists in that the "atoms" in EM-field structure have zeroth rest mass. The origin of wave
in given structure is determined by the mechanism, quite
analogous to Bloch waves' formation in solid state of condensed matter. They are harmonic trigonometric functions
for Maxwellian EM-field (real or imaginary parts by description in complex form), which determine wave character of quantized Maxwellian EM-field. At the same
time, there are simultaneously the corpuscules, propagating along given EM-field "boson-atomic" 1D-lattice structure,
that is, chargeless spin 1/2 topological relativistic solitons -
photons, formed in usual conditions (or spinless charged
solitons in so-called "doped" EM-field structure). It becomes now to be understandable, that the display of the
corpuscular or wave nature of EM-field will be dependent
on experimental conditions.

\section{Conclusions}

 Therefore,  a number of new achievement in classical and quantum theory of electrodynamics in comparisin with experimental data are revieved in given paper. they are the following.

It is shown on the basis of complex number theory,  that any quantumphysical quantity is complex quantity.

It has been established the partition of
linear space $\left\langle\mathcal F,+,\cdot \right\rangle$ over the ring of scalars and pseudoscalar set, the vectors in which are sets of contravariant (or covariant) EM-field tensors and pseudotensors $\left\{{F}^{\mu\nu}\right\}$, $\left\{\tilde{F}^{\mu\nu}\right\}$, (or $\left\{{F}_{\mu\nu}\right\}$, $\left\{\tilde{F}_{\mu\nu}\right\}$), into 4 subspaces $\left\langle\mathcal F^{(i)},+,\cdot \right\rangle$, $i = \overline {1,4}$. 
In subspaces  $\left\langle \mathcal F^{(1)},+,\cdot \right\rangle$,  $\left\langle \mathcal F^{(2)},+,\cdot \right\rangle$  vector $\vec {E}$ is polar vector, and vector $\vec {H}$ is axial.  At the same time,  the components of vector $\vec {E}$ in the second subspace correspond to pure space components of  field tensor, and the components of  vector $\vec {H}$ correspond  to time-space mixed components of given field tensor. Arbitrary element of subspace  $\left\langle \mathcal F^{(3)},+,\cdot \right\rangle$  is 4-\textit{pseudo}tensor. Its space components are in fact the components of antisymmetric 3-\textit{pseudo}tensor, which determines polar magnetic field vector $\vec {H}$, (that is vector, which is dual to given tensor), while time-space mixed components are the components of axial 3-vector $\vec {E}$ of electric field.   
Therefore, the symmetry properties of the components of the vectors $\vec {E}$ and $\vec {H}$ under improper rotations in the third subspace  will be opposite to those ones in the first subspace. The symmetry properties of the components of $\vec {E}$ and $\vec {H}$ under improper rotations in the fourth subspace  will be opposite to those ones in the second subspace.

The algebraic properties of the set of functionals, determined on the space $\left\langle F,+,\cdot \right\rangle$ were established. In particular, the statement, indicating, that the set of  functionals 
 $\left\{\Phi [\tilde {F}^{\mu \nu}(x)]\right\}$, $\left\{\tilde{\Phi}[\tilde {F}^{\mu \nu}(x)]\right\}$, preassigned on the space $\left\langle F,+,\cdot \right\rangle$, produces  linear space $\left\langle \Phi',+,\cdot \right\rangle$ over a field of scalars $P$, which is dual to the space $\left\langle F,+,\cdot \right\rangle$, however it is nonselfdual,
 is established.
Given result is in fact the consequence of the existence of two kinds of independent tensor and scalar characteristics  of EM-field, that is, genuine tensor and scalar functions on the one hand  and pseudotensor and pseudoscalar functions on the other hand.
The practical significance of given result consists in the 
necessity, if some physical phenomenon with participation of EM-field is studied, to take into consideration always both the spaces, that is   $\left\langle F,+,\cdot \right\rangle$ and  $\left\langle \Phi,+,\cdot \right\rangle$. More strictly, known Gelfand triple, which includes together with spaces $\left\langle F,+,\cdot \right\rangle$ and  $\left\langle \Phi,+,\cdot \right\rangle$ the Hilbert space with topology, determined in the proper way, has to be taken into consideration.

Additional gauge invariance of complex relativistic fields was studied. It has been found, that conserving quantity, corresponding to invariance of generalized relativistic equations under the operations of additional gauge symmetry group - multiplicative group $\mathfrak R$ of all real numbers (without zero) - is purely imaginary charge. So, it was shown, that complex fields are characterized by complex charges. It gives  the key for correct generalization of field equations, in particular, for electrodynamics.  
 
 Additional hyperbolic dual symmetry of Maxwell equations is established, which includes Lorentz-invariance to be its particular case. The essence of additional hyperbolic dual symmetry of Maxwell equations is that, that Maxwell equations along with dual transformation symmetry, established by Rainich, given by (\ref{eq1b}) - (\ref{eq1c}), are invariat under the dual transformations of another kind.  Hyperbolic dual transformations for electric and magnetic field strengh vector functions are
\begin{equation}
\label{eq1bbca}
 \left[\begin{array} {*{20}c}  \vec {E{''}} \\ \vec {H{''}} \end{array}\right] = \left[\begin{array} {*{20}c} \cosh\vartheta& i\sinh\vartheta  \\ -i\sinh\vartheta&\cosh\vartheta \end{array}\right]\left[\begin{array} {*{20}c}  \vec {E} \\ \vec {H} \end{array}\right],
\end{equation}
where  $\vartheta$ is arbitrary continuous parameter,
$\vartheta \in [0,2\pi]$. 

Generalized Maxwell equations are obtained on the basis of both dual and hyperbolic dual symmetries of  EM-field. It is shown, that in general case both scalar and vector quantities, entering in equations, are quaternion quantities, four components of which have different parities under improper rotations.  

Invariants for  EM-field, consisting of  dually symmetric parts, for both the cases of dual symmetry and hyperbolic dual symmetry are found. 
It is concluded, that Maxwell equations with all quaternion vector and scalar variables give concrete connection between dual and gauge symmetries of EM-field.

 It is shown, that there exists physical conserving
quantity, which is simultaneously invariant under  both Rainich dual and additional hyperbolic dual symmetry transformation of Maxwell
equations. It is spin in general case and spirality in the geometry, when vector $\vec{E}$ is directed along abscissa axis,and $\vec{H}$ is directed along ordinate axis in  $(\vec{E}, \vec{H})$ functional space. It is additional proof for quaternion four component 
structure of EM-field to be a single whole. Spin takes on special leading significance among the physical characteristics of EM-field, since the only spin (spirality  in the geometry considered) combine two subsystems of photon fields,  which have definite $P$-parity  (even and uneven) with the subsystem of two fields, which have definite $t$-parity (also even and uneven) into one system. It is considered to be the proof for four component structure of EM-field to be a single whole, that is, it is the confirmation along with the possibility of the representation of EM-field in four component quaternion form, given by (\ref{eq5abcce}), (\ref{eq6bcdde}), (\ref{eq7abccde}), (\ref{eq8abccde})[sufficient condition],
the necessity of given representation. It
extends the overview on the nature of EM-field itself. It seems to be remarkable, that given result on the special leading significance of spin is in agreement with result in \cite{D_Yearchuck_A_Dovlatova}, where was shown,  that spin is quaternion vector of the state in Hilbert space, defined under ring of quaternions, of any quantum system (wthin the
frame of the chain model considered) interacting with EM-field.

The  connection between symmetry of dynamical systems, in particular, between gauge symmetry and analytical properties of quantities, which are invariant under corresponding symmetry qroup has been established. For example, the
 analicity of complex-valued function $Q(z)$ = $Q_1(z) + iQ_2(z)$ of complex variable $z = \vec{r} + ict$, representing itself the complex charge of EM-field   
is proved.

The main postulate of quantum mechanics: "To any mechanical quantity can be set
up in the correspondence the Hermitian
matrix by quantization" was proved.
 
New principle of EM-field  quantization is proposed. It is development of canonical Dirac quantization method, which  is realized in two aspects. The first aspect consist in  choosing of field functions, which are immediately  observable quantities - 4-vector-functions  $E_\mu(\vec{r},t)$ and $H_\mu(\vec{r},t)$, the first three components of which are the components of 3-vector-functions $\vec{E}(\vec{r},t)$ and $\vec{H}(\vec{r},t)$ of electric and magnetic field strengths correspondingly, the fourth component is $\frac{i c \rho_e(\vec{r},t)}{\lambda}$, $\frac{i c \rho_m(\vec{r},t)}{\lambda}$, where $\rho_e(\vec{r},t)$ is electric charge density,  $\rho_m(\vec{r},t)$ is magnetic charge density, which is equaled to zero in the case of single-charge electrodynamics, $\lambda$ is medium conductivity, which in the case of free field in vacuum is $\lambda_v = \frac{1} {Z_0} = \frac{1} {120\pi} Ohm^{-1}$. The second  aspect is the realization along with well known time-local quantization of space-local quantization and space-time-local quantization, which allows to establish,  correspondingly, the time of photon creation (annihilation), the space coordinate of photon creation (annihilation) and the the space and time coordinates simultaneously of photon creation (annihilation). 

It is shown, that Coulomb field can be quantized in 1D- and 2D-systems, that is, it is radiation field in given low-dimensional systems.

New model of photons is proposed. The photons in quantized EM-field are main excitations  in oscillator structure of EM-field, which is equivalent to spin S = 1 "boson-atomic" structure,  like  matematically to well known spin S = 1 boson matter  structure - carbon atomic backbone chain structure in many conjugated polymers. They have two kind nature.  The photons of the first kind  represent themselves neutral chargeless EM-solitons of  SSH-soliton family. The photons of the second kind  represent themselves charged spinless EM-solitons, which also can be referred  to  SSH-soliton family.

Substantially more clear the
nature of corpuscular-wave dualism is represented.

\section{Aknowledgement} Authors are grateful to Y.Yerchack for discussions and the help in the work.


\begin{thebibliography}{16}
\bibitem{Heaviside} Heaviside O, Electrical Papers, London, 1
(1892), Phil.Trans.Roy.Soc.A,\textbf{183} (1893)  423-430, Electromagnetic Theory, London, 1893
\bibitem{Larmor} Larmor J, Collected papers, London, 1928
\bibitem{Rainich} Rainich G Y, Trans.Am.Math.Soc.,\textbf{27} (1925) 106–136
\bibitem{Tomilchick} Strazhev V.I, Tomilchick L.M, Electrodynamics with Magnetic Charge, Minsk, Nauka i Tekhnika, 1975, 336 pp
\bibitem{Berezin} Berezin A.V, Kurochkin Yu.A, Tolkachev Ye.A, Quaternions in Relativistic Physics, Minsk, Nauka i Tekhnika, 1989, 199 pp
\bibitem{Yearchuck_Doklady} Yearchuck D, Yerchak Y, Red'kov V, Doklady NANB,\textbf{51} N 5 (2007) 57-64
\bibitem{Yearchuck_Yerchak} Yearchuck D, Yerchak Y, Kirilenko A, Popechits V, Doklady NANB, \textbf{52}, N 1 (2008) 48-53
\bibitem{Yearchuck_PL} Yearchuck D, Yerchak Y, Alexandrov A, Phys.Lett.A, \textbf{373} (2009) 489-495
\bibitem{D_Yearchuck_A_Dovlatova} D.Yearchuck, Y.Yerchak, A.Dovlatova,  Optics Communications, \textbf{283} (2010) 3448-3458
\bibitem{Akhiezer} Akhiezer A.I, Berestetskii V.B, Quantum Electrodynamics, M., Nauka, 1969, 623 pp 
\bibitem{Bateman} Bateman H, Proc.London Math.Soc., \textbf{8} (1909)  223—264  
\bibitem{Ibragimov} Ibragimov N, Group Properties of Some  Differential Equations, Novosibirsk, Nauka, i967, 59 pp
\bibitem{Fushchich} Fushchich V I, Doklady AS USSR,  \textbf{246} (1979)  846—850
\bibitem{Yearchuck_Alexandrov_Dovlatova} D.Yearchuck, A.Alexandrov, A.Dovlatova,  Applied Mathematical and Computational Sciences, \textbf{3}, N 2 (2011) 169-200
\bibitem{Dovlatova_Yerchuck} Alla Dovlatova and Dmitri Yerchuck, Journal of Physics: Conference Series, \textbf{343}, 012133, DOI:10.1088/1742-6596/343/1/012133 
\bibitem{Planck} Planck M, Ann.d.Phys., \textbf{4}  (1901) 553 
\bibitem{Einstein} Einstein A, Ann.d.Phys., \textbf{17}, (1905) 132 
\bibitem{Broglie} De Broglie L.V., Comp.Rend., \textbf{177} (1923) 507 
\bibitem{Lewis} Lewis G.N, Nature, \textbf{118}(1926) 874 (1926)
\bibitem{Speziali} Speziali P, Ed.Albert Einstein-Michele Besso Correspondence (1903-1955), Herman,
Paris p.453 (1972)
\bibitem{Landau} Landau L.D, Lifshitz E.M, Physikalische Zeitschrift der Sowjet Union, \textbf{8} (1935) 153-169
\bibitem{Abragam} A.Abragam, Nuclear Magnetism, M., Iz.In.Lit., 1963, 553 pp
\bibitem{Bloch} F.Bloch, Phys.Rev., \textbf{70}, N 7, 8 (1946) 460-474
\bibitem{Dirac1928} Dirac P.A M, Proceedings of the Royal Society, \textbf{117A} (1928) 610-624
\bibitem{Ertchak_J_Physics_Condensed_Matter} Ertchak D.P, Kudryavtsev Yu.P, Guseva M.B, Alexandrov A.F et al, J.Physics: Condensed Matter, \textbf{11}, N3 (1999) 855-870 
\bibitem{Heimann} Carbyne and Carbynoid Structures,  Dordrecht, Boston, London, Edited by Heimann R.B, Evsyukov S.E, Kavan L, 1999, 446 pp
\bibitem{Yerchuck_Dovlatova_Alexandrov} Yerchuck D, Dovlatova A, Alexandrov A, in press
\bibitem{A_Dovlatova_D_Yerchuck} Yerchuck D, Dovlatova A, Alexandrov A, in press
\bibitem{Winter_Kuzmany} Winter J, Kuzmany H, Phys.Rev.B, 53, N 2 (1996 -II) 655-661
\bibitem{Ertchak_Carbyne_and_Carbynoid_Structures} Ertchak D.P, In "Carbyne and Carbynoid Structures", pp 357-369, Dordrecht, Boston, London, Edited by Heimann R.B, Evsyukov S.E, Kavan L, 1999, 446 pp
\bibitem{Dirac P.A M} Dirac P.A M, Proc.Roy.Soc.A, 133 (1931) 60 - 72
\bibitem{Ph.Enc.} Physical Encyclopedia,(ed.in-Chief Prokchorov A.M) v.5, pp 540 - 545, M., 1998, 688 
\bibitem{Burinskii}  Burinskii A, The VIIth International Conference Quantum Theory and Symmetries, August 7-13, 2011, Prague, Abstracts, p.24
\bibitem{Tolkachev} Tolkachev E.A, Tomilchick L.M, Covariant Methods in Theoretical Physics, Minsk, 1981, pp 44-48
\bibitem{Landau_Lifshitz_Field_Theory} Landau L.D, Lifshitz E.M, Field Theory, M., Nauka, 504 pp
\bibitem{Stephani} Stephani H, Relativity. An Introduction to Special and General 
Relativity, 2004, Cambridge University Press, 395 pp
\bibitem{Rashevskii} Rashevskii P K, Riemann Geometry and Tensor Analysis, 2003, M., Editorial URSS, 664 pp
\bibitem {Bohm} Bohm A, Quantum Mechanics: Foundations and Applications, Springer-Verlag, N.-Y., Berlin, Heidelberg, Tokyo, 1986, M., Mir, 1990, 720 pp
\bibitem{Noether}  N\"{o}ther E, Nachrichten von der Gesellschaft der Wissenschaften zu G\"{o}ttingen, Math-phys.Klasse, (1918) 235–257
\bibitem{Bogush} Bogush A A, Moroz L G, Introduction to the Theory of Classical Fields, M., Editorial URSS, 2004, 384 pp
\bibitem{Fedorov} Fedorov F.I, Lorentz Group, M., Nauka, 1979, 384 pp
\bibitem{Slepyan_Yerchak} Slepyan G.Ya, Yerchak Y.D, Hoffmann A,  Bass F.G, Phys.Rev.B, \textbf{81} (2010) 085115 
\bibitem{Angot} Andre Angot, Complements de Mathematiques, Paris, 1957, 778 pp
\bibitem{Baryshevsky} Baryshevsky V.G, arXiv:hep-ph/9912438v3 23 Feb 2000
\bibitem{Schmidt} Thomas L.Schmidt, Adilet Imambekov and Leonid I.Glazman, Phys.Rev.B, \textbf{82} (2010) 245104 
\bibitem{Scully} Scully M O, Zubairy M S, Quantum Optics, Cambridge University Press, 1997, 650 pp
\bibitem{Born} Born M, Jordan P, Zeitschrift f\"{u}r Physik,  \textbf{34} (1925) 858-888
\bibitem{Born_Heisenberg} Born M, Heisenberg W, Jordan P, Zeitschrift f\"{u}r Physik,
\textbf{35} (1926) 557-606
\bibitem{P.Dirac} Dirac P.AM,  Proc.Roy.Soc.,A, \textbf{114} (1927) 243-265, 
\bibitem{Galapon} Galapon E, The VIIth Int.Conf. "Quantum Theory and Symmetries", Abstracts, p.51, August 7-13, 2011, Prague
\bibitem{Pauli} Pauli W, Handbuch der Physik (ed.H.Geiger and K.Scheel), 1st edn, vol.23, Springer, 1926, pp.1–278
\bibitem{Galapon_E} Galapon E, Proc.Roy.Soc.,A \textbf{458} (2002)  451-472
\bibitem{Shirkov} Bogoliubov N N, Shirkov D V, Introduction in the theory
of quantized fields, M., Nauka, 1973, 414 pp
\bibitem{Dutra} Dutra S M, Cavity Quantum Electrodynamics, John Wiley and Sons, Inc., Hoboken, New Jersey, 2005, 389 pp
\bibitem{Poole} Poole Charles P, Jr., Technique of EPR Spectroscopy, M., Mir, 1970, 557 pp   
\bibitem {Dovlatova_Yearchuck} Dovlatova A, Yearchuck D, Chem.Phys.Lett., \textbf{511} (2011) 151-155 
\bibitem{Yerchuck_Dovlatova} Yearchuck D, Dovlatova A, Journal of Physical Chemistry C, \textbf{116}, N 1 (2012) 63-80, DOI:10.1021/jp205549b
\bibitem{Kramers} Kramers H A, Quantum Mechanics, North-Holland, Amsterdam, 1958 
\bibitem{Power} Power E A, Introductory Quantum Electrodynamics, Longman, London, 1964 
\bibitem{Dovlatova_Yerchuck_Borovik} Dovlatova A, Yerchuck D, Borovik F, in press 
\bibitem{Luther} Emery V J and Luther A, Phys.Rev.Lett., \textbf{33} (1974) 589
\bibitem {Jackiw} Jackiw R and Rebbi C, Phys.Rev.D \textbf{13} (1976) 3398 
\bibitem {SSH} Su W.P., Schrieffer J.R, and Heeger A.J, Phys.Rev.Lett., \textbf{42} 1698 (1979)
\bibitem {SSH_PRB} Su W.P., Schrieffer J.R, and Heeger A.J, Phys.Rev.B, \textbf{22}, (1980) 2099 
\bibitem{Anderson} P. W. Anderson, Science 235, 1196 (1987)
\bibitem{Anderson_P_W} Anderson P W, In: Frontiers and Borderlines in Many Particle Physics, edited by P.A. Broglia and J.R. Schrieffer, North-Holland, Amsterdam, 1987 
\bibitem{Berestecky} Berestecky V.B, Lifshitz E.M, Pitaevsky L.P, Quantum Electrodynamics, M., Nauka, 1989, 728 pp
\end{thebibliography}
\end{document}